\documentclass[review,sort&compress]{elsarticle}

\journal{Acta Materialia}

\usepackage[english]{babel}
\usepackage[utf8x]{inputenc}
\usepackage{amsmath}
\usepackage{graphicx}
\usepackage{caption}
\usepackage{subcaption}
\usepackage{mathtools}
\usepackage{amsfonts}
\usepackage{amsthm}
\usepackage[final]{listings}
\usepackage{float}

\usepackage{tikz}
\usepackage{pgfplots}
\usetikzlibrary{calc}
\usetikzlibrary{backgrounds}
\usepgflibrary{shapes}
\usetikzlibrary{through}
\usetikzlibrary{intersections}
\usepackage{placeins}

\usepackage[hyphens]{url}

\usepackage{xifthen}

\usepackage{hyperref}
\usepackage{esdiff}
\usepackage{cleveref}
\usepackage{ifthen}

\usepackage{esdiff}

\usepackage{standalone}
\usepackage{siunitx}
\usepackage{nicefrac}

\usepackage[
changebar,
todonotes,
authormarkup=none,
]{changes-pro}

\presetkeys{todonotes}{inline}{}

\renewcommand{\a}{\ensuremath{_{\hat \alpha}}}

\newcommand{\rz}{{\if mm {\rm I}\mkern -3mu{\rm R}\else \leavevmode
\hbox{I}\kern -.17em \hbox{R} \fi}}
\newcommand{\nz}{{\if mm {\rm I}\mkern -3mu{\rm N}\else \leavevmode
\hbox{I}\kern -.17em \hbox{N} \fi}}

\newcommand{\grad}[1]{\ensuremath{\nabla{#1}}}
\renewcommand{\div}[1]{\ensuremath{\nabla\cdot{#1}}}

\newcommand{\vv}[1]{\boldsymbol{#1}}

\newcommand{\vphi}{\ensuremath{\vv{\phi}}}
\newcommand{\vc}{\ensuremath{\vv{c}}}

\newcommand{\vmu}{\ensuremath{\vv\mu}}

\newcommand{\phia}{\ensuremath{\phi_{\hat \alpha}}}

\newcommand{\ha}{\ensuremath{h_{\hat \alpha}}}

\newcommand{\x}{\ensuremath{\vv{\xi}}}

\newcommand{\systemAlCu}{\textsf{Al-Cu}}
\newcommand{\systemAlAgCu}{\textsf{Al-Ag-Cu}}

\newcommand{\pace}{{\textsc{Pace3D}}}
\newcommand{\calphad}{{\textsc{Calphad}}}

\newcommand{\compAl}{\textsf{Al}}
\newcommand{\compCu}{\textsf{Cu}}
\newcommand{\AlCu}{$\theta$}
\newcommand{\AltwoCu}{$\mathrm{Al_2Cu}$}

\newcommand{\Liq}{\textsf{liq}}

\newdimen\CdotAxis
\newcommand*{\CdotAux}[3]{%
  {%
    \settoheight\CdotAxis{$#2\vcenter{}$}%
    \sbox0{%
      \raisebox\CdotAxis{%
        \scalebox{#1}{%
          \raisebox{-\CdotAxis}{%
            $\mathsurround=0pt #2#3$%
          }%
        }%
      }%
    }%
    \dp0=0pt %
    \sbox2{$#2\bullet$}%
    \ifdim\ht2<\ht0 %
      \ht0=\ht2 %
    \fi
    \sbox2{$\mathsurround=0pt #2#3$}%
    \hbox to \wd2{\hss\usebox{0}\hss}%
  }%
}

\def\addlegendimage{\csname pgfplots@addlegendimage\endcsname}

\DeclareSIUnit{\molpc}{mol\text{-}\%}
\newcommand{\atp}{at.\%}

\newcommand{\cor}{\textcolor{black}}

\biboptions{sort&compress}
\usepackage{lineno,hyperref}

\begin{document}

\begin{frontmatter}
\title{Simulation of dendritic-eutectic growth with the phase-field method}


\author[1]{Marco Seiz\corref{cor1}}\ead{marco.seiz@kit.edu}
\author[1]{Michael Kellner}
\author[1,2]{Britta Nestler}
\cortext[cor1]{Corresponding author}
\address[1]{Institute of Applied Materials, Karlsruhe Institute of Technology, Stra\ss{}e am Forum 7, 76131 Karlsruhe, Germany}
\address[2]{Institute of Digital Materials,  Karlsruhe University of Applied Sciences, Moltkestr. 30, 76133 Karlsruhe, Germany}

\date{\today}
\begin{abstract}
Solidification is an important process in many alloy processing routes.
The solidified microstructure of alloys is usually made up of dendrites, eutectics or a combination of both.
The evolving morphologies are largely determined by the solidification process and thus many materials properties are dependent on the processing conditions.
While the growth of either type of microstructure is well-investigated, there is little information on the coupled growth of both microstructures.
This work aims to close this gap by formulating a phase-field model capable of reproducing dendritic, eutectic as well as dendritic-eutectic growth.
Following this, two-dimensional simulations are conducted which show all three types of microstructures depending on the composition and processing conditions.
The effect of the dendritic-eutectic growth on the microstructural lengths, which determine materials properties, is investigated and the morphological hysteresis between eutectic growth and dendritic-eutectic growth is studied by employing solidification velocity jumps.
Further, the influence of primary crystallization is investigated in large-scale two-dimensional simulations.
Finally, qualitative three-dimensional simulations are conducted to test for morphological changes in the eutectic.

\end{abstract}

\end{frontmatter}

\section{Introduction}


Tailoring the properties of materials to suit the intended application is common nowadays.
One part of this tailoring is determining the appropriate phases for the application and thus the chemical composition of the chosen material.
The other part is achieving a microstructure which further enhances the desired properties.
The microstructure depends on the material and its composition as well as on the processing conditions.
Predicting the microstructure by the given material information and processing conditions can be done theoretically as well as numerically.
In the context of solidification, analytical theories provide solutions for e.g. isolated dendrite growth \cite{Langer1977,Lipton1987b,Pelce1987}, arrays of dendrites \cite{Kurz1986,Gandin1996,Spencer1997} or eutectic growth \cite{Jackson1966,Hunt1970,Coriell1987,Himemiya1999}.
However, these theories have limits of applicability, like rapid solidification or coupled growth between dendrites and eutectics.
Numerical investigations such as simulations do not necessarily share these limitations.
The most prominent simulation method for solidification is the phase-field method.
While the phase-field method has been shown to correctly reproduce both dendrite growth \cite{Karma1998,Karma1999,Choudhury2012_2,Cogswell2015, Pineau2018Gandin} as well as eutectic growth \cite{Lewis2004,Akamatsu2015,Nestler2011,Hoetzer2016}, simulations combining both at the same time are few \cite{Eiken2015,Wang2021b,Gu2022}.
The focus of this work is to simulate the coupled growth of both types of microstructures with the phase-field method.
For this purpose, the material system \systemAlCu{} is employed, as many experiments~\cite{Jordan1971,Jordan1971a,Cadirli2013,Liotti2016} as well as simulations~\cite{Choudhury2012,Ferreira2017,Zhang2019} have been conducted investigating the microstructure formation.
In these works both type of microstructures are observed to evolve separately, thus the system \systemAlCu{} is predestined for the investigation of their combined growth.
Jordan and Hunt~\cite{Jordan1971} for example studied in their experimental work the growth of dendrites within an eutectic structures with off-eutectic composition by increasing the solidification velocity.


The paper is structured as follows:
First, approximate theories relating the growth conditions to the front undercooling of dendrites and eutectics are determined, based on literature~\cite{Jackson1966,Burden1974,Kurz1981}.
These will allow the calculation of boundary curves between purely eutectic microstructures and those with a mix of eutectic and dendrites.
Next, the employed phase-field method will be detailed including an empirical nucleation mechanism, followed by the thermodynamic description of the system.
The nucleation mechanism is then validated by evaluating solid fractions in an isolated domain as well as the results of eutectic solidification with and without nucleation.
In the final section, two-dimensional simulations of coupled dendritic-eutectic growth are performed for various directional growth conditions:
These include variations in the growth velocities (constant and abrupt changes), temperature gradients and melt compositions.
Additionally, the time spent in primary crystallization is varied. 
Finally, three-dimensional simulations of dendritic-eutectic growth are conducted in order to test for morphological changes in the eutectic.



\section{Theory}
\subsection{Microstructural evolution}
\label{sec:theory}

A qualitative, theoretical model for the necessary conditions separating coupled dendritic-eutectic growth from eutectic growth is developed in this section.
The purpose of this model is to give accurate predictions once simulative or experimental data of morphological operating points ($c_0, \Delta T, v, G, \ldots$) is known.
Thus it is not formulated in terms of materials properties such as surface energy, but rather with general parameters which are determined from these data.
Key to the separation of the morphologies is the determination of the undercooling of both morphologies, as it is assumed  that the morphology with the highest temperature is dominant.
Furthermore, the undercooling models will allow for testing whether the coupled growth changed the growth conditions of the individual morphologies.

The dendritic front temperature $T_{df}$, inspired by \cite{Burden1974} and \cite{Kurz1981}, is modelled as
\begin{align}
T_{df} &= T_l(c_0) - \Delta T_d\\
\Delta T_d &= A\frac{G}{v} + B(c_0v)^{0.5} + Cc_0 \label{eq:ucD}
\end{align}
with the liquidus temperature $T_l(c_0)$, the temperature gradient $G$, the front velocity $v$, the concentration of solute in the melt $c_0$ and the material dependent constants $A$, $B$ and $C$.
The melt concentration dependence is usually contained within the constant $B$ as well as the liquidus temperature via linear phase diagram approximation \cite{Burden1974,Kurz1981}.
The inclusion of melt concentration $c_0$ in this form is motivated by the undercooling expressions in \cite{Kurz1981} and significantly improves the fit to simulation data as will be shown later.
For completeness, the model without including the melt concentration $c_0$ is written as $ \Delta T_d = A\frac{G}{v} + Bv^{0.5}$.
%

For eutectics, $\Delta T_e = Ev^{0.5}$ is assumed to estimate the eutectic front undercooling $T_{ef} = T_e - \Delta T_e$, again with a materials dependent constant $E$.
This is motivated by the general scaling law discovered by Jackson and Hunt\cite{Jackson1966} 
\begin{align}
 \Delta T_e &= K_1 \lambda v + \frac{K_2}{\lambda} \label{eq:ucE}
\end{align}
with material constants $K_1, K_2$ and the wavelength $\lambda$.
This describes experimental observations well if the minimal undercooling is assumed to describe the dominant eutectic wavelength $\lambda_{JH}$.
Employing this assumption and doing a bit of algebra yields $E=2\sqrt{K_1 K_2}$.
Furthermore, the eutectic growth constant is then given by $\lambda^2 v = \frac{K_2}{K_1}$.
$E$ will not be directly fitted, but rather $K_1$ and $K_2$ in \cref{eq:ucE}.
This allows the inclusion of simulations not growing at the optimal lamellar spacing for the determination of the constants.
Strictly speaking the constants $K_1$ and $K_2$ also depend on the melt concentration via the phase fractions.
In Jackson and Hunt's paper\cite{Jackson1966} the constants end up affine and nonlinearly dependent on the melt concentration, which makes it harder to include than for the dendritic undercooling.
Hence the constants $K_1, K_2$ are determined for various off-eutectic compositions and fitted to functions of $c_0$.

The boundary curve separating coupled dendritic-eutectic growth from eutectic growth is then described by
\begin{align}
 T_d - A\frac{G}{v} + B(c_0v)^{0.5} + Cc_0 &= T_e - E(c_0)v^{0.5}
\end{align}
which will be solved numerically in a later section after the constants have been determined.


\subsection{Phase-field model}
A thermodynamically consistent phase-field model, based on a grand potential functional and an Allen-Cahn-type variation, is used \cite{Plapp2011,Choudhury2012_2,Hoetzer2015}.
The $N=4$ order parameters $\phi_{\hat \alpha}$, describe the local volume fractions of two $\alpha$-Al phases, the $\theta$-\AltwoCu{} phase and the liquid $l$ melt.
Two different order parameters are introduced for the $\alpha$ phase in order to distinguish an isotropic and anisotropic variant.
To differentiate the phases  $\alpha$ and $\theta$ from their indices, the indices are represented by $\hat\alpha$ and $\hat\beta$.
The chemical potential vector~$\vv \mu$ consists of a parameter $\mu_\mathrm{i}$ for each component (i=\compAl, \compCu) and is derived from the mass balance of the ~$K=2$ concentrations and from Fick's law.
The coupling of the~$N$ phase fields, the~$K$ chemical potentials and the imprinted temperature~$T$ results in the following set of evolution equations:
 \begin{align}
\tau(\vv\phi,\nabla\vv\phi) \varepsilon \diffp{\phia}{t}=&
-\underbrace{
\varepsilon \left(\diffp{a(\vv\phi,\nabla\vv\phi)}{\phia} - \nabla \cdot \frac{\partial a(\vv\phi,\nabla\vv\phi)}{\partial\nabla\phia} \right)-\frac{1}{\varepsilon}\diffp{\omega(\vv\phi)}{\phia}
}_{:=rhs_{1,\hat \alpha}}\nonumber\\  
& -\underbrace{
\diffp{\psi(\vv\phi,\vv\mu,T)}{\phia} + \xi_\alpha
}_{:=rhs_{2,\hat \alpha}}  
- \underbrace{
\frac{1}{N}\sum^N_{\hat \beta=1} (rhs_{1,\hat \beta} + rhs_{2,\hat \beta})
}_{:=\Lambda}\,,
\label{eq:phi_evolution} \\
 \diffp{\vmu}{t}=&\quad\left[ \sum_{\hat\alpha=1}^N \ha(\vphi) \left(\frac{\partial \vc_{\hat\alpha}(\vmu, T)}{\partial \vmu} \right)\right]^{-1} 
\nonumber \\
&
\Biggl( \div{} \Big(\vv{M}(\vphi,\vmu,T)\grad{\vmu} - \vv{J}_{at}(\vphi,\vmu,T) \Big) \nonumber\\
 & -\sum_{\hat \alpha=1}^N \vc\a(\vmu,T) \frac{\partial \ha(\vv\phi)}{\partial t} 
     -\sum_{\hat \alpha=1}^N 
  \ha(\vphi) \left(\frac{\partial \vc\a(\vmu, T)}{\partial T} \right) \diffp{T}{t} \Biggr),
\label{eq:chem_pot_evolution}\\
 \diffp{T}{t} =&\quad \diffp{}{t} \left(T_0 + G(y - vt)\right) =-Gv\label{eq:t_evolution}.
\end{align}
The interested reader is referred to \cite{Choudhury2012_2,Hoetzer2015} for a complete description of the model.
Here only the pertinent parameters will be explained:
The relaxation parameter $\tau$ and the gradient energy $a$ are modelled as isotropic or anisotropic, depending on the phase:
\begin{align}
 \tau(q_{\hat \alpha\hat \beta}) &= \sum_{\hat \alpha<\hat \beta}^N A_{\hat\alpha\hat\beta}^\tau(q_{\hat \alpha\hat \beta}) \tau_{\hat\alpha\hat\beta} \\
 a(q_{\hat \alpha\hat \beta}) &= \sum_{\hat \alpha<\hat \beta}\gamma_{\hat \alpha\hat \beta} \left(A_{\hat\alpha\hat\beta}^{\gamma} ( q_{\hat \alpha\hat \beta}  ) \right)^2 \left| q_{\hat \alpha\hat \beta} \right|^2\\
 A_{\hat\alpha\hat\beta}^\tau &= A_{\hat\alpha\hat\beta}^{\gamma},
\end{align}
with the interface orientation given by the generalized gradient vector $q_{\hat \alpha\hat \beta} =  \phi_{\hat \alpha}\nabla\phi_{\hat \beta}-\phi_{\hat\beta}\nabla\phi_{\hat \alpha}$.

One $\alpha$-Al variant will be modelled with a four-fold anisotropy w.r.t the liquid phase, yielding dendritic morphologies, with the remaining phases modelled as isotropic, employing the following (an)isotropy functions
\begin{align}
A_{iso}^{\tau,\gamma}( q_{\hat \alpha\hat \beta}) &= 1 \label{eq:isotropic}\\
A_{four}^{\tau,\gamma} ( q_{\hat \alpha\hat \beta})&=1-\zeta_{\hat\alpha\hat\beta}\left(3-4\frac{\left|q_{\hat \alpha\hat \beta}\right|^4_4}{\left|q_{\hat \alpha\hat \beta}\right|^4}\right)
\label{eq:fourfold}
\end{align}
and the definitions $\left|v\right|^4_4 = \sum_i v_i^4$ and  $\left|v\right|^4 = (\sum v_i^2)^2$ \cite{Nestler2005} with the index $i$ running over the spatial dimensions.
The parameter $\zeta_{\hat \alpha\hat \beta}$ describes the strength of the anisotropy.
Furthermore, a stochastic noise term $\xi_\alpha$ following \cite{Karma1999} is added to the phase-field equation~\cref{eq:phi_evolution} in order to enhance dendritic side branching.


The driving force for the phase transitions is described by the differences of the grand potentials $\psi_{\hat\beta}$, which are stored in the grand potential vector $\psi$.
The grand potentials are derived from the Gibbs energies of the different phases~\cite{Kellner2017}, which are obtained from the thermodynamic \calphad{} database of Witusiewicz et. al~\cite{Witusiewicz2004} for the ternary system \systemAlAgCu{}.
To reduce the computational effort, the Gibbs energies are approximated by a parabolic approach of the form~\cite{Kellner2022}:
\begin{align}
 g_{\hat\alpha}(\vv c,T)&=\sum_{i=1}^{K-1}\sum_{\substack{j=1 \\ i \leq j}}^{K-1} A^{ij}_{\hat\alpha}(T)\,\prescript{i}{}c\prescript{j}{}c + \sum_{l=1}^{K-1}B^l_{\hat\alpha}(T)\, \prescript{l}{}c +C_{\hat\alpha}(T)\,.\label{eq:Gibbs_parabolic}
\end{align}


The present phase-field model employs an obstacle potential type, yielding a diffuse interface outside of which the phase-fields take on the values of either 0 or 1.
This allows the skipping of phase-field calculations in these so-called bulk regions, but also precludes using the phase-field noise $\xi_\alpha$ as a way to include homogeneous nucleation.
While the phase-field noise within the interface could lead to heterogeneous nucleation, the higher order terms in $\omega(\vv\phi)$ which remove third-phase contributions from two-phase interfaces\cite{Nestler2005} will remove the newly nucleated phase-fields quickly.
Thus in order to enable the evolution of new phases within the simulation, an explicit nucleation mechanism is implemented into the phase-field model.
The goal of this mechanism is to allow the system to pick the evolutionary favorable phases and morphologies without affecting the operating point in steady state of both dendrites and eutectics.
Each cell containing a liquid interface is assumed to be capable of nucleating phases which it does not already contain.
If the nucleation potential of a randomly picked phase is above a threshold, the liquid phase-field is recolored to the nucleated phase, with the remaining phase-fields being held constant.
This is accompanied by a jump in chemical potential, as the concentration must be held constant during this transformation.
This mechanism is applied on the entire domain at a set interval of time steps, as to allow the system to relax between nucleations.
The interval is chosen as $n = \frac{W}{v\Delta t}$, i.e. the number of time steps after which the front has moved an interface width $W$.
The velocity can either be estimated in-situ or is directly given by \cref{eq:t_evolution}.

\begin{figure}
 \includegraphics[width=\textwidth]{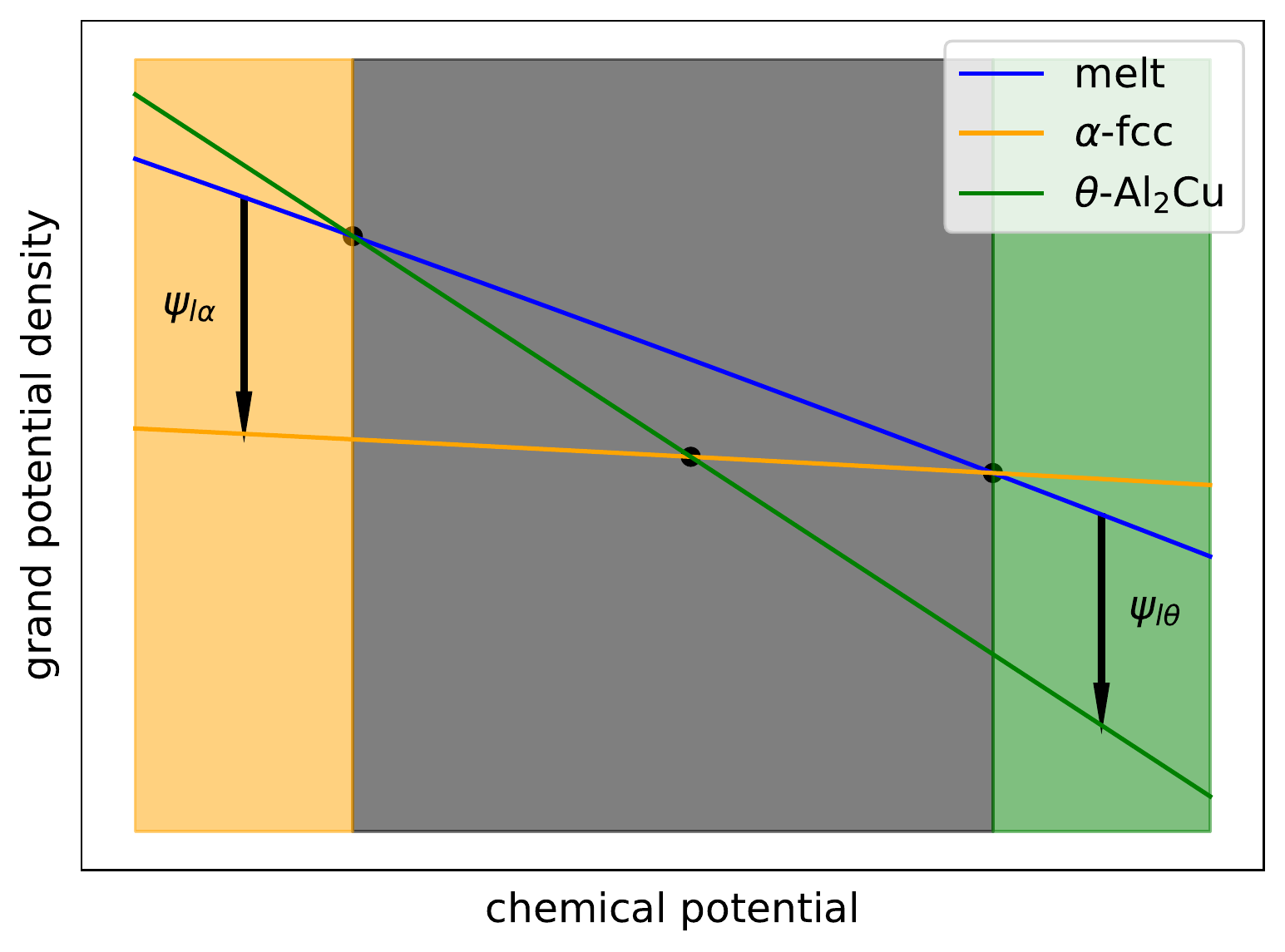}
 \caption{The grand potentials of the phases over the chemical potential for a constant temperature are depicted.
 The shaded grey region in the center describes the space in which eutectic growth is possible, with the colored shaded regions indicating where nucleation of the respectively colored phase is possible.
 The driving force for nucleation of either phase is depicted by arrows for two chosen chemical potentials.}
 \label{fig:nuc-expl}
\end{figure}

The nucleation potential for a liquid interface not containing the phase $\alpha$ is written as
\begin{align}
 \psi_{l\alpha}(\mu, T) &= (\psi_l(\mu, T) - \psi_\alpha(\mu, T))h_l(\phi)
\end{align}
i.e. simply the difference between the two grand potentials at the local chemical potential $\mu$, representing the bulk driving force interpolated with the transformed liquid volume via the weighting function $h_l(\phi)$.
In order for $\alpha$ to nucleate on this interface, it should have a driving force capable of growth which exceeds a threshold value, i.e.
\begin{align}
 \psi_{l\alpha}(\mu,T) &> \psi_{barrier}(\phi, \mu, T)
\end{align}
where an additional nucleation barrier $\psi_{barrier}(\phi, \mu, T)$ is introduced.
Since nucleation in the interface is considered and the scale of the simulations is far above that of classical nucleation theory, the nucleation barrier is determined in an ad-hoc manner suited to eutectic solidification:
If a eutectic structure is advancing sufficiently far from its optimal spacing, its constituent phases will tend to oscillate and exhibit concave regions along the front.
In these concave regions an excess of insoluble components will tend to accumulate, i.e. in front of an Al-rich $\alpha$ crystal, Cu in excess of the equilibrium melt composition will accumulate.
This can eventually lead to stagnant or even melting interface.
Nucleating a phase capable of dissolving these components in these regions would prevent this and allow the re-establishment of a convex front, thereby possibly reducing the grand potential of the system.
Thus the state in which a solid-liquid interface begins to melt is assumed to describe when a new phase can be nucleated.
This state is approximated by the equilibrium chemical potential of the present interface, with the associated barrier being the nucleation potential at this chemical potential.
\Cref{fig:nuc-expl} shows this in more detail with sketch of the grand potentials at constant temperature and the relevant regions:
The equilibrium points are marked by black dots and their bounding polygon (grey) describes the space in which eutectic growth is possible.
Outside of this region, one of the solid phases begins to melt, corresponding to moving across its liquidus line in the phase diagram.
But since the temperature is below the eutectic temperature, the liquid phase should be unstable w.r.t a combination of both solid phases.
Hence in these regions the opposite phase is allowed to nucleate, given that it has a driving force ($\psi_{l\alpha}(\mu, T) > 0$) for growth w.r.t the liquid phase.
If the latter condition were not enforced, nucleation would also happen when it would increase the grand potential.
It is also tacitly assumed that the nucleation barrier due to surface energy is reduced to zero for this case of heterogeneous nucleation.
Since it is not included, phases can nucleate and then die off due to surface energy.
Thus an improvement in the model might be adding this to the nucleation barrier while at the same time including the induced change in the equilibrium chemical potential.

In full, the nucleation condition for a phase $\alpha$ on a $l\beta$ interface then reads
\begin{align}
 \psi_{l\alpha}(\mu, T) &> 0 \\
 \psi_{l\alpha}(\mu, T) &> \psi_{l\alpha}(\mu_{eq,l\beta}(T), T)
\end{align}
with the chemical potential in equilibrium $\mu_{eq,l\beta}(T)$. 

The approach is similar in spirit to that of \cite{Kellner2020}, as it was developed in tandem with their work.
The key difference is the usage of driving forces for determining when to nucleate phases instead of employing concentration differences.
This trivially includes a dependence on temperature which was missing in \cite{Kellner2020}.
It also leaves no open parameters for the nucleation barrier as this is entirely determined by the energetics of the system.
Thus the mechanism requires no knowledge of the phase diagram shape, only of the grand potentials which are already necessary for the phase-field simulations.
The mechanism is also extendable to homogeneous nucleation in which case classical nucleation theory provides information about the nucleation rate and nucleation barrier, but this is not considered in the present paper.


\paragraph{Computational aspects}
All simulations are conducted in the massively parallel phase-field solver \pace{} \cite{Hoetzer2018}.
The time derivative is resolved with an explicit first order Euler scheme and spatial derivatives with second order finite differences.
The time step width is chosen based on a von Neumann analysis of the equations in order to keep the explicit time integration stable.
The parallelization is done with MPI.
The HAWK supercomputer is used for the majority of the simulations, with the employed core counts ranging from 128 to 2048, depending on the computational requirements.
The runtime of individual simulations ranged from a few hours to about a week for the slowest solidification conditions and largest domains.
The eutectic validation studies were calculated on a local machine on up to three cores for up to two days.
Within most of the simulations a moving window technique is employed in order to allow for a quasi-infinite domain without excessively huge computational domains.
This is achieved by regular checks on the position of the interface.
If the interface is above a certain point, henceforth called the moving window cutoff, all fields are shifted below this cutoff.
Since only integer shifts are employed, no interpolation between positions is necessary and simple copy operations can be employed to implement the field shift.
With this one can ensure a minimum distance between the solidification front and the boundary of the domain.
Generally this distance is set to be at least 5 diffusion lengths $l_d = \frac{D}{v}$ such that the concentration far field is not dominated by the boundary condition but rather behaves as in an infinite melt.



\section{Parametrization}
\label{sec:fitting}

The coupled growth of eutectic and dendritic structures is simulated in this work for the binary material system \systemAlCu{}.
In order to approximate this material system in the phase-field simulations, the energies describing the material system are approximated based on the thermodynamic \calphad{} database from Witusiewicz et al.~\cite{Witusiewicz2004} and by using the parabolic approach described in~\cref{eq:Gibbs_parabolic}.
The input data includes both Gibbs free energy and chemical potential values as well as phase equilibrium points, both determined via \calphad{}, resulting in a procedure similar to \cite{Seiz2021}.
All concentrations employed are in atomic fraction or equivalently mole fraction of copper, with the assumption of equal molar volumes.
The following equations give the resulting functions with 8 significant digits in dimensionless units:


\begin{align}
g_\alpha(c,T)  &=  \left(147.73532 T - 128.37484\right) c^{2} \nonumber\\
          &+ \left(3.5000629 T - 53.205937\right) c \nonumber\\
          &- 57.867925 T + 27.198937  \label{eq:G_alpha}\\
g_\theta(c,T)  &= \left(294.11794 T - 254.29651\right) c^{2}  \nonumber\\
          &- \left(170.96673 T - 96.996795 \right) c  \nonumber\\
          &- 28.930239 T +  2.260627\label{eq:G_delta} \\
g_l(c,T)  &= \left(21.442726 T - 17.807343\right) c^{2}  \nonumber\\
          &+ \left(5.587987 T - 55.592733\right) c  \nonumber\\
          &- 58.655641 T +  28.085635\label{eq:G_liq}
\end{align}

\Cref{tab:Eutectic_reaction} shows the temperatures and equilibrium concentrations of the eutectic reaction for the system from~\cite{Witusiewicz2004} and from the approximated system, respectively.
\begin{table}[h]
\begin{center}
\caption{Temperatures and equilibrium concentrations of the eutectic reaction $\Liq \rightleftharpoons \alpha + \theta$ for the binary \systemAlCu{} system from~\cite{Witusiewicz2004} and from the approximated system}\label{tab:Eutectic_reaction}
\small
\begin{tabular}{|l||c|c|c|c|}
\hline
       & $T_{\textrm{e}}$ & $c_{\textrm{eq.}}$ of $\alpha{}$ & $c_{\textrm{eq.}}$ of \AlCu{} & $c_{\textrm{eq.}}$ of $\Liq{}$ \\
 & in K &  in \atp{}~\compCu{} &  in \atp{}~\compCu{} &  in \atp{}~\compCu{} \\
\hline
\calphad{} PD \cite{Witusiewicz2004} & 820  & 2.54   & 31.8  & 17.5      \\
reconstructed PD  & 816  & 2.59   & 31.8  & 18.1      \\
\hline
\end{tabular}
\end{center}
\end{table}

\begin{figure}[h]
\centering
 \includegraphics[scale=1.0, width=\linewidth]{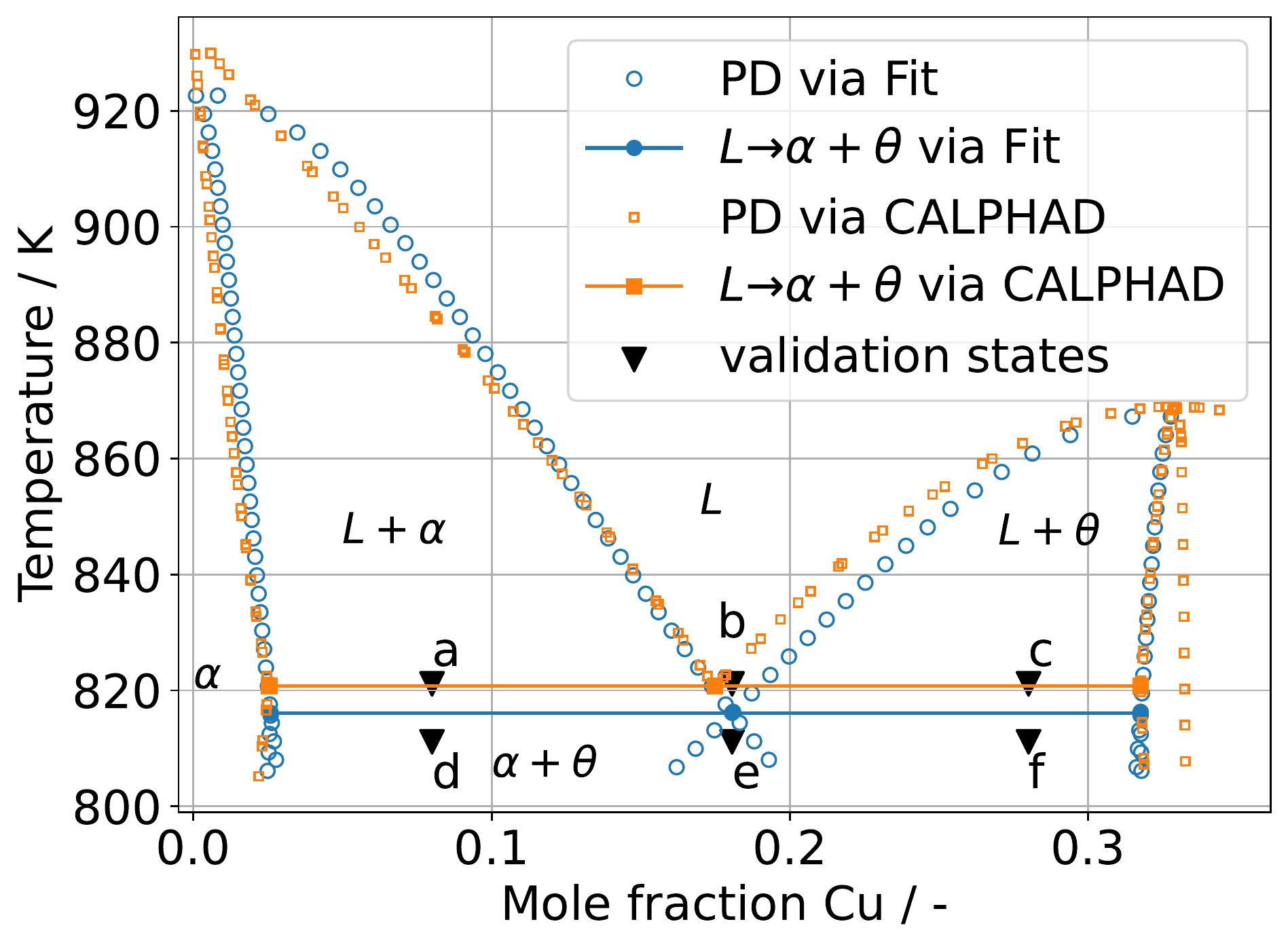}
\caption{Al-rich side of the \systemAlCu{} phase diagram, calculated via CALPHAD based on \cite{Witusiewicz2004} as well as by the fitted free energies.
\cor{For the CALPHAD calculation, only the $\alpha$-Al, $\theta$-\AltwoCu{} and liquid phases are considered.}
The fitted free energies show good accordance given the large temperature range.
The states which will be investigated as part of the validation are marked by the black triangles (a-f).
}
\label{fig:PD_AlCu}
\end{figure}

\Cref{fig:PD_AlCu} shows the Al-rich side of the \systemAlCu{} phase-diagram calculated from~\cite{Witusiewicz2004} (orange) compared with the reconstructed phase-diagram derived from the approximated Gibbs energies of~\cref{eq:G_alpha,eq:G_delta,eq:G_liq} (blue).
Excepting conditions close to the melting point of $\alpha$-Al, good accordance of the phase-transition lines as well as of the position of the eutectic reaction can be found.

The employed nondimensionalization parameters are listed in \cref{tab:nondim} and the remaining physical parameters in \cref{tab:params}.
These are generally based on literature values for \systemAlCu{}, except the surface energy, which was chosen much larger in order to allow for high driving forces without suffering from a mushy interface\cite{Fleck2022}.

\begin{table}[h]
\centering
\caption{nondimensionalization parameters}
\label{tab:nondim}
 \begin{tabular}{ll}
scale   & value \\
length &  $\SI{1e-7}{m}$  \\
time & $\SI{5e-6}{s}$ \\
diffusivity & $\SI{2e-9}{m^2/s}$\\
velocity & $\SI{0.02}{m/s}$\\
temperature & $\SI{820}{K}$ \\
energy density & \SI{1e8}{J/m^3} \\
surface energy & \SI{1e1}{J/m^2} \\
molar volume & \SI{1e-5}{m^3/mol} \\
\end{tabular}

\end{table}
\begin{table}[h]
\centering
\caption{Employed physical and numerical parameters for the simulations.}
\label{tab:params}
 \begin{tabular}{lll}

parameter   & simulation value & physical value \\

\multicolumn{3}{c}{\textit{Numerical parameters}}\\
grid spacing $\Delta x$   &    1  &   $\SI{1e-7}{m}$  \\
time step $\Delta t$   &    0.125  & $\SI{0.625e-6}{s}$  \\
interface parameter $\epsilon$   &    $3\Delta x$  &   $\SI{3e-7}{m}$  \\
interface width $W$   &    $7.5\Delta x$  &   $\SI{7.5e-7}{m}$  \\
\multicolumn{3}{c}{\textit{Physical parameters}}\\
surface energy $\gamma_{\alpha\beta}$     & 0.08 & $\SI{0.8}{J/m^2}$ \\
diffusivity in the melt      & 1               & $\SI{2e-9}{m^2/s}$ \\
diffusivity in the solids           & $\num{1e-3}$               & $\SI{2e-12}{m^2/s}$ \\
kinetic coefficient $\tau_{\alpha l}$       & 0.138 & $\SI{6.92e8}{Js/m^4}$ \\
kinetic coefficient $\tau_{\theta l}$       & 0.0968 & $\SI{4.84e8}{Js/m^4}$ \\
kinetic coefficient $\tau_{\alpha \theta}$       & 0.417 & $\SI{2.08e9}{Js/m^4}$ \\
anisotropy strength $\zeta$ & 0.04 & 0.04 \\
\end{tabular}

\end{table}

\FloatBarrier{}


\section{Validation}
Before simulating the combined growth of eutectic and dendritic structures within a single phase-field simulation, \cor{the implementation is tested, followed by the individual simulation of the processes in order to validate their separate growth.}

\paragraph{Implementation test}
\cor{The implementation of the phase-field model is qualitatively validated by sampling the test states from \cref{fig:PD_AlCu} and observing the resulting microstructure.
The phase fractions given by the phase diagram should be approximated, given that the phases are present in the domain.
In order to ensure the latter, nucleation is allowed.
Furthermore, the morphology of the microstructure should depend on the (an)isotropy of the phases:
States within a solid-liquid region of the phase diagram should yield dendrites with anisotropy, but seaweed without.
Below the eutectic temperature, both solid phases should be present, with a well-developed eutectic at the eutectic composition, but only second-phase linings in the channels between the primary phases for larger deviations from the eutectic composition.}

\cor{The simulation starts with a seed of either anisotropic $\alpha$ or isotropic $\theta$, depending on which side of the eutectic composition the point lies.
For the eutectic composition an anisotropic $\alpha$ seed is set.
The initial seed concentration is set to the respective phases' eutectic equilibrium concentration, with the melt being set with $c_0 \in \{0.08, c_e, 0.28\}$ respectively and $c_e=0.181$ being the eutectic composition.
The temperature is set to $T_e \pm \SI{5}{K}$, with $T_e = \SI{816}{K}$ being the eutectic temperature.}
All boundaries in the simulation domain are assumed to be no-flux boundaries.
The simulation domain is resolved with 1000 cells in each direction, corresponding to a $\SI{100}{\um}\times\SI{100}{\um}$ physical domain.

  \begin{table}[h]
 \centering
  \caption{Comparison of mass fractions $X_i$ between the phase diagram (PD) and the simulation results (Sim) in the converged state.}
 \begin{tabular}{c | c | c || c | c || c | c}
      & \multicolumn{2}{c||}{$X_\alpha$} & \multicolumn{2}{c||}{$X_\theta$} & \multicolumn{2}{c}{$X_l$} \\
      & Sim & PD & Sim & PD & Sim & PD \\ \hline
(a) & 0.631 & 0.633 & 0.000 & 0.000 & 0.369 & 0.367 \\
(b) & 0.000 & 0.000 & 0.000 & 0.000 & 1.000 & 1.000 \\
(c) & 0.000 & 0.000 & 0.702 & 0.701 & 0.298 & 0.299 \\
(d) & 0.817 & 0.814 & 0.183 & 0.186 & 0.000 & 0.000 \\
(e) & 0.473 & 0.473 & 0.527 & 0.527 & 0.000 & 0.000 \\
(f) & 0.127 & 0.130 & 0.873 & 0.870 & 0.000 & 0.000 \\
 \end{tabular}
 \label{tab:massfractions}
\end{table}

\begin{figure}[p]
\begin{center}
\begin{subfigure}[b]{0.7\textwidth}
 \includegraphics[width=\textwidth]{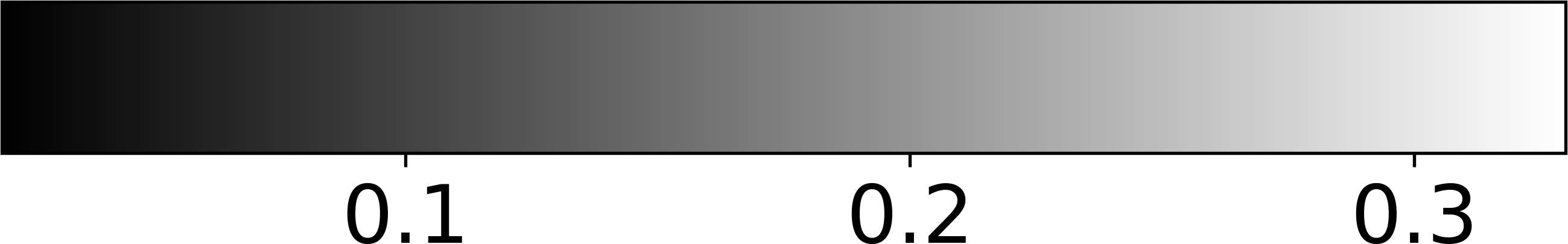}
 \caption*{molar fraction Cu}
\end{subfigure}

  \begin{subfigure}[b]{0.3\textwidth}
    \includegraphics[width=\textwidth]{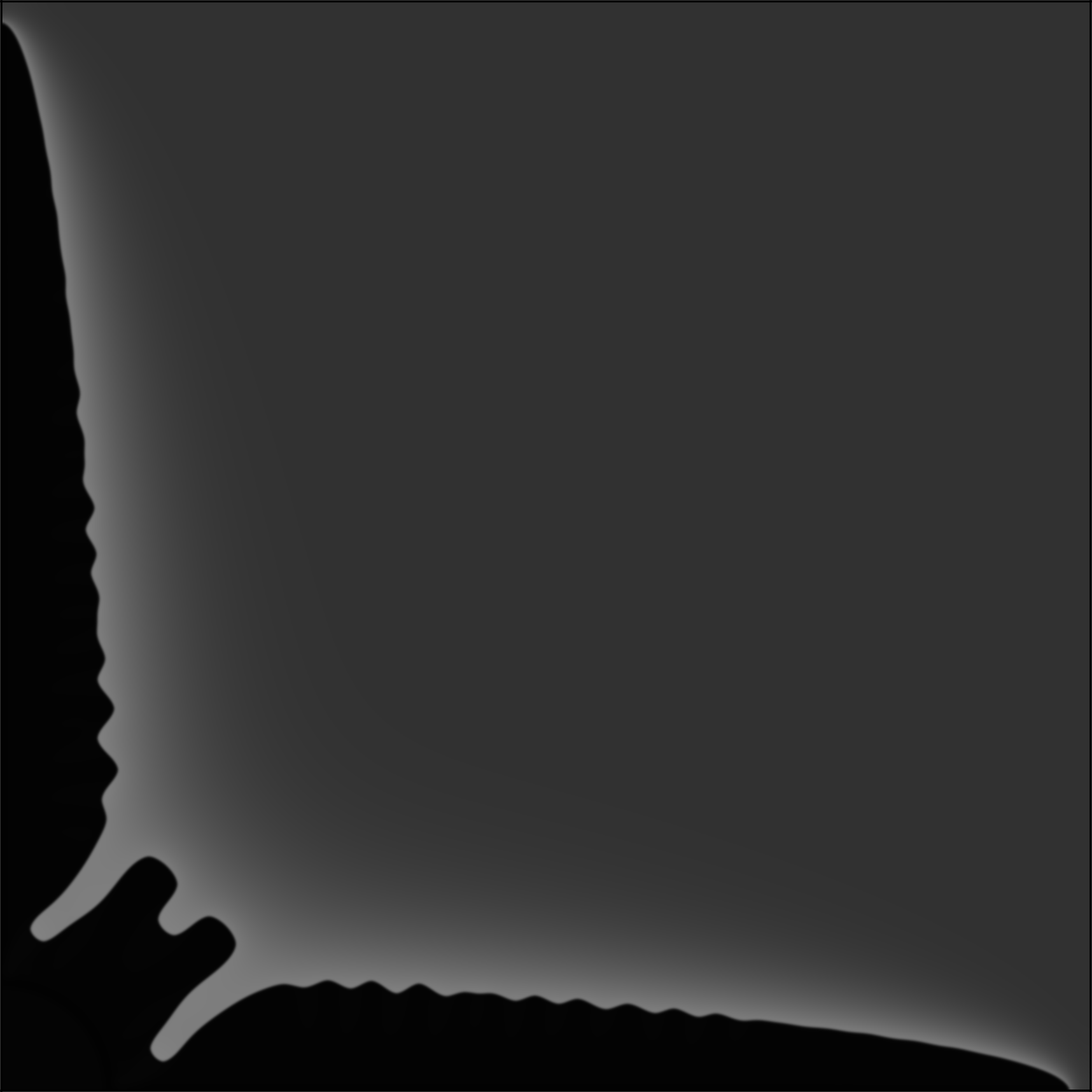}
    \caption{$c_0 = 0.08, T = T_e + \SI{5}{K}$}
  \end{subfigure}
  \begin{subfigure}[b]{0.3\textwidth}
    \includegraphics[width=\textwidth]{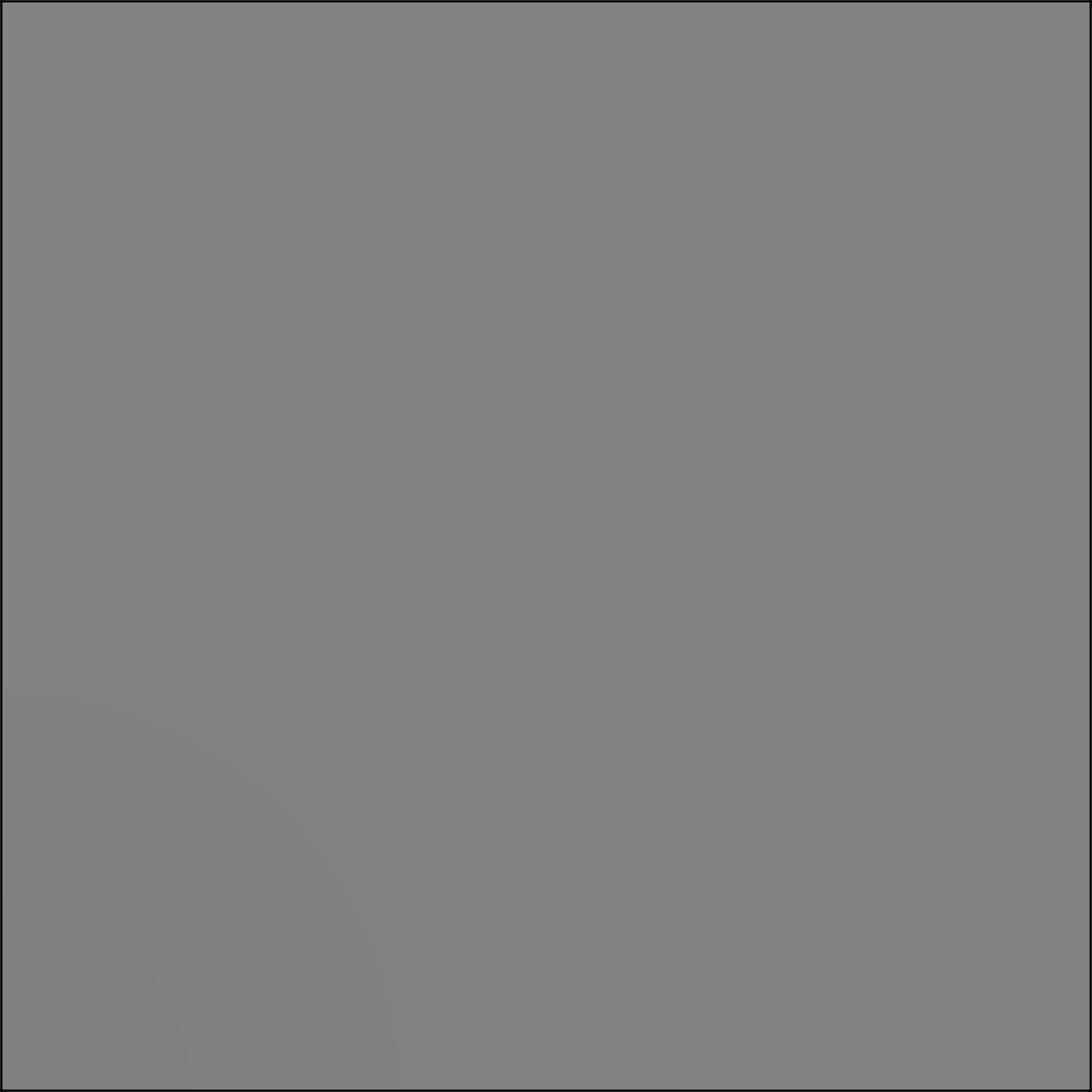}
    \caption{$c_0 = c_e, T = T_e + \SI{5}{K}$}
  \end{subfigure}
    \begin{subfigure}[b]{0.3\textwidth}
    \includegraphics[width=\textwidth]{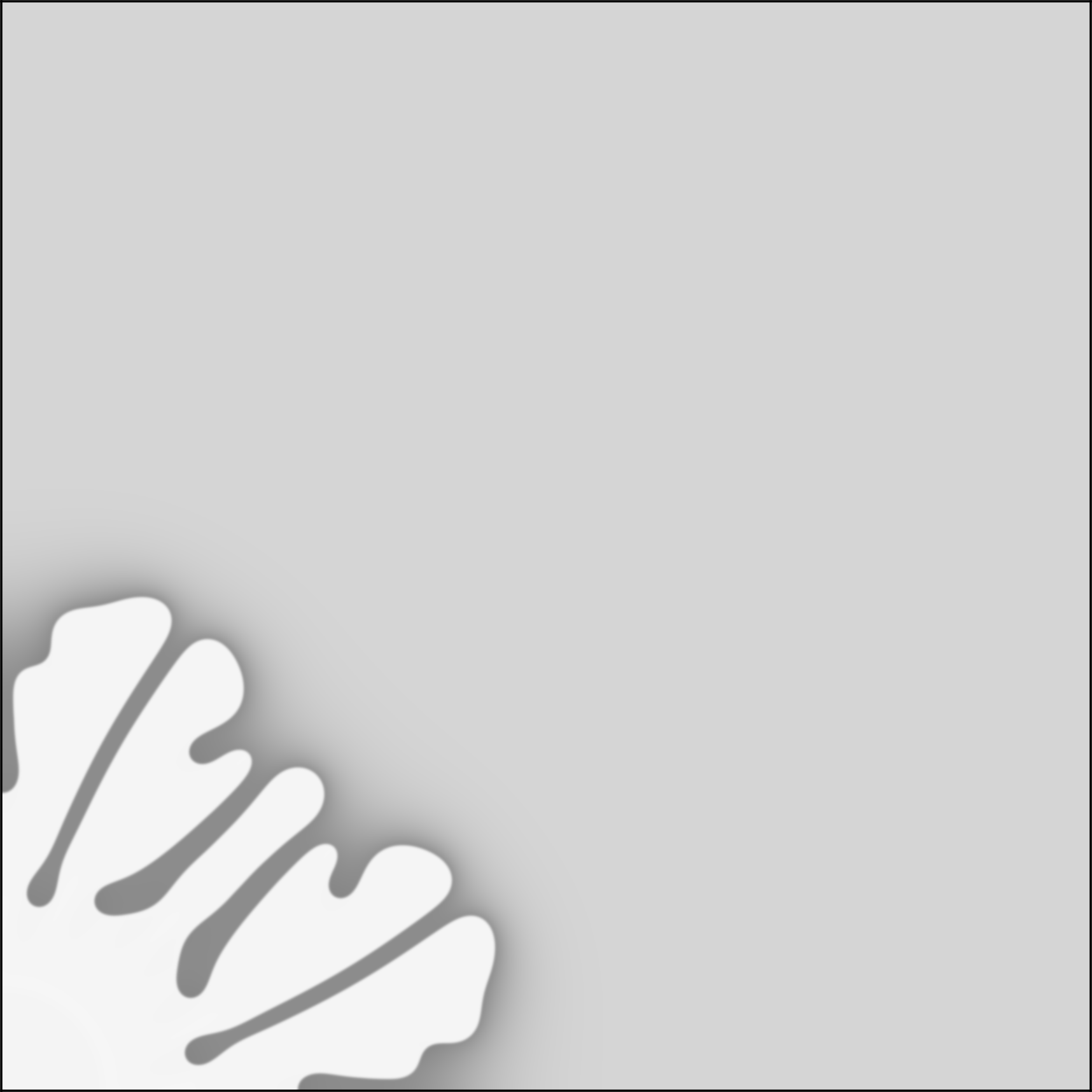}
    \caption{$c_0 = 0.28, T = T_e + \SI{5}{K}$}
    \end{subfigure}
    
  \begin{subfigure}[b]{0.3\textwidth}
    \includegraphics[width=\textwidth]{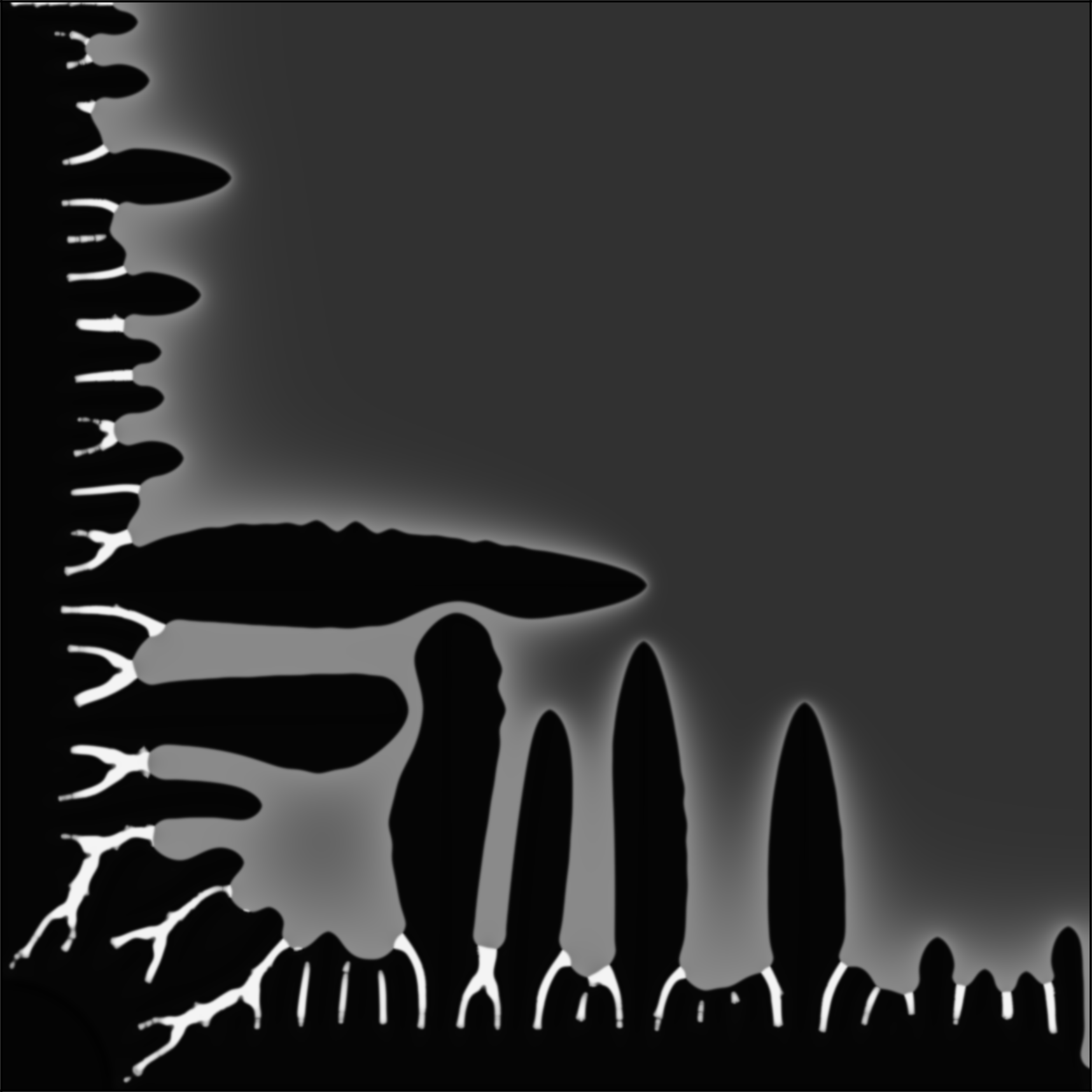}
    \caption{$c_0 = 0.08, T = T_e - \SI{5}{K}$}
  \end{subfigure}
  \begin{subfigure}[b]{0.3\textwidth}
    \includegraphics[width=\textwidth]{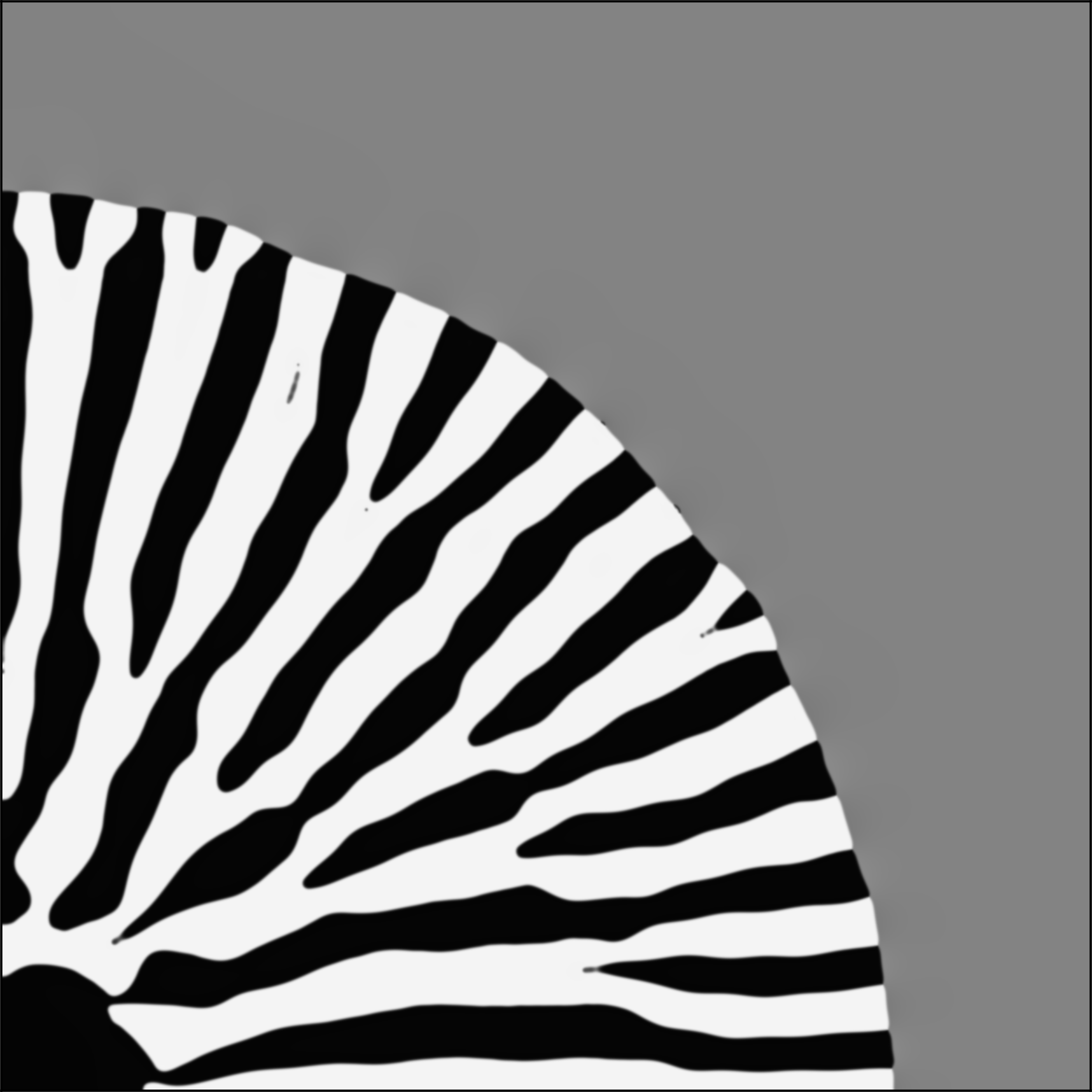}
    \caption{$c_0 = c_e, T = T_e - \SI{5}{K}$}
  \end{subfigure}
  \begin{subfigure}[b]{0.3\textwidth}
    \includegraphics[width=\textwidth]{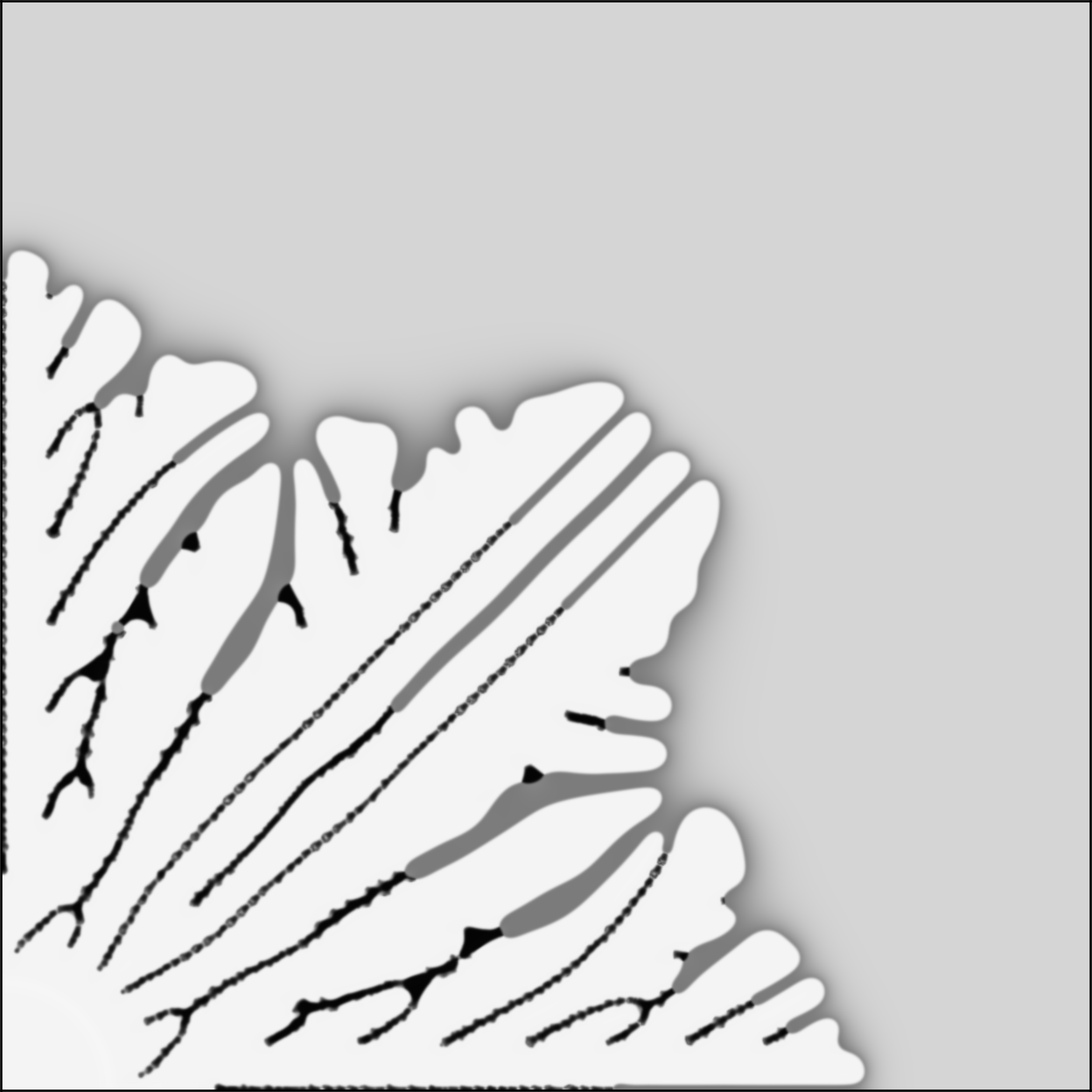}
    \caption{$c_0 = 0.28, T = T_e - \SI{5}{K}$}
  \end{subfigure}
  \caption{Various intermediate morphologies observed in the simulations.
  Dendritic, seaweed and eutectic growth is observed as well as second-phase lining of interdendritic/cellular spaces if below the eutectic temperature (d)-(f).
  All depicted states except for (b,e) were observed at $t=\SI{37.5}{ms}$.
  In (b) the initial seed vanished around  $t=\SI{843}{ms}$, and in (e) the eutectic only started nucleating at around $t=\SI{37.5}{ms}$, hence a later time ($t=\SI{938}{ms}$) was used to show the eutectic pattern.
  }
  \label{fig:morphos}
  \end{center}
  \end{figure}

The simulations are run until the volume fractions of all present solid phases change by less than $1\%$ when calculated over a $\SI{100}{ms}$ period.
A comparison of theoretical and observed mass fraction is given in \cref{tab:massfractions}, showing a good agreement for all investigated states.
The composition field for intermediate states of the simulations are shown in  \cref{fig:morphos}.
Black corresponds to pure $\alpha$, whitish-grey to $\theta$ whereas dark grey corresponds to the melt.
This color scheme will also be used in the remaining simulation images.
The morphology of the phases fits with theoretical expectations, i.e. the anisotropic $\alpha$ grows as a four-sided dendrite (a,d), whereas the isotropic \AlCu{} phase grows in a seaweed-like pattern (c,f).
In both cases a lower temperature also increases the growth rate.
As expected, the solid phase completely vanishes in (b) since it is in the monophasic liquid region of the phase diagram.
For state (e) a radially patterned eutectic is observed since the eutectic nucleates along the circumference of the seed.

%

\paragraph{Validation of model for eutectic growth simulations}
\label{sec:eutvali}
Satisfactorily matching simulation studies of the eutectic growth have been shown previously by several authors for this kind of phase-field model without using a nucleation mechanism~\cite{Choudhury2011,Steinmetz2016,Kellner2017}.
Thus the focus in this section is on validating the proposed nucleation mechanism similar to Kellner et al.~\cite{Kellner2020}.
In their work it is shown that simulations at arbitrary domain lengths including nucleation can be mapped back onto a normalized Jackson-Hunt curve for the lamellar spacing.
In effect this probes whether the steady-state growth point is recovered even in a nucleating system.
This computational experiment is reproduced for the investigated \systemAlCu{} system.

\begin{figure}
\begin{center}
    \begin{subfigure}[t]{0.2\textwidth}
    \includegraphics[width=\textwidth]{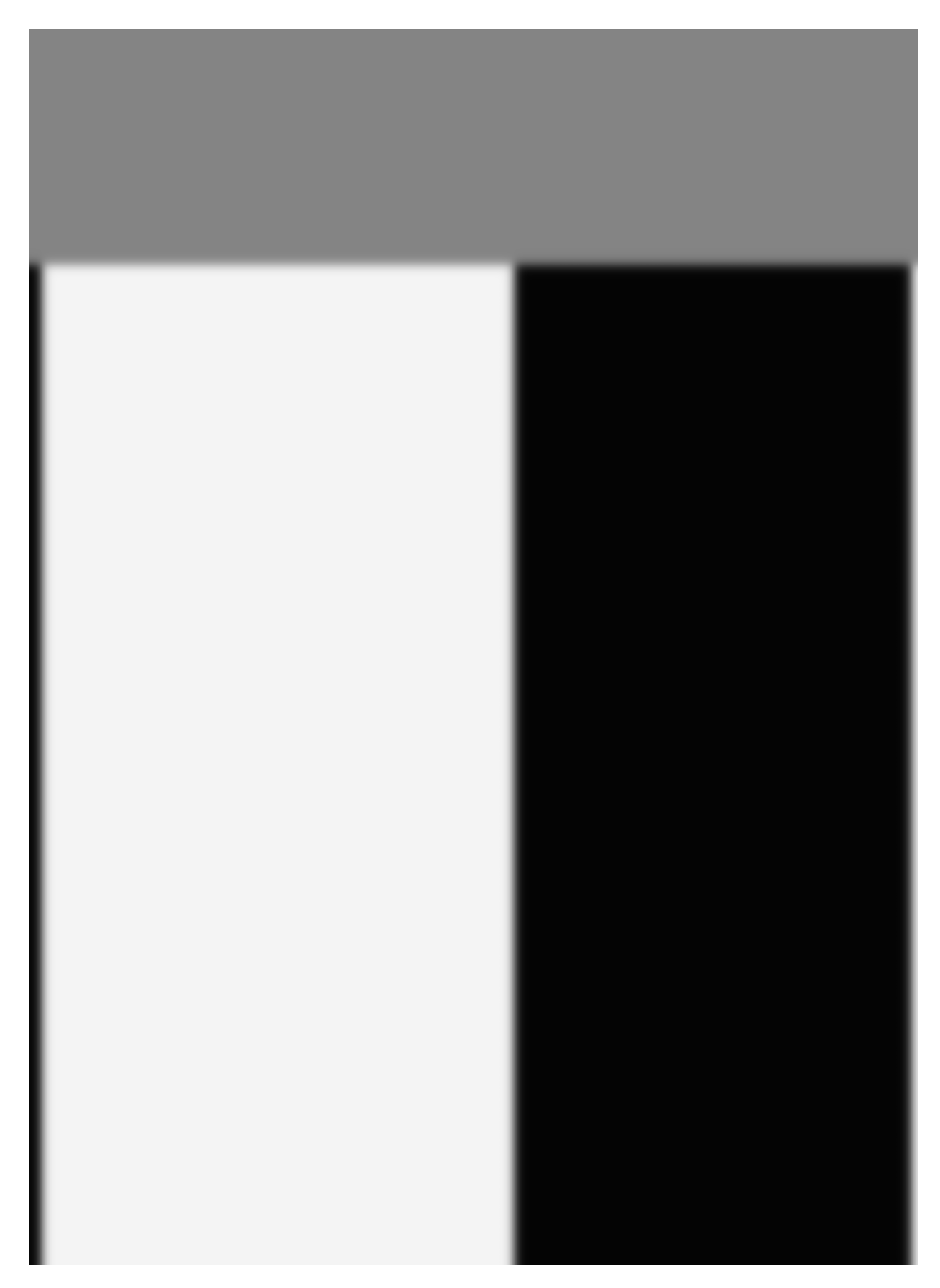}
    \caption{initial}
    \label{fig:eut_ini}
  \end{subfigure}
\begin{subfigure}[t]{0.2\textwidth}
    \includegraphics[width=\textwidth]{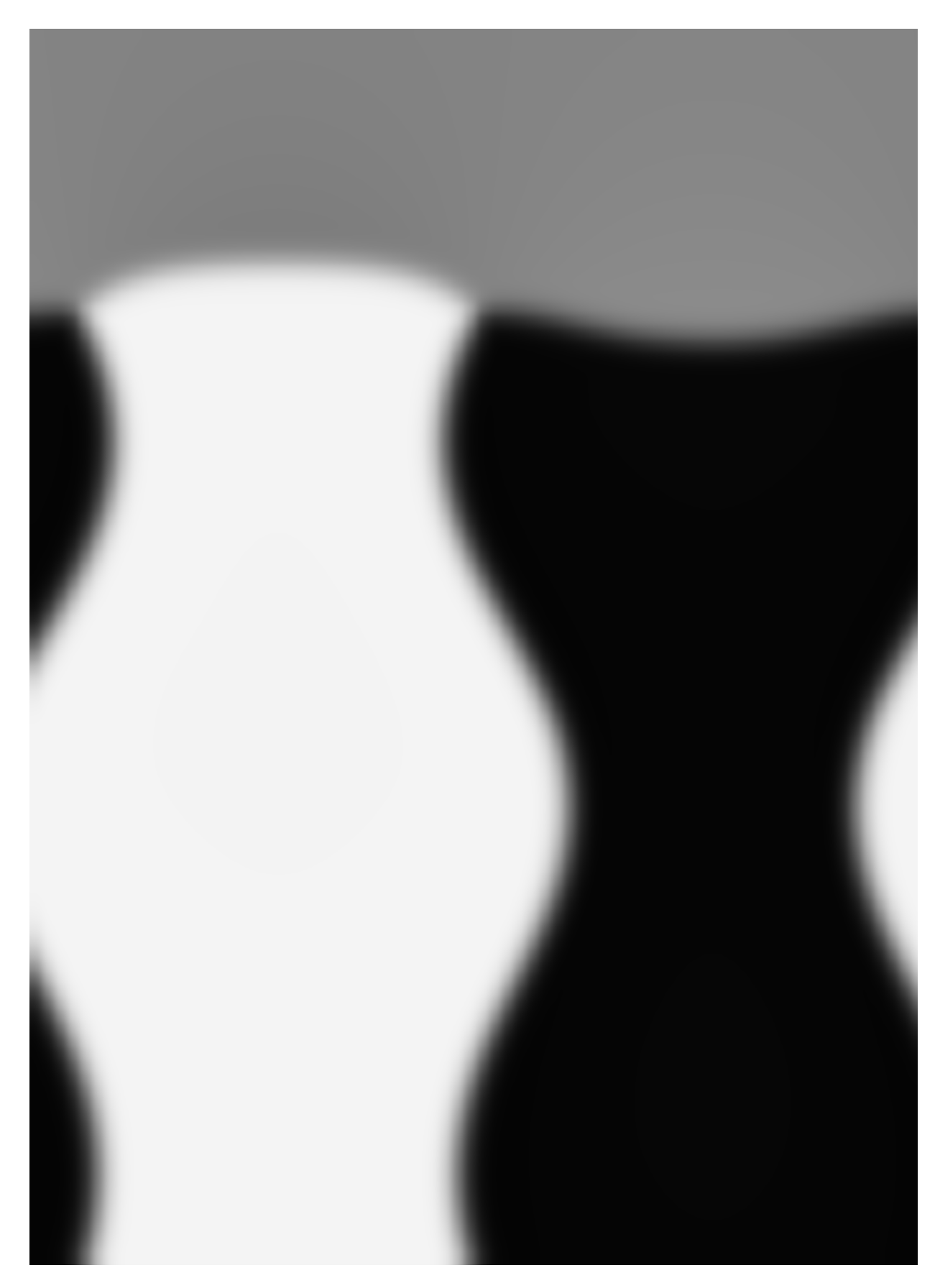}
    \caption{oscillating}
    \label{fig:eut_instab}
  \end{subfigure}
\begin{subfigure}[t]{0.2\textwidth}
    \includegraphics[width=\textwidth]{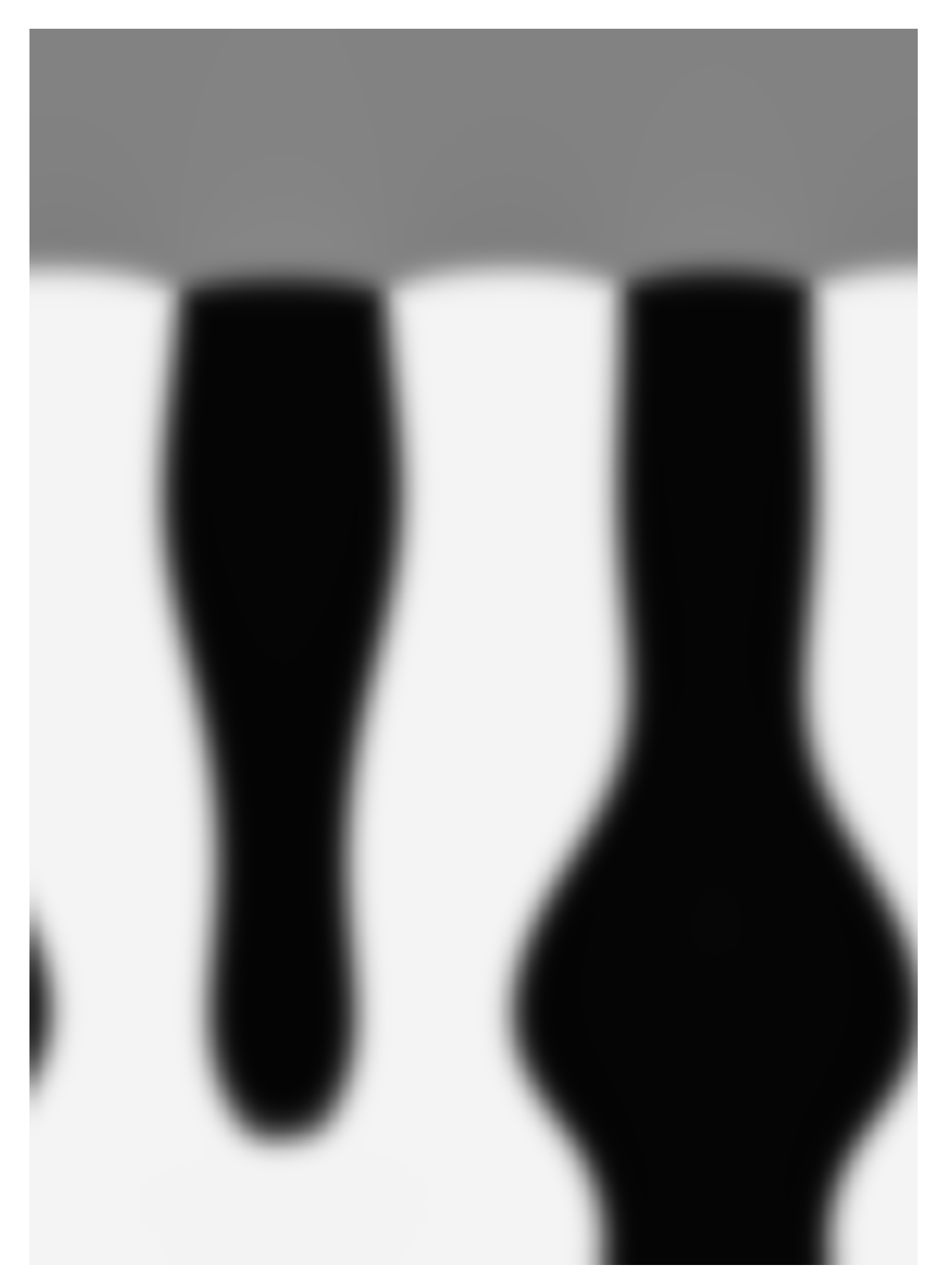}
    \caption{shortly after nucleation}
    \label{fig:eut_nuc}
  \end{subfigure}
\begin{subfigure}[t]{0.2\textwidth}
    \includegraphics[width=\textwidth]{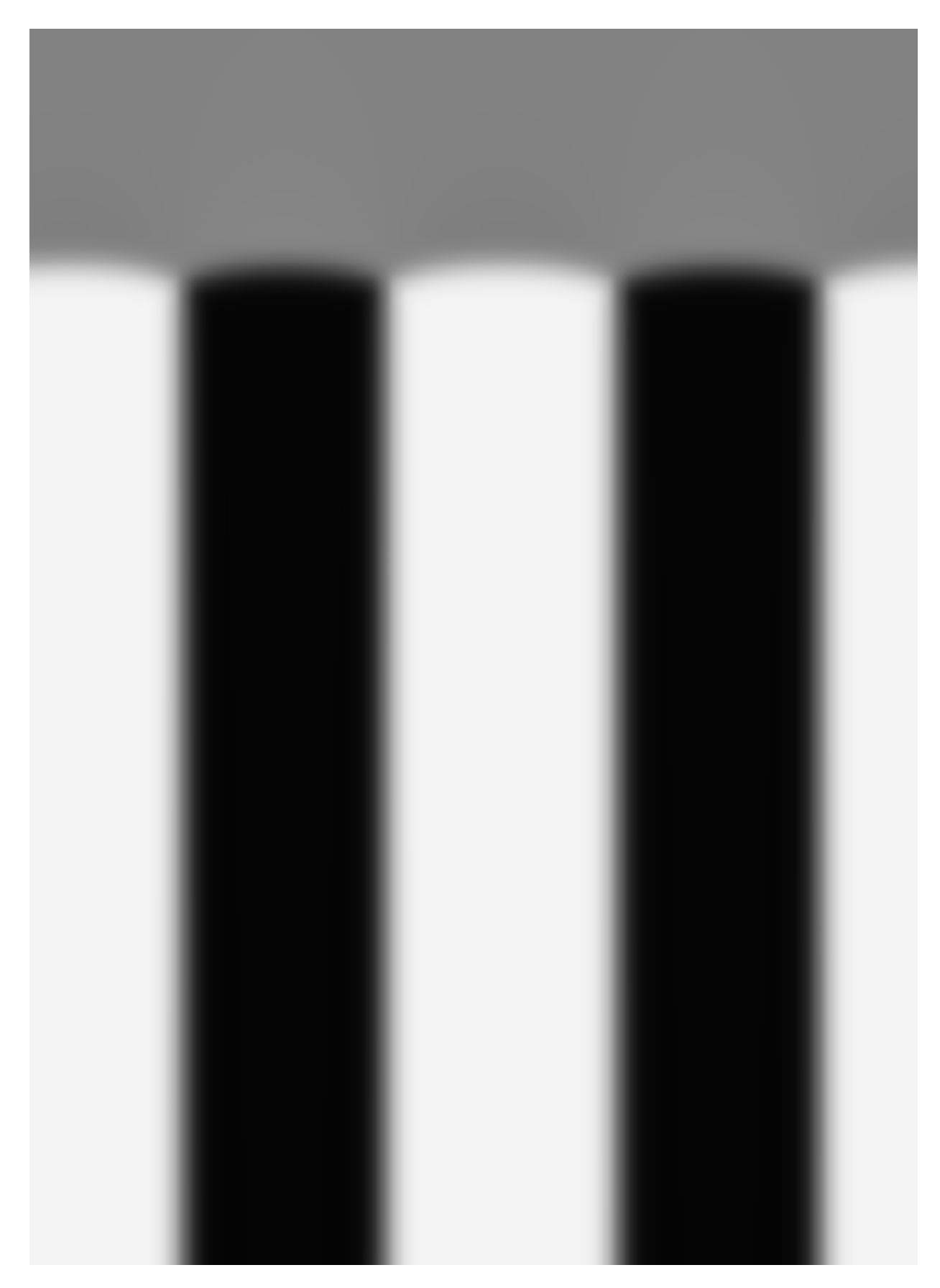}
    \caption{long past nucleation}
    \label{fig:eut_nucfin}
  \end{subfigure}
%
 \caption{Initial setup as well as exemplary evolutionary states during eutectic growth.
 The domain is cut off slightly above the moving window cutoff in order to emphasize the solid phases.}
 \label{fig:eutsetup}
 \end{center}
\end{figure}

The principal setup of the simulation study is shown in \cref{fig:eutsetup}, along with typical evolutionary states:
An initial pair of isotropic $\alpha$-Al and $\theta$ phases is set at the bottom of the domain with the fractions determined by the lever rule (a).
The top part of the domain is filled with melt at the eutectic composition $c_e$, with this composition also being imposed as a Dirichlet condition at the top.
At the bottom no-flux conditions are employed, whereas on the sides periodic boundary conditions are applied.
The temperature is assumed to be homogeneous.
If the spacing $\lambda$ is sufficiently above the dominant lamellar spacing $\lambda_{JH}$, oscillations can be observed (b).
Without nucleation, these persist and may lead to one phase overgrowing the other, in which case the simulation is aborted and the data is not taken into account.
Nucleation will occur in the concave parts of the front with the present mechanism, leading to a refinement of the spacing and less oscillatory growth (c,d).

\begin{figure}
\begin{center}
    \begin{subfigure}[]{0.45\textwidth}
    \includegraphics[width=\textwidth]{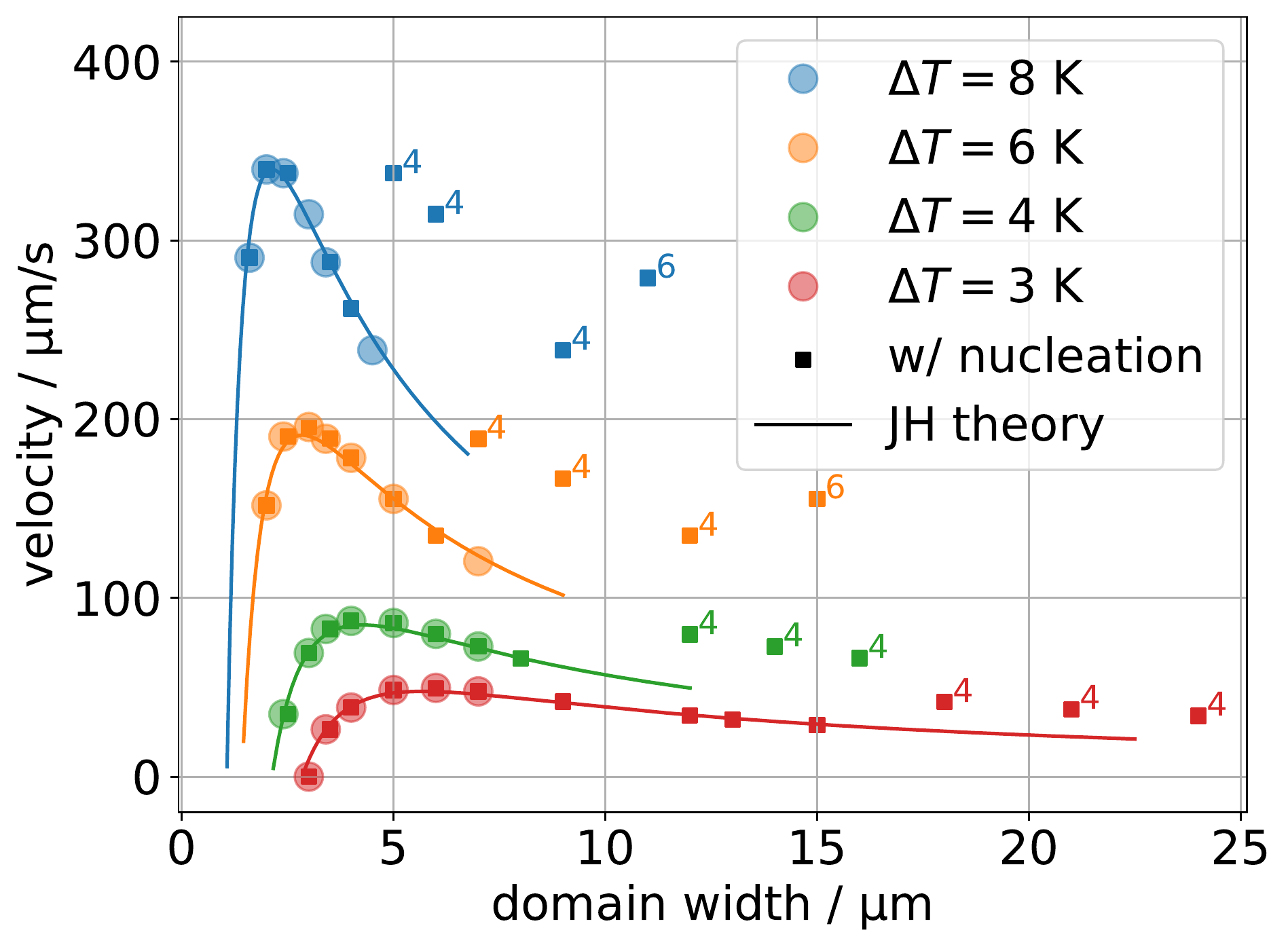}
  \end{subfigure}
~
\begin{subfigure}[]{0.45\textwidth}
    \includegraphics[width=\textwidth]{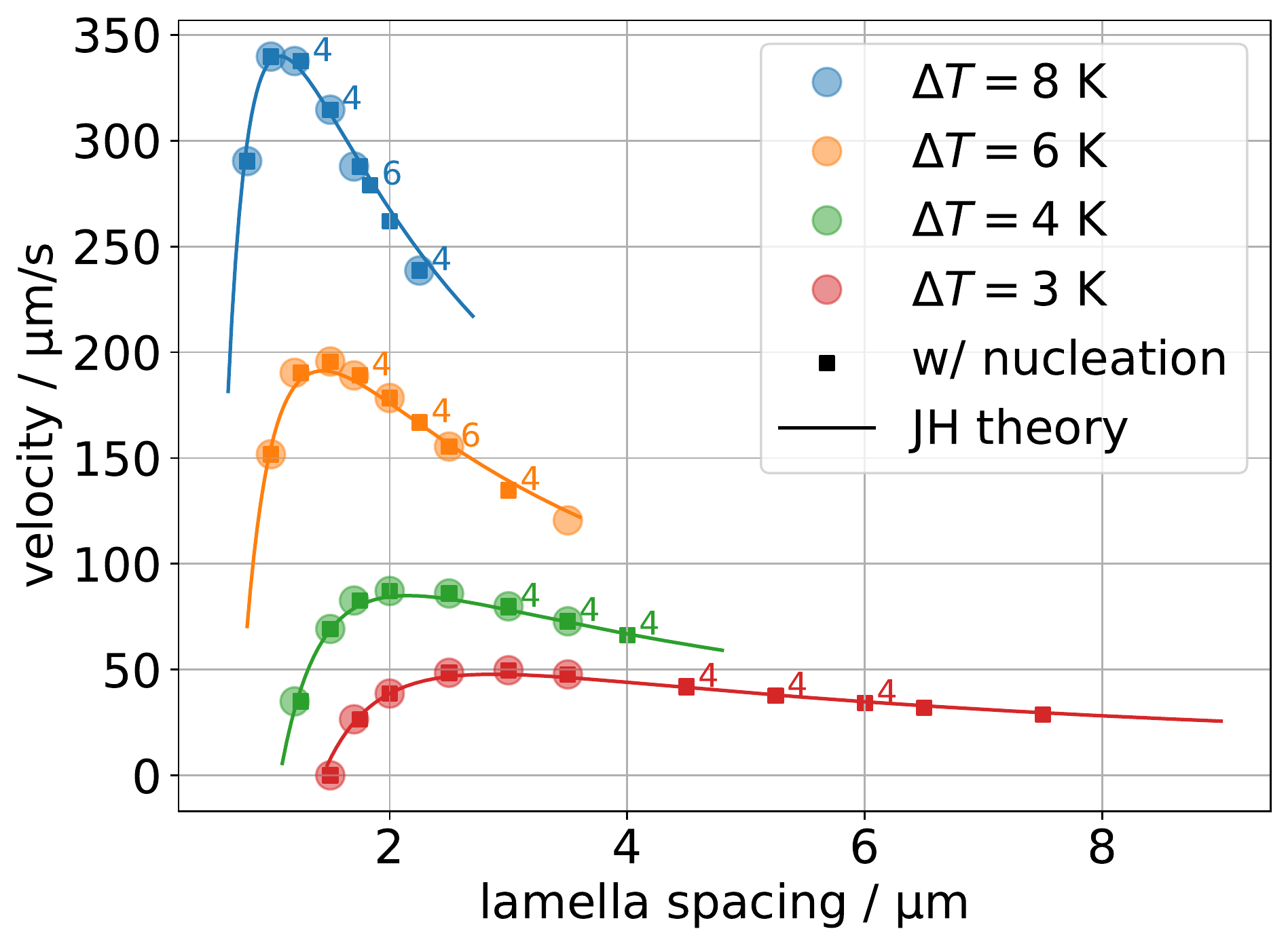}
  \end{subfigure}

 \caption{Comparison of eutectic theory (lines) and simulations with (squares) and without (circles) nucleation for various undercoolings.
 \cor{
 The left plot shows the same data as the right plot, but plotted over the domain width instead of the lamellar spacing.
 The initial configuration always consists of a single pair of $\alpha$ and $\theta$, representing two lamellas.
 The number besides the squares indicates how many lamellas are observed in steady-state, with no number indicating two lamellas.
 }
 Matching behavior between theory and simulation is observed over the entire undercooling range.
 Furthermore, the simulations with nucleation fall onto the curve described by JH theory and achieve similar steady-state velocities to simulations without nucleation.
 }
 \label{fig:lamv-lam}
 \end{center}
\end{figure}

First, several undercoolings $\Delta T \in \{3, 4, 6, 8\}\si{K}$ will be investigated without nucleation activated.
For each considered undercooling, a range of domain lengths is employed to allow different lamellar spacings $\lambda$ and thus front velocities $v$.
The values for the domain lengths are determined iteratively starting from an estimated dominant lamellar spacing.
Following the theory of Jackson and Hunt\cite{Jackson1966}, the curve $v(\lambda)$ should contain a global maximum which represents the dominant lamellar spacing $\lambda_{JH}$.
Thus if no maximum is observed, additional domain lengths are added in the direction of the slope of the curve.
Once a maximum is observed, the set of domain lengths is frozen.
Based on these simulations the concentration-independent model of \cref{eq:ucE} is fitted, yielding $K_1 = 0.02696$, $K_2 = 0.05197$ in nondimensional units.
Next, simulations with nucleation activated are conducted for each undercooling and its corresponding set of domain lengths, with additional simulations at significantly larger domain sizes than the observed $\lambda_{JH}$ in order to allow multiple pairs of lamellas to nucleate from a single pair.
In total this yields \cref{fig:lamv-lam}, showing the solid front velocity \cor{over domain width and the final lamellar spacing} for all conducted eutectic simulations.
The transparent circles denote the nucleation-less simulations, whereas the squares represent the simulations with nucleation active.
The solid line is the analytical Jackson-Hunt result, based on the previously calculated $K_1, K_2$.
First, the circles match the theory without a selection criterion well, suggesting that the main features of Jackson-Hunt theory are captured with the simulations.
Second, the squares map back closely to the curve, suggesting that steady-state growth is not significantly affected by the nucleation mechanism.
It should be noted that herein simulations growing at $\geq 2\lambda_{JH}$ did not necessarily exhibit strong oscillations in their growth.
This leads to only minor solute excess in front of the solid phases which inhibits nucleation.
Hence the squares will tend to cluster not around $\lambda_{JH}$ but rather around a \cor{spacing} somewhat larger, similar to \cite{Kellner2020}.

\begin{figure}[h]
\begin{center}
   \begin{subfigure}[t]{0.45\textwidth}
    \includegraphics[width=\textwidth]{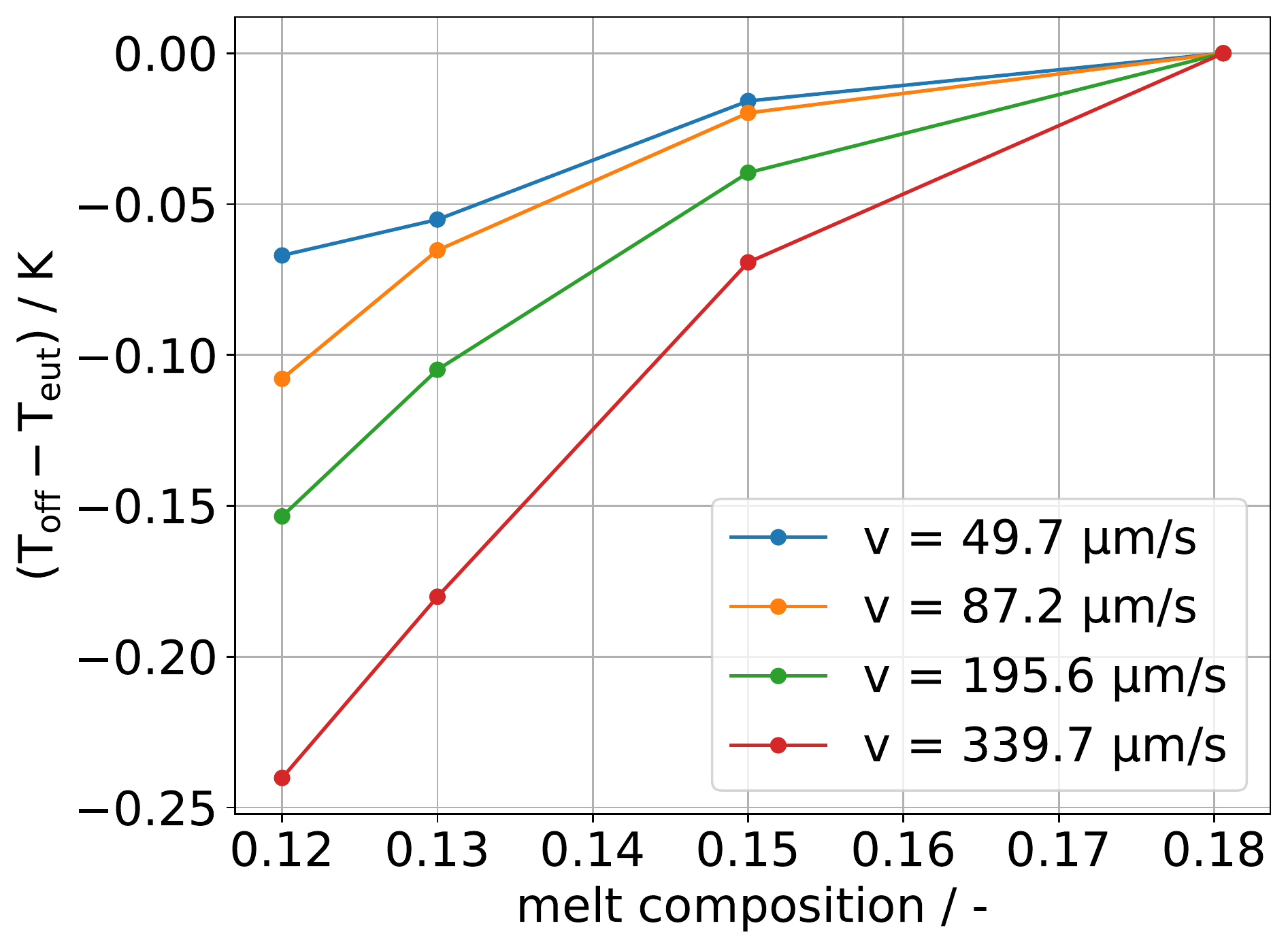}
    \caption{Difference of front undercooling for the off-eutectic simulations to the eutectic simulation.
    With increasing distance from the eutectic composition, the front grows at an increasingly lower temperature.}
    \label{fig:ucE-off}
  \end{subfigure}
\hspace{0.5cm}
  \begin{subfigure}[t]{0.45\textwidth}
    \includegraphics[width=\textwidth]{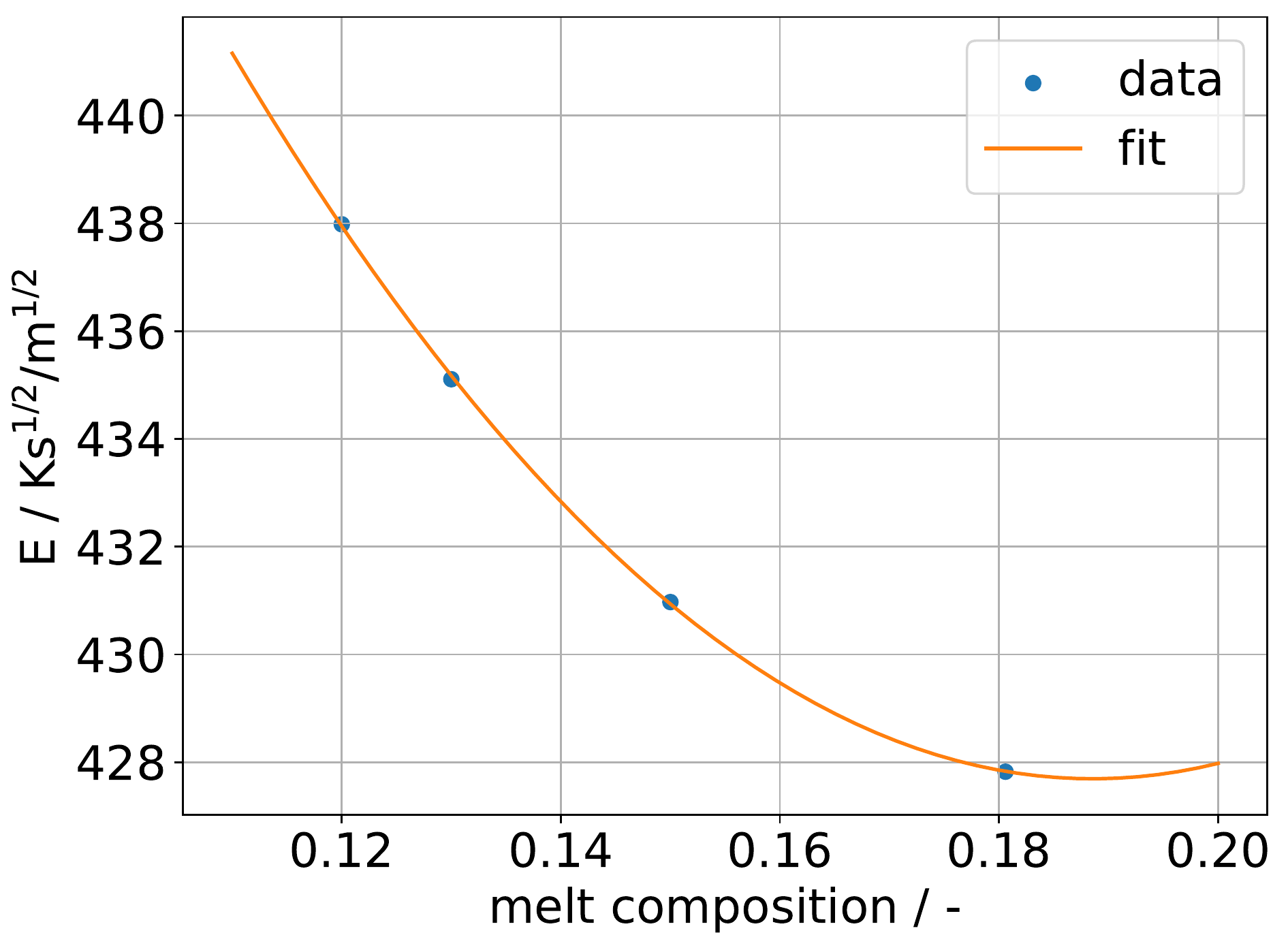}
    \caption
    {
    The concentration dependence of the growth constant $E$ in $\Delta T_e = Ev^{0.5}$.
    A quadratic polynomial seems to describe the dependence satisfactorily.
    }
    \label{fig:AEfit-x0}
  \end{subfigure}
  \end{center}
  \caption{Results of the off-eutectic simulations.}
  \label{fig:offE-results}
\end{figure}

In order to determine the influence of off-eutectic compositions on the undercooling, further simulations are conducted.
For these, the frozen temperature approximation \cref{eq:t_evolution} is employed.
The velocities and domain lengths are based on the maxima from the previous study and the melt concentrations $\{ 0.12, 0.13, 0.15, c_e\}$ are employed, i.e. three hypoeutectic concentrations and the eutectic concentration.
The gradient is chosen to be $\SI{99}{K/mm}$ in order to speed up convergence of the temperature field.
The simulations are run until the velocity differs by less than $2\%$ from the imposed velocity.
Plotting the difference of the off-eutectic front temperature to the eutectic front temperature for these simulations yields \cref{fig:ucE-off}.
It is easy to see that the front temperature is decreasing with increasing distance from the eutectic composition.
The eutectic constant $E$ is calculated for each composition and then a parabola is fit to this data, with \cref{fig:AEfit-x0} showing that the fit matches the data well.
Thus the eutectic undercooling model reads $\Delta T_e = (0.376c_0^2 - 0.142c_0 + 0.08714) v^{0.5}$.
The effective value of $E$ at the eutectic composition is $\SI{428}{Ks^{0.5}/m^{0.5}}$ which compares well with the investigations at the eutectic composition, which yields a value of  $\SI{434}{Ks^{0.5}/m^{0.5}}$.

\begin{figure}[h]
\hspace{3cm}
    \begin{subfigure}[]{\textwidth}
    \includegraphics[width=0.5\textwidth]{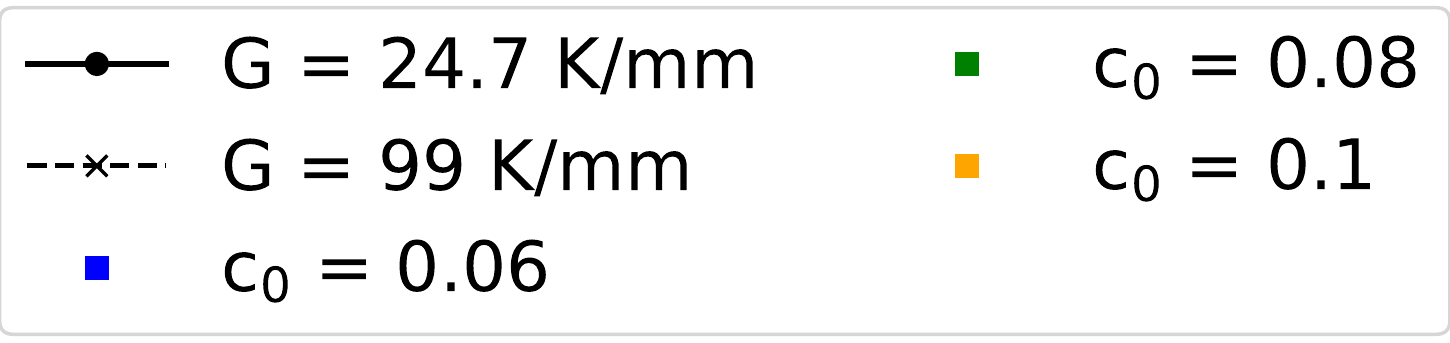}
  \end{subfigure}
  \begin{center}
    \begin{subfigure}[]{0.45\textwidth}
    \includegraphics[width=\textwidth]{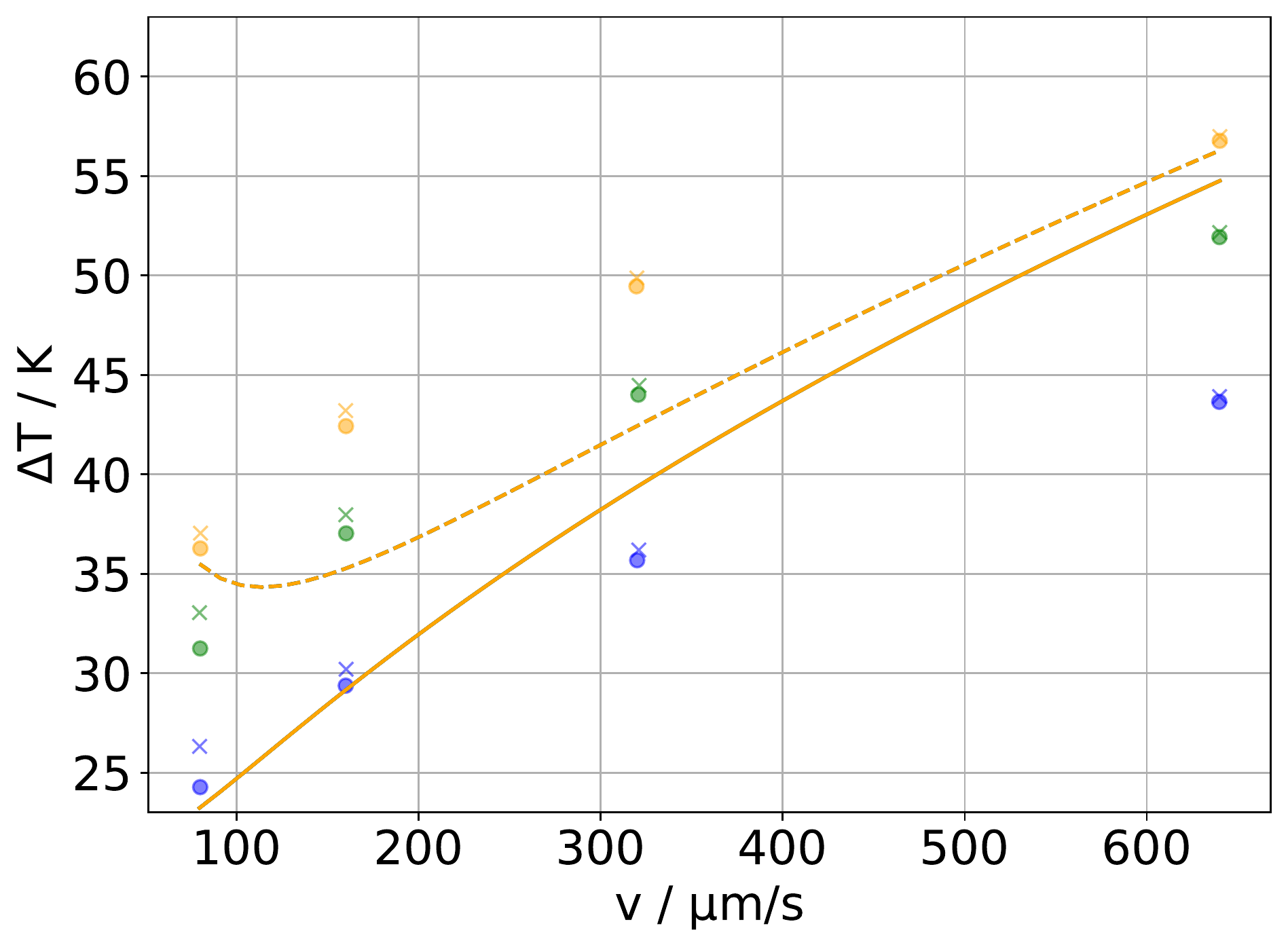}
\caption{fit without $c_0$-dependence}
  \end{subfigure}
~
\begin{subfigure}[]{0.45\textwidth}
    \includegraphics[width=\textwidth]{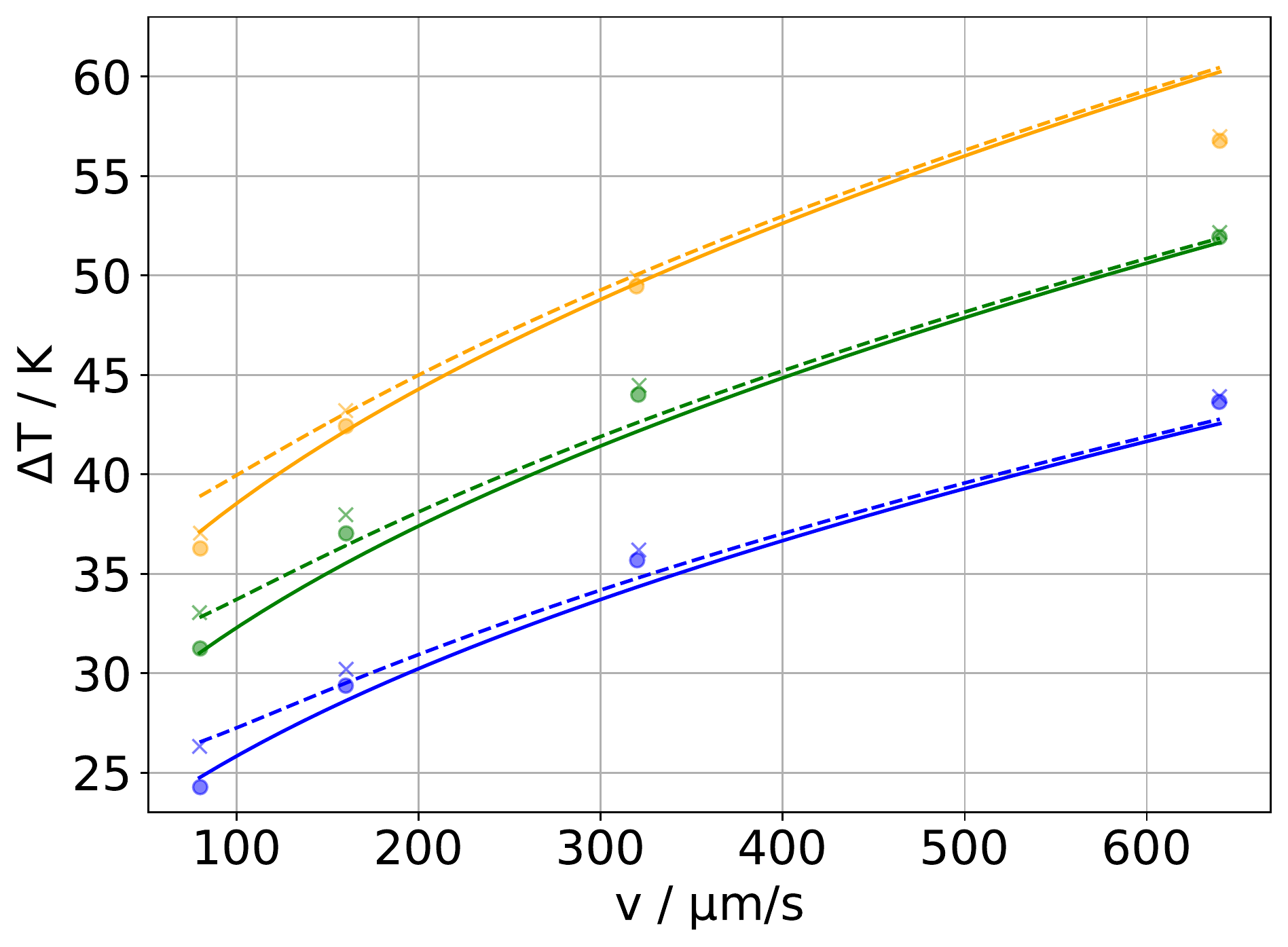}
    \caption{fit with $c_0$-dependence}
  \end{subfigure}
   \caption {
   \cor{
  A scatter plot of the interfacial undercooling over the imposed velocity is depicted.
  The observed undercooling (markers) rises with velocity, composition (color) and gradient (marker type).
  The same data is shown in both plots, with the lines indicating the predictions of the different models for the dendrite tip undercooling, varying similarly for velocity, composition (color) and gradient (line style).
  The composition dependent model generally matches the data better than the composition independent model.
  Note that the concentration independent model only produces two lines, as any choice of $c_0$ will lead to the same line for the same $G$.
  }
  }
 \label{fig:dT-v-scatter}
  \end{center}
\end{figure}

\paragraph{Determination of dendrite model parameters}
The simulations for the determination of the constants within the dendrite tip undercooling model \cref{eq:ucD} will now be described.
An initial periodic, anisotropic $\alpha$-seed is placed at the bottom of the domain inside of a homogeneous melt of concentration $c_0$.
The frozen temperature approximation \cref{eq:t_evolution} is employed again.
A quasi-infinite domain is simulated by employing the moving-window technique.
Various temperature gradients $G \in \{24.7, 99.0\}\si{K/mm}$, velocities $v \in \{80, 160, 320, 640\}\si{\um/s}$ as well as melt concentrations $c_0 \in \{0.06, 0.08, 0.1\}$ are employed.
Nucleation was allowed for all simulations, but no nucleation was observed since it is energetically unfavorable for the investigated parameters.
The simulations are run until the front velocity differed by less than $2\%$ from the imposed velocity.
This yields tuples of $(T_i, v, G, c_0)$ values which are used to fit the undercooling formulation of \cref{eq:ucD}, with the interfacial undercooling $T_l(c_0)-T_i$ as the dependent variable.
The nondimensionalized coefficients are given by $A = 0.957, B = 0.788, C = 0.288$ for the melt concentration dependent model and $A = 6.58, B=0.370$ for the model without an explicit melt concentration dependence.
A scatter plot of the measured undercoolings over the velocity is shown in \cref{fig:dT-v-scatter}, with the two models indicated as lines.
\cor{The non-monotonic behavior of the concentration independent model at lower velocities is expected\cite{Burden1974}.
The concentration dependent model shows this as well for even lower velocities.}
The mean unsigned error defined by $\sum \frac{|\Delta T_{observed}-\Delta T_{estimated}|}{N}$ is $\SI{5.81}{K}$ for the concentration-independent model and $\SI{1.05}{K}$ for the concentration-dependent model.
In total one can observe that the explicit inclusion of melt concentration increases the model accuracy significantly.


\paragraph{Boundary curve of the coupled zones}
\begin{figure}[h]
  \includegraphics[width=\textwidth]{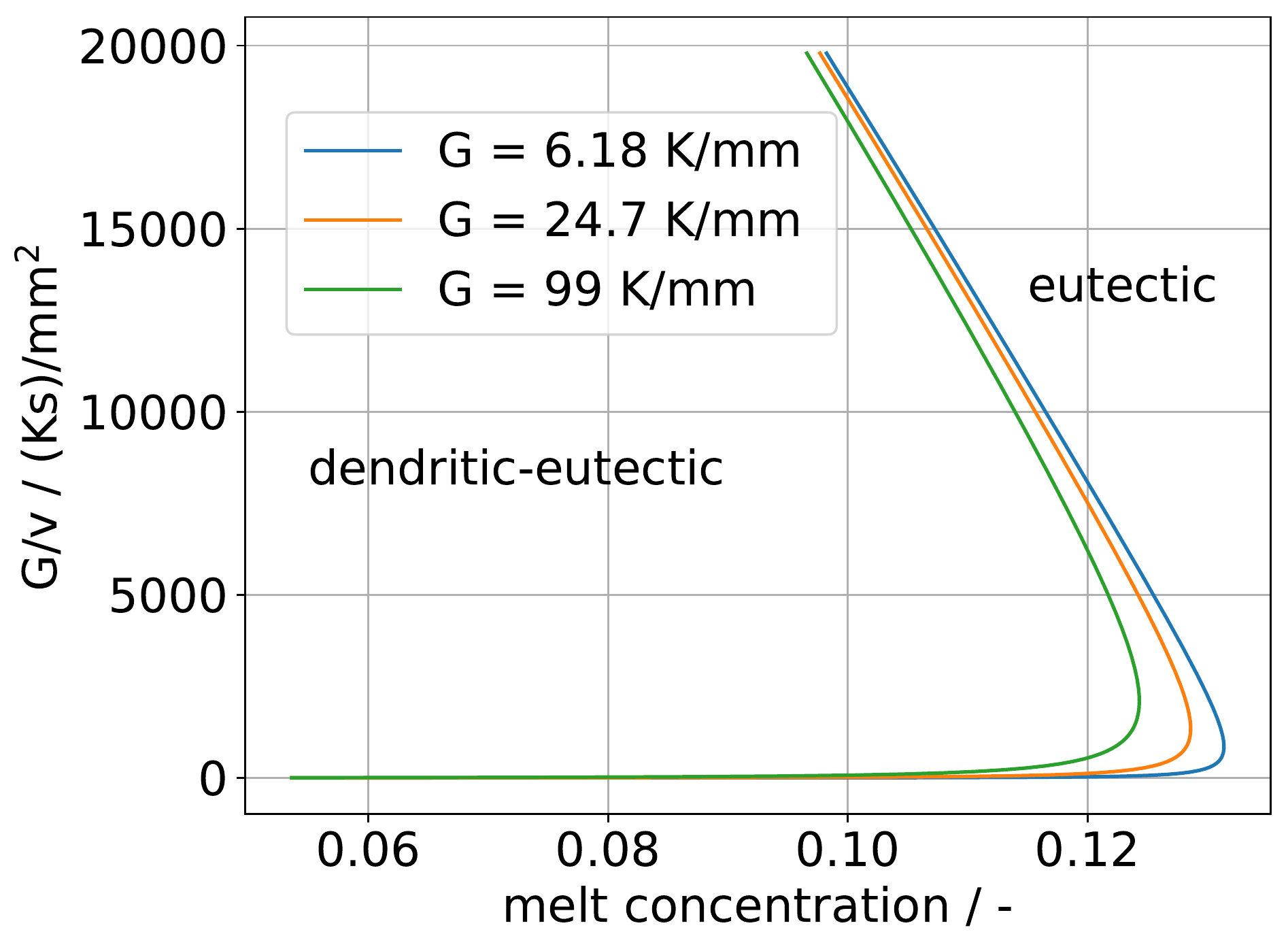}
  \caption{Numerically calculated boundary curves between pure eutectics and a mixed dendritic-eutectic microstructure.}
  \label{fig:boundarycurves}
\end{figure}

Now that the undercooling models for dendrites and eutectics are fully specified, the boundary curve between the two morphologies can be calculated.
For each $(G,v)$ point, the resulting nonlinear equation in $c_0$ is solved numerically.
Three gradients ( $G \in \{6.18, 24.7, 99.0\}\si{K/mm}$ ) are chosen, for which the range of cooling rates $Gv$ from \SI{3e-2}{K/s} to \SI{40}{K/s} is sampled.
The resulting set of points is plotted as a $c_0 - G/v$ diagram in \cref{fig:boundarycurves} as suggested by \cite{Burden1974}.
The curves separate the eutectic range to the right from the coupled dendritic-eutectic range to the left.
The eutectic range is always increased by increasing the gradient.
If $G/v$ is sufficiently high, i.e. at low velocities, the influence of gradient diminishes and the extent of the eutectic range is only weakly dependent on the gradient.
In the high velocity regime there is a significant effect of the gradient on the eutectic range.
Further to the left one would expect a purely dendritic microstructure once the melt composition is around the solubility limit.
This microstructure will not be separately considered in the present paper, but can also be easily simulated with the present model.
The majority of the simulations will be conducted around the ``knee'' of these curves in order to probe the minimal extent of the eutectic range.


\section{Results \& Discussion}
In this section novel results investigating the conditions for dendritic-eutectic growth and its influence on the microstructure are presented and discussed.

\paragraph{Boundary curve validation \& microstructural influences}
Given that the boundary curve is now known, processing conditions which are likely to yield dendritic-eutectic growth can be set.
Specifically, simulations with gradients $G \in \{ 6.18, 24.7, 99.0\}\si{K/mm}$, pulling velocities $v \in \{ 80, 160, 320\}\si{\um/s}$ and melt compositions $c_0 \in \{ 0.1, 0.11, 0.12, 0.13\}$ are conducted.
The initial and boundary conditions are similar to the setup of pure dendritic growth in the previous section.
The starting temperature $T_0 = T_e - \SI{2}{K}$ is now below the eutectic temperature.
The domain height of 5000 cells corresponds to \SI{500}{\um} and the width of 2500 cells corresponds to \SI{250}{\um}.
The moving window cutoff is set at \SI{250}{\um}, i.e. there are at least \SI{250}{\um} between the front and the boundary at all times.
The diffusion length for the smallest velocity corresponds to \SI{25}{\um} and thus there are at least 10 diffusion lengths between the front and the boundary, mimicking an infinite melt.
Unless mentioned otherwise, the simulation images always depict a region of size $\SI{280}{\um}\times\SI{250}{\um}$, i.e. the whole width of the domain and slightly above the solidification front in terms of height.
This is done to emphasize the solid structure.
The simulations are continued until either the eutectic is shifted outside of the domain, a eutectic front stabilizes or the height difference between the dendrite tip and the eutectic becomes constant.
The former two conditions are based on the observation that once one of the morphologies becomes dominant, the other morphology will not appear without external influence again.
The latter condition is employed instead of a velocity convergence criterion as multiple fronts are advancing at different velocities.
Usually, the primary dendrite will reach a converged velocity first, with the eutectic still adjusting its position w.r.t the dendrite tip.

\begin{figure}
    \begin{subfigure}[b]{0.3\textwidth}
    \includegraphics[width=\textwidth]{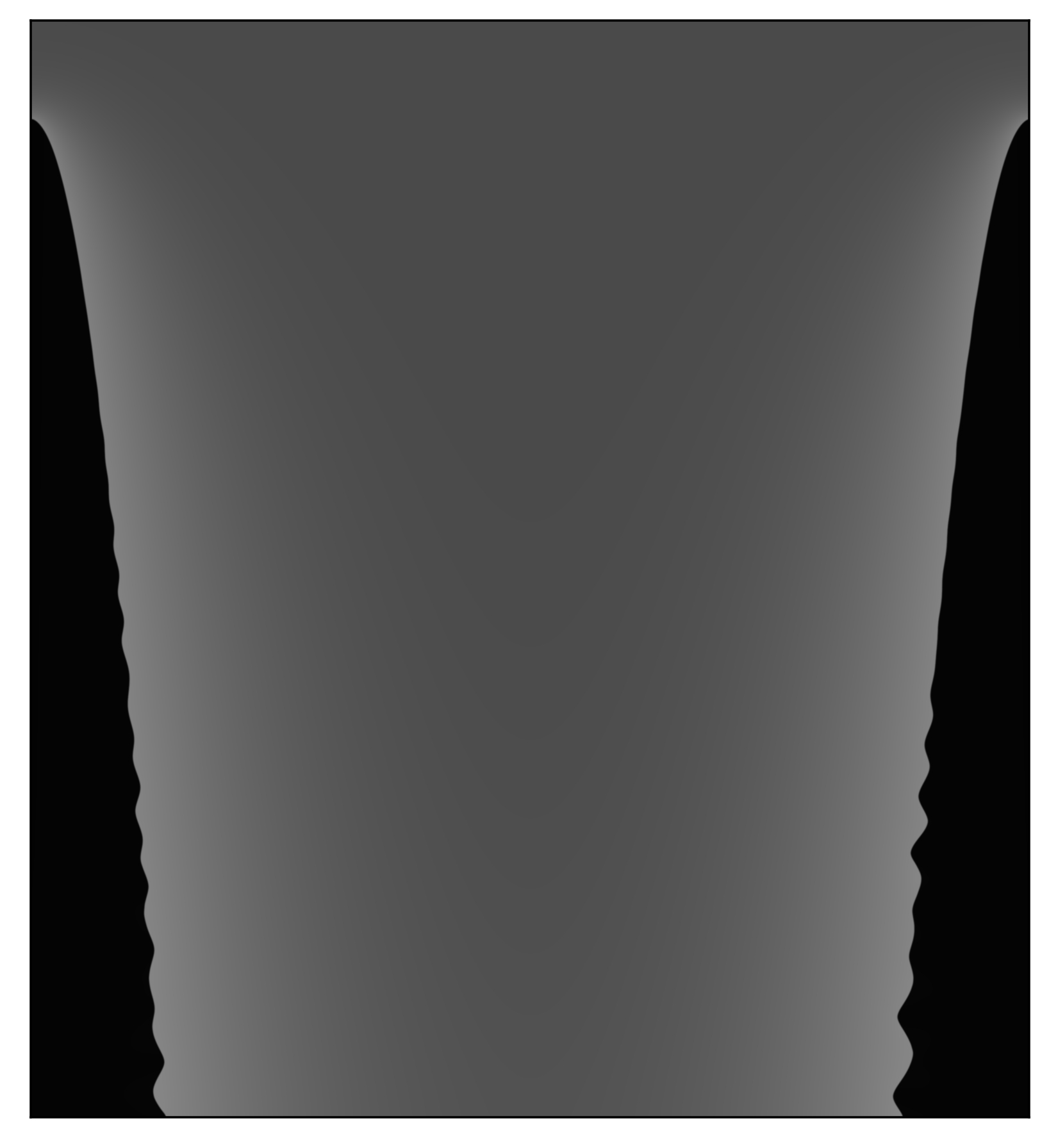}
    \caption{$c_0 = 0.11$}
  \end{subfigure}
  \begin{subfigure}[b]{0.3\textwidth}
    \includegraphics[width=\textwidth]{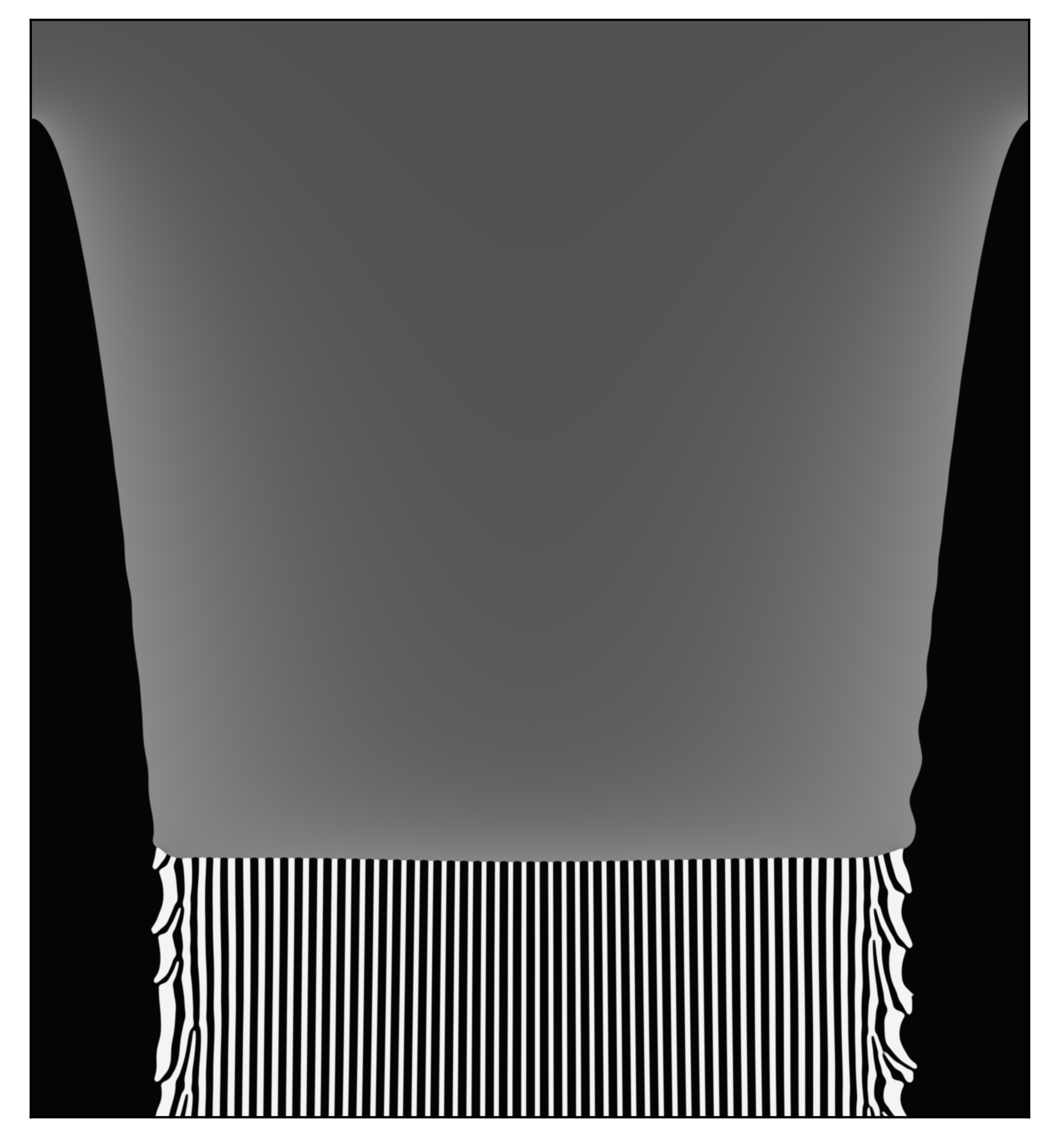}
    \caption{$c_0 = 0.12$}
    \label{fig:desim}
  \end{subfigure}
    \begin{subfigure}[b]{0.3\textwidth}
    \includegraphics[width=\textwidth]{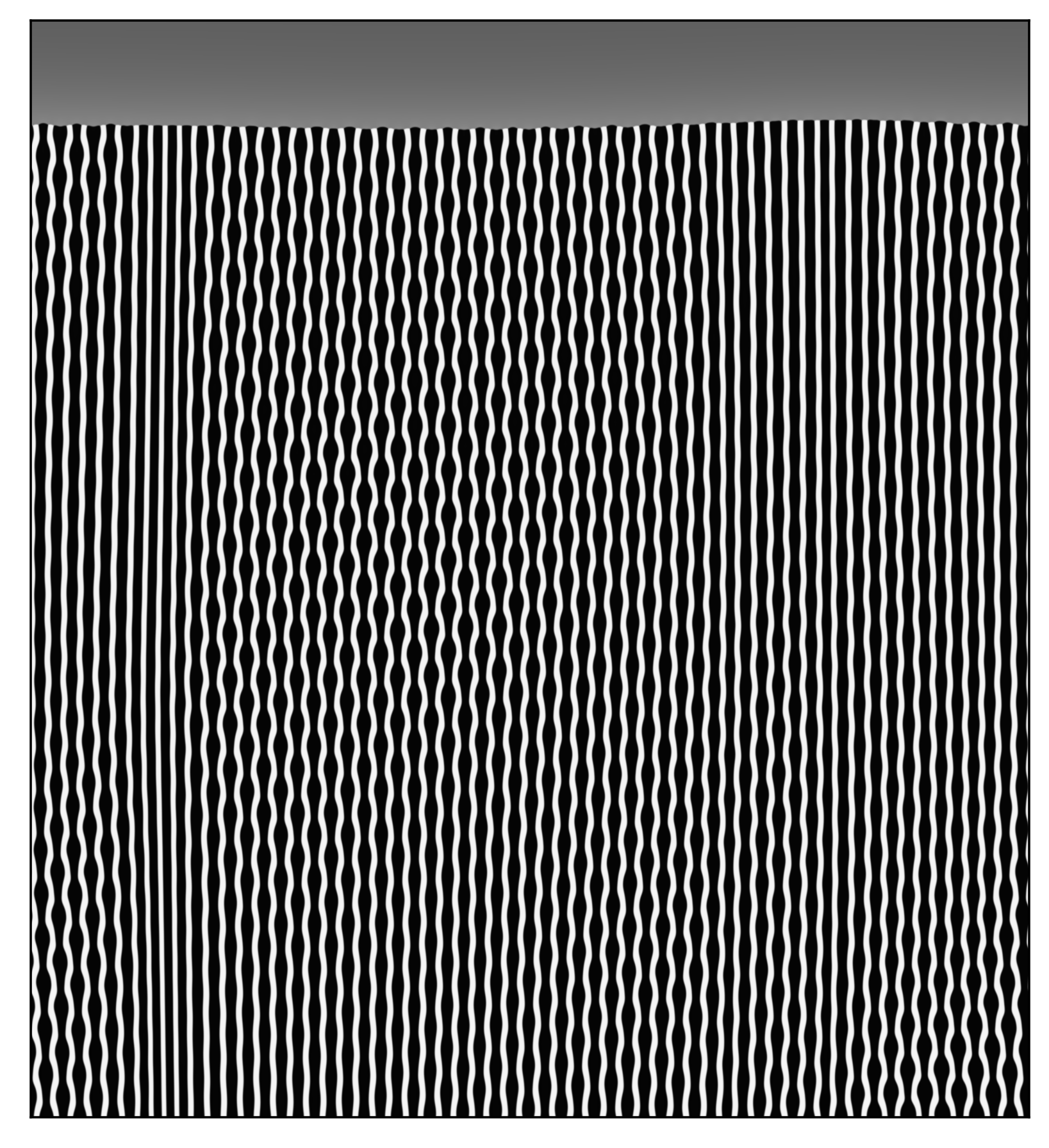}
    \caption{$c_0 = 0.13$}
    \end{subfigure}
    
    \caption{
    Observed microstructures for $v=\SI{160}{\um/s}$, $G=\SI{24.7}{K/mm}$ and various melt compositions.
    Both purely dendritic as well as eutectic structures are found as well as simulations in which both morphologies grow within the moving window concurrently.
    }
    \label{fig:withG}
\end{figure}

\Cref{fig:withG} shows exemplary simulation results.
Purely dendritic (D), dendritic-eutectic (D+E) and purely eutectic (E) structures are observed, depending on the melt composition $c_0$.
Note that in the case of dendritic-eutectic structures, the $\theta$ lamellas close to the dendrite are thicker than in the middle.
This is due to the melt composition close to the dendrite being enriched in Cu which is rejected by the dendrite, which is also easily observed with the composition field being slightly brighter (more Cu) closer to the dendrite.
Simulations in which only dendrites remain will be counted as dendritic-eutectic in the following.
This is due to the fact that if a sufficiently higher moving cutoff were to be used, the eutectic would not be shifted out of the domain and hence both morphologies would be observed, as long as the melt composition is larger than the corresponding solidus composition.
Generally, if dendritic-eutectic growth is the goal of the simulation, then the simulation needs to be able to span the temperature difference between the dendrite front temperature $T_{df}$ and the eutectic front temperature $T_{ef}$.
With the frozen temperature approximation (\cref{eq:t_evolution}) this suggests that the physical domain up to the moving window cutoff should be at least $L = \frac{T_{df} - T_{ef}}{G}$.
If this length is negative, it also implies that the eutectic should be the dominant morphology.
Note that this is a necessary but not sufficient condition, as the initial conditions have an effect on the resulting morphology as will be shown later.

\begin{figure}
 \includegraphics[width=\textwidth]{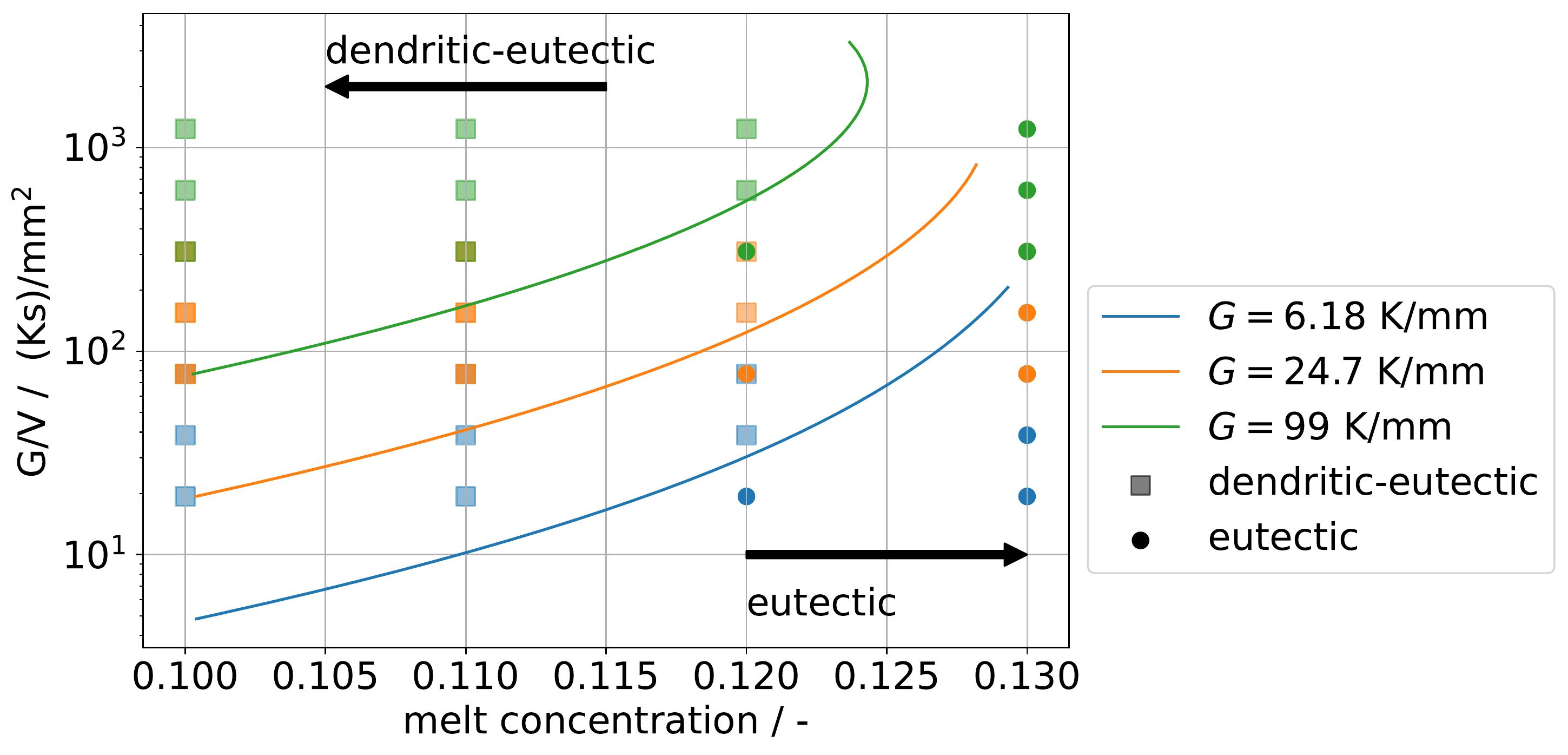}
 \caption{
 The microstructure map differentiating the eutectic range from the dendritic-eutectic range.
 The theoretical boundary curve clearly separates the two observed morphology regimes.
 }
 \label{fig:micromap}
\end{figure}

The results can be displayed succinctly in a $\{ c_0-G/v \}$ plot as suggested by \cite{Burden1974}.
This is done in \cref{fig:micromap}, displaying the results for all simulations at once along with the boundary curves calculated based on the theory described in \ref{sec:theory}.
All eutectics, represented by circles, lie to the right of their respective boundary curves.
Similarly, the dendritic-eutectic structures are observed to the left of the curves, suggesting that the maximum temperature condition for the transition between eutectic and dendritic-eutectic morphologies describes the boundary curve well.
This also implies that the front undercooling of the individual morphologies is either not significantly changed compared to their isolated growth or changed by the same value.
Due to the choice of $G-v$ pairs, several points result in the same $G/v$ value but with different gradients and different morphologies.
Thus the full specification of solidification conditions ($\{v, G, c_0\}$) is necessary to determine the morphology.

\begin{figure}
\begin{center}
\begin{subfigure}[h]{0.45\textwidth}
 \includegraphics[width=\textwidth]{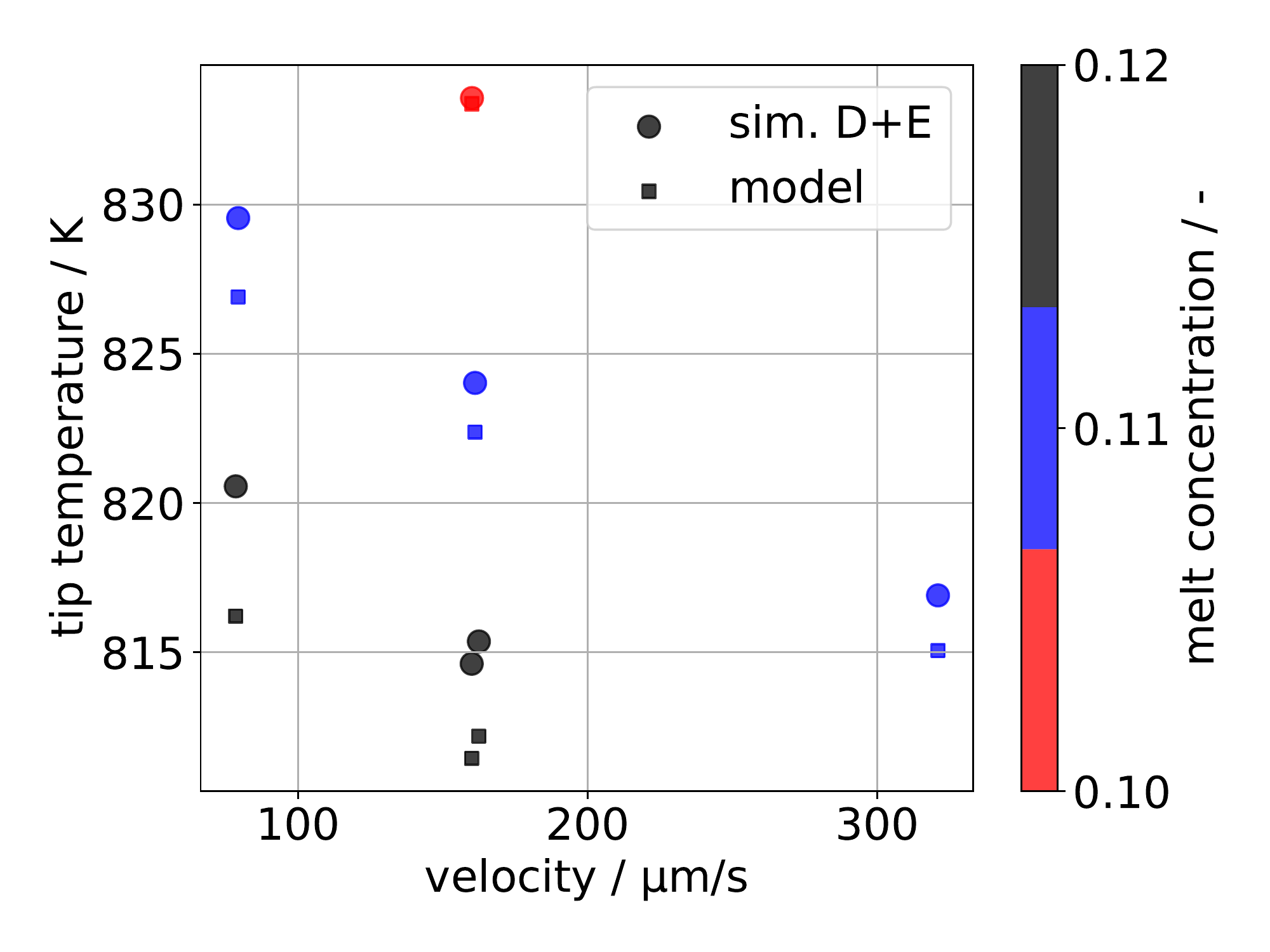}
 \end{subfigure}
\hspace{0.2cm}
 \begin{subfigure}[h]{0.45\textwidth}
  \includegraphics[width=\textwidth]{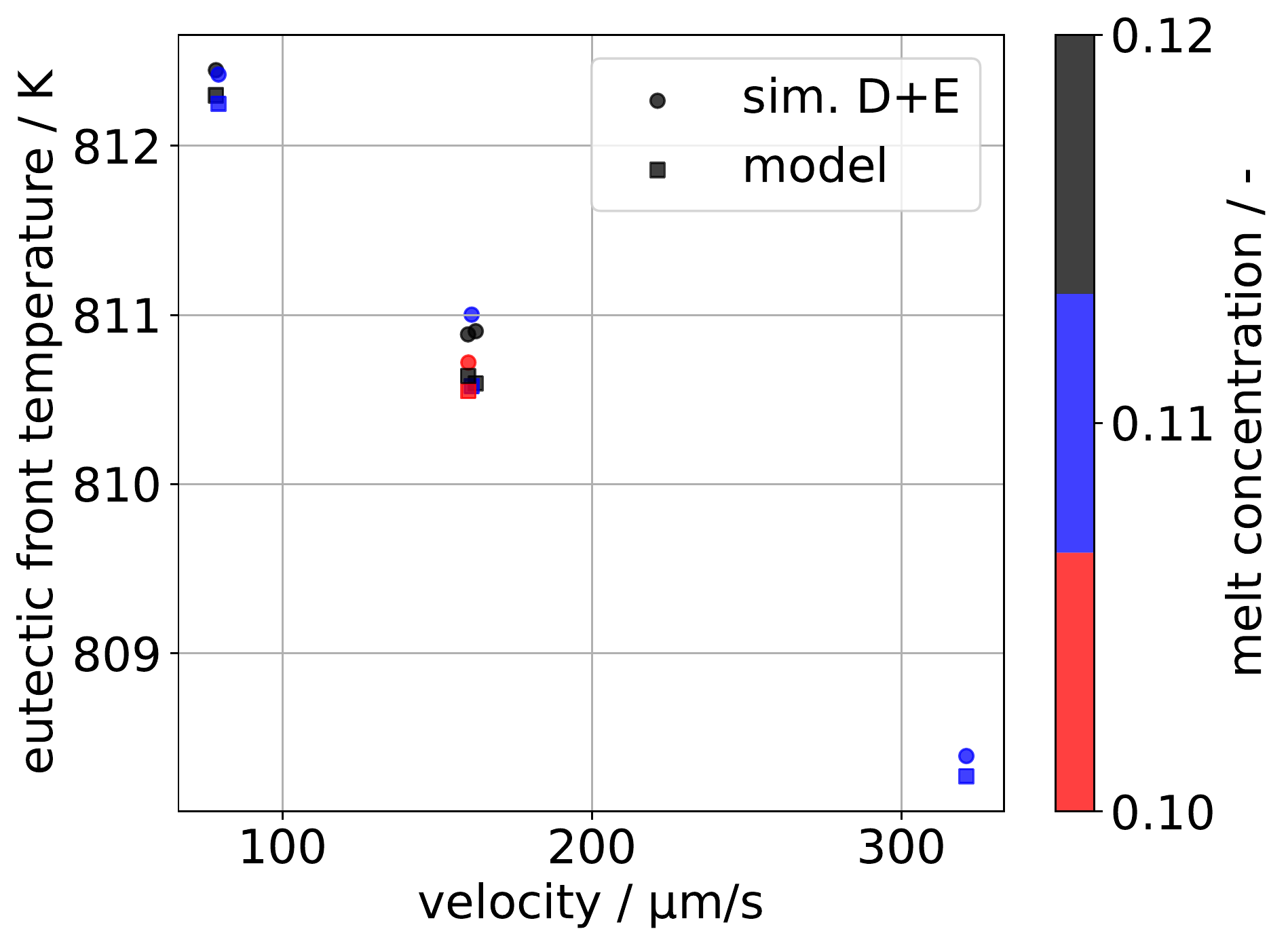}
 \end{subfigure}
 \caption{Comparison of observed front temperatures during dendritic-eutectic growth and the prediction of the respective isolated growth models.
 There is a systematic underprediction of front temperature, but of similar magnitude as the earlier deviations between data and the model.
 Thus the coupled growth does not seem to affect the undercooling-velocity relationship significantly.}
 \label{fig:gc-compare}
 \end{center}
 
\end{figure}

The observed growth conditions ($\Delta T-v$) can be compared to the models which were determined earlier.
This is shown in \cref{fig:gc-compare}.
While there is a systematic underprediction of the undercooling by the model, it is of similar magnitude as to the isolated growth conditions which were used to determined the model parameters.
Thus there is no significant effect of coupled growth on the underlying undercooling-velocity relationship.

Next, the influence of dendritic-eutectic growth on the microstructural lengths is investigated.
The relevant microstructural lengths of the dendrite are the primary dendrite arm spacing (PDAS) and secondary dendrite arm spacing (SDAS).
In the present setup one cannot make statements about the PDAS as usually only a single dendrite is contained within the simulation domain.
However, a qualitative statement regarding the SDAS is possible:
If the eutectic grows sufficiently close to the dendrite tip, secondary arms cannot develop fully before being enveloped by the eutectic.
Thus the SDAS will tend to be smaller than for purely dendritic growth.

The eutectic spacing however can be easily investigated for the present simulations, as large numbers of lamellas are contained within the eutectic and dendritic-eutectic simulations.
A bit of preprocessing is necessary for dendritic-eutectic simulations in order to exclude the dendrite and its closest neighboring $\theta$ lamellas from the analysis:
Specifically, the $\alpha$ and $\theta$ phases are separated and segmented\cite{cc3d} on their own.
For the $\alpha$ phase, the isotropic and anisotropic variants are added together.
It is assumed that any segments larger than four times the median are dendrites, which are henceforth excluded from the analysis.
Furthermore, small segments of e.g. failed nucleation are excluded as well by using a minimum segment size of 100 cells.
For the $\theta$ phase the lamellas close to the dendrite need to be excluded as these are severely thicker.
Since a simple size threshold is hard to define for these, only the $\theta$ segments past the second and before the second to last $\alpha$ lamella are analyzed, with the same small segment filter applied as for the $\alpha$ phase.
The remaining segments are put together to form an image of a ``well-formed'' eutectic, which is analyzed with the same procedure as for purely eutectic simulations.
In the present case, the individual phase widths $w_\alpha, w_\theta$ perpendicular to the growth direction are analyzed, with their sum being the spacing $\lambda$.

\begin{figure}
  \includegraphics[width=\textwidth]{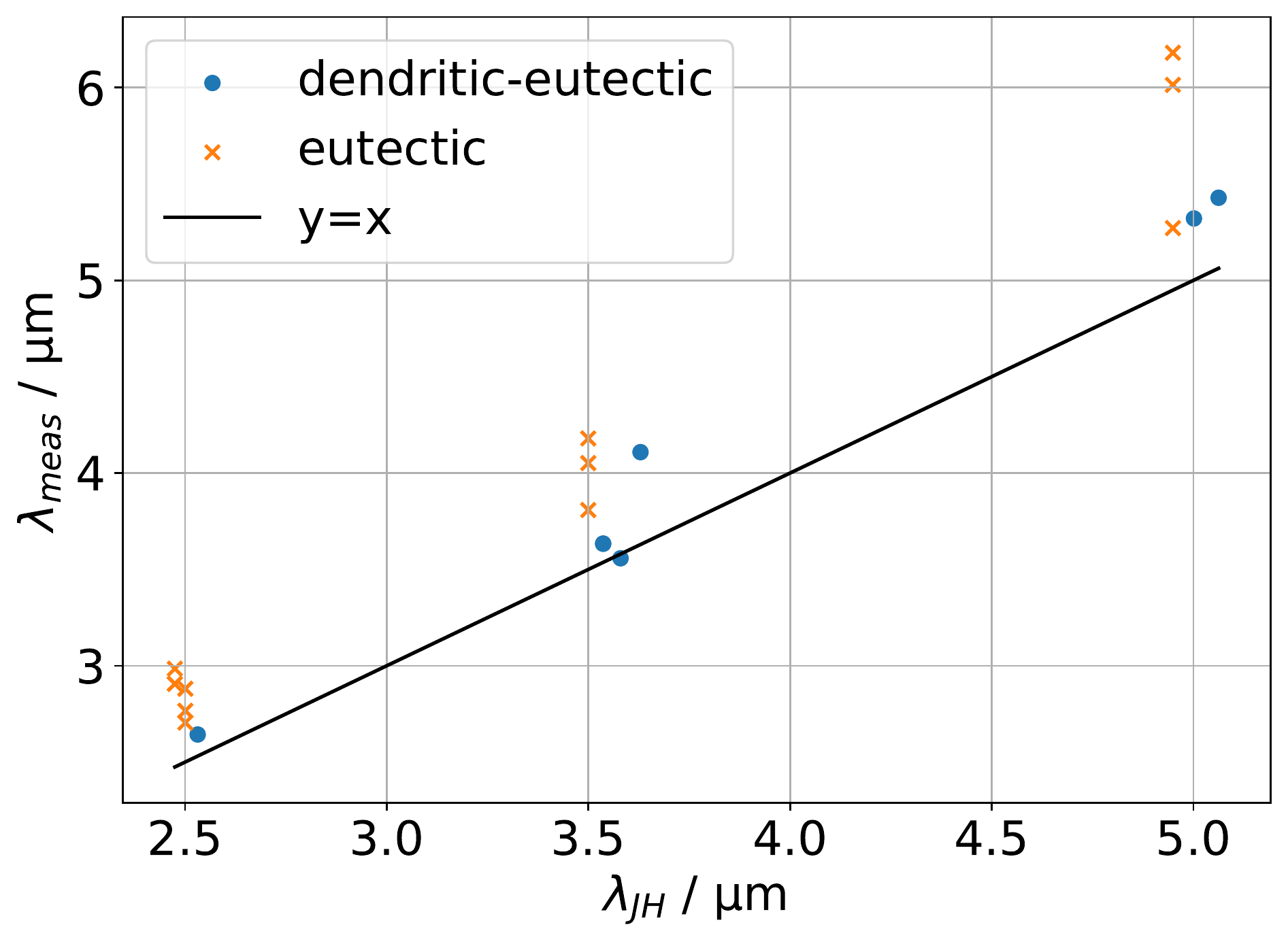}
  \caption{
  A comparison between the theoretically expected spacings $\lambda_{JH}$ and the measured spacings $\lambda_{meas}$.
  The black line serves as a guide for the eye.
  The dendritic-eutectic simulations tend to be above this line but roughly parallel to it.
  The eutectic simulations tend to deviate more.
  }
   \label{fig:comparison-spacing}
\end{figure}

The results of analyzing the simulations containing a eutectic are shown in \cref{fig:comparison-spacing} with a scatter plot of the theoretically calculated and measured spacings.
If there is no influence of the dendrite on the growing eutectic, then the results should cluster around the line $y=x$.
This is generally observed, with a slight scatter upwards.
The eutectic simulations tend to be above the line, due to a combination of factors:
First, many of the $\alpha$ lamellas are represented by the dendritic phase, as these lamellas originally branched off from the dendrite.
Thus these have a different surface energy and also triple point angles.
Second, as explained in the validation, the nucleation mechanism tends to generate slightly larger spacings than predicted by the minimum undercooling criterion in the JH theory.
When comparing the dendritic-eutectic to the purely eutectic simulations, the presence of a dendrite tends to slightly decrease the spacing.
One possible explanation for this is that the dendrite itself tends to increase the Cu content in the melt ahead of the eutectic, altering the far-field the eutectic is growing against.
In order to estimate the effective far-field concentration, the fraction of $\theta$ within the eutectic is evaluated.
The total composition leading to this fraction is then iteratively determined and thus an estimate for the effective far-field concentration obtained.
This would theoretically lead to refinements on the order of $\SI{0.01}{\um}$ to $\SI{0.1}{\um}$ for the present simulations, with the actual refinement ranging from $\SI{0.1}{\um}$ to $\SI{0.5}{\um}$.
Thus only a part of the observed deviations can be explained with far-field effects.
The remaining effect might be due to structural effects of the dendrite on the eutectic, which will be the subject of further research.

\begin{figure}
  \includegraphics[width=\textwidth]{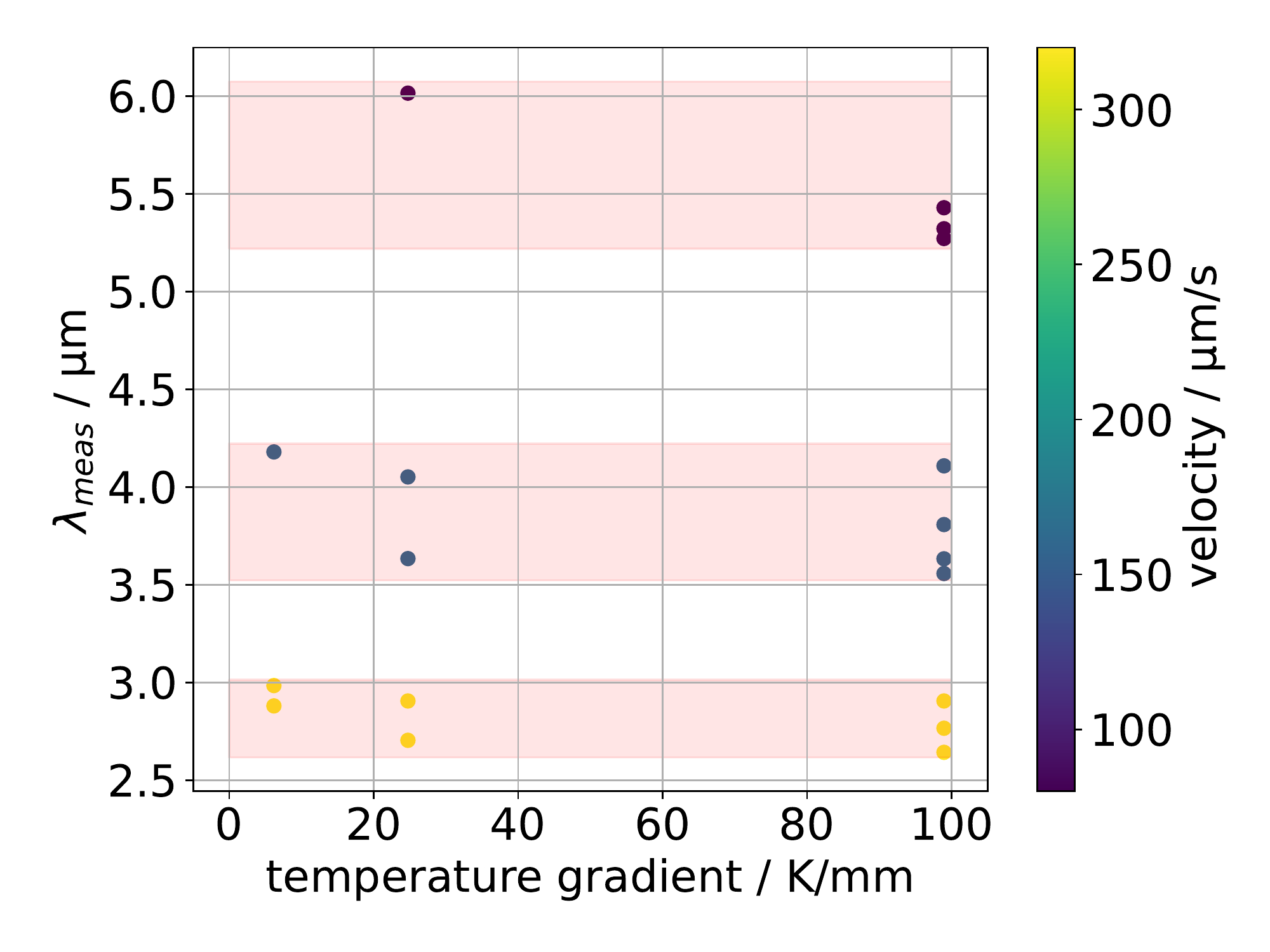}
  \caption{
  The measured lamellar spacing for all simulations containing eutectic is plotted over the employed temperature gradients.
  For each employed velocity, a band of spacings is spanned by the system, indicated by the shaded regions.
  Excepting the smallest velocity, there is little difference between spacings at the lowest and highest gradients.
  }
   \label{fig:comparison-gradient}
\end{figure}

Furthermore, the present data can also be analyzed as to whether the temperature gradient has any influence on the eutectic spacing relationship, since this is excluded in the theoretical considerations.
Plotting the spacings for the eutectic simulations over the gradient yields \cref{fig:comparison-gradient}, which shows individual bands of spacings for each velocity.
Excepting the slowest velocity, there is little difference between spacings obtained at the highest and lowest gradients.
The smaller velocities and temperature gradients tend to show larger oscillations in the lamellar structure, making the measurement less reliable for these.
In total however there seems to be no significant influence of the temperature gradient on the spacing within the simulations.

\begin{figure}
   \begin{subfigure}[b]{0.3\textwidth}
    \includegraphics[width=\textwidth]{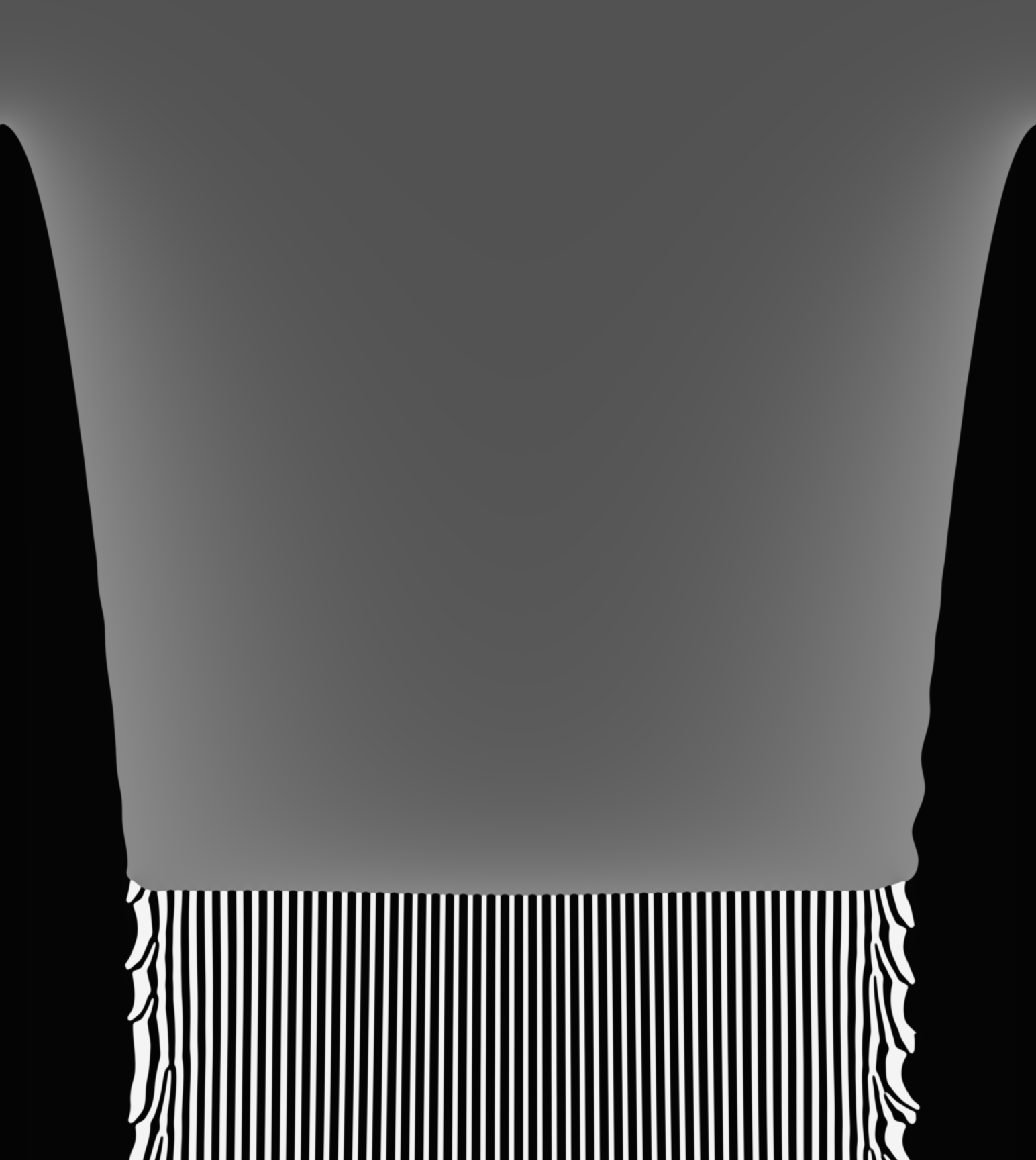}
    \caption{\centering $t = \SI{0}{s}$}
    \label{fig:vjump-ini}
  \end{subfigure}
  \begin{subfigure}[b]{0.3\textwidth}
    \includegraphics[width=\textwidth]{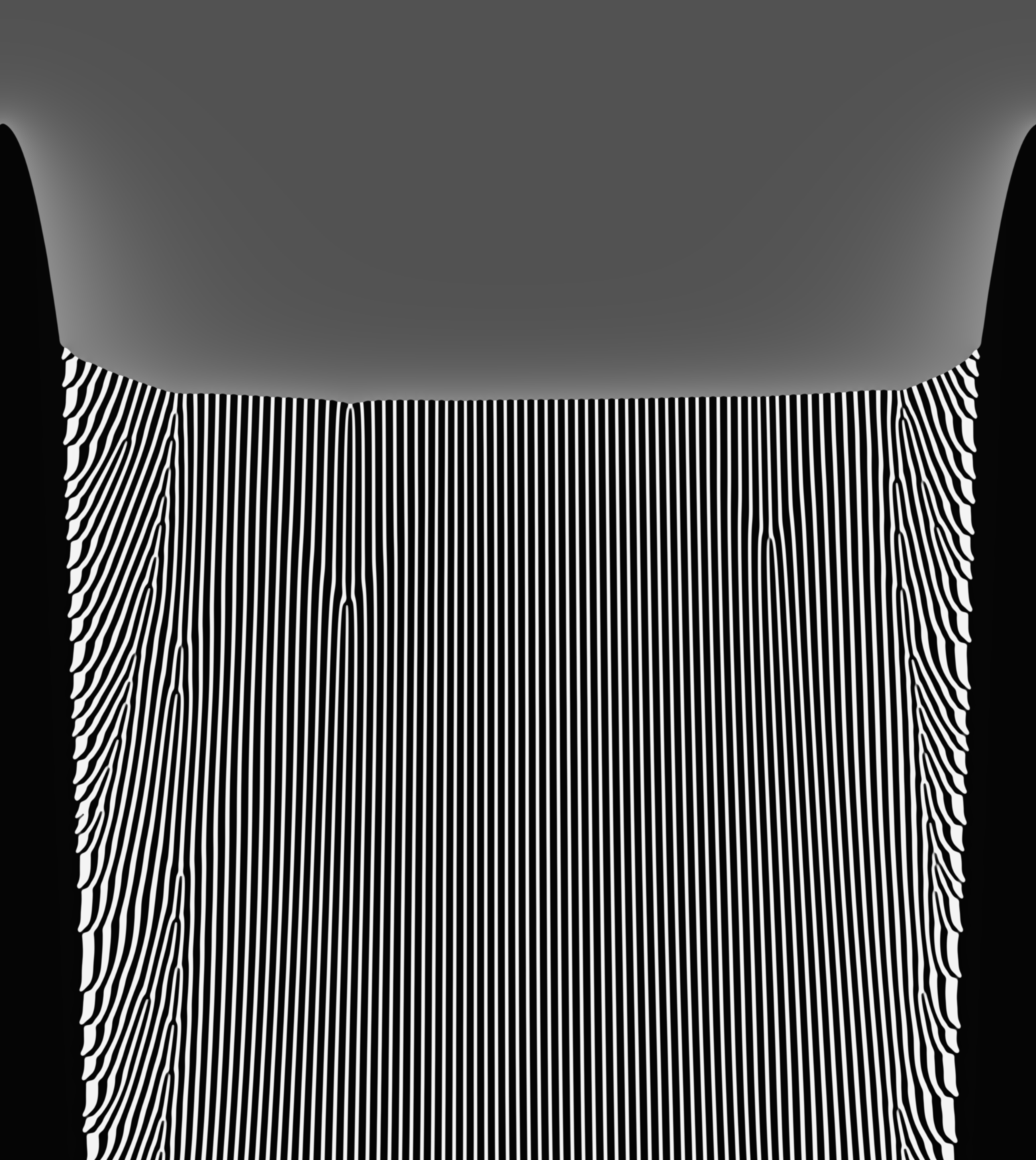}
    \caption{\centering $t = \SI{1.875}{s}$}
    \label{fig:vjump-mid}
  \end{subfigure}
    \begin{subfigure}[b]{0.3\textwidth}
    \includegraphics[width=\textwidth]{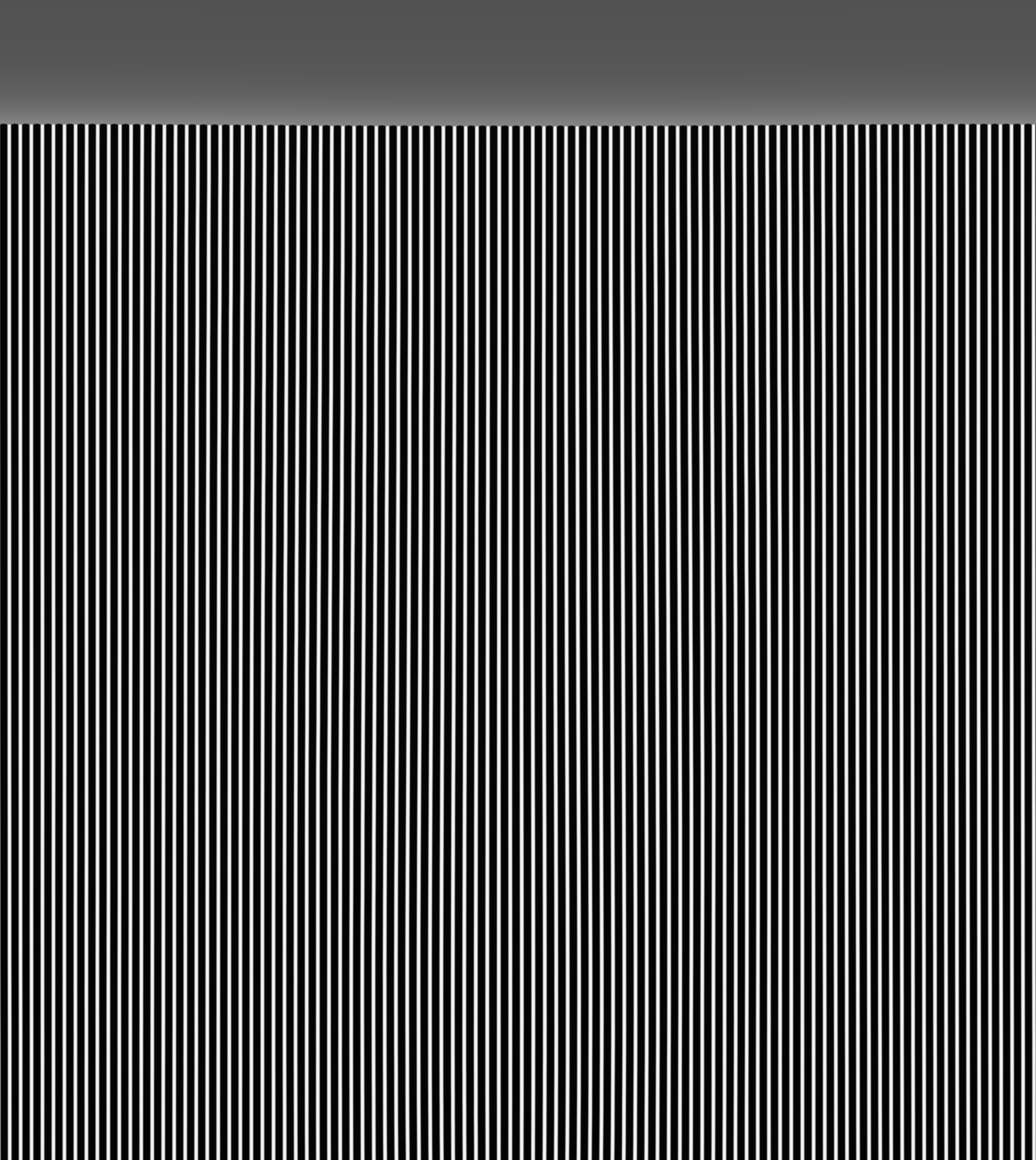}
    \caption{\centering $t = \SI{5.5}{s}$}
    \label{fig:vjump-last} 
    \end{subfigure}
    
    \begin{subfigure}[b]{0.9\textwidth}
   \includegraphics[width=\textwidth]{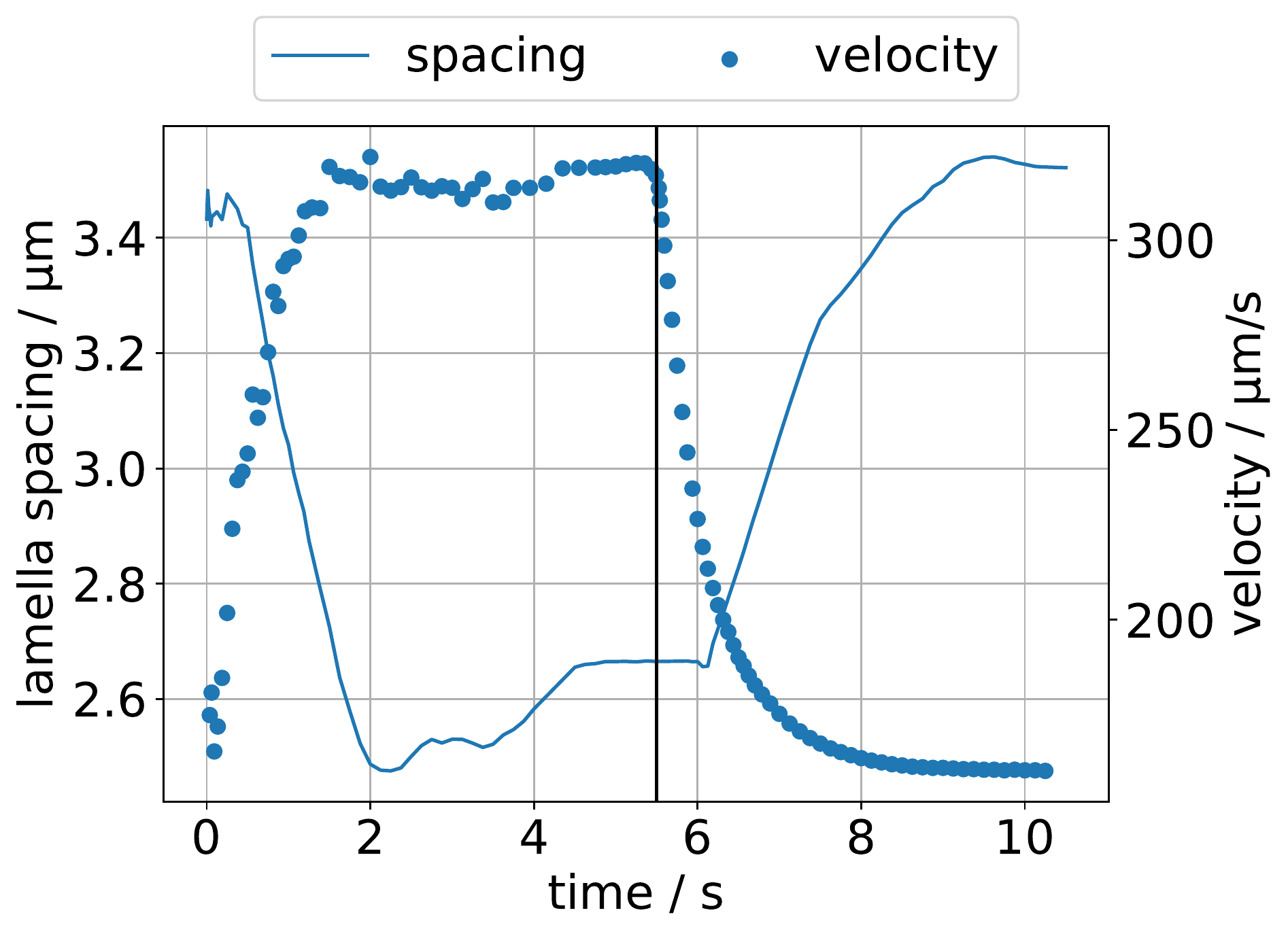}
   \caption{Lamella spacing and velocity over time.}
   \label{fig:vlam-vjump}
   \end{subfigure}
  \caption{
  The top row shows simulation states for a jump from $\SI{160}{\um/s}$ to  $\SI{320}{\um/s}$, up to the point where the jump is reverted.
  The eutectic grew at a constant distance from the dendrite tip prior to the jump.
  After the jump, it slowly creeps upwards towards the dendrite tip before enveloping it and establishing a flat eutectic front.
  At the bottom, the lamellar spacing and eutectic velocity during the entire process is shown, with the black vertical line separating the two velocity regimes.
  The velocity begins adjusting almost immediately, with the lamellar spacing lagging behind in its adjustment.
  There tends to be an over/undershoot in the spacing before a stable spacing is reached.
  }
   \label{fig:de-to-e-full}
\end{figure}

\paragraph{Influence of velocity variation}
Next, simulations will be conducted in order to investigate transitions between the morphologies by abruptly changing the velocity of the temperature field.
The first transition is for a gradient of $\SI{24.7}{K/mm}$ and a melt concentration of $0.12$,  with the velocity jump being from $\SI{160}{\um/s}$ to $\SI{320}{\um/s}$.
This should move the simulation from a dendritic-eutectic growth regime into a purely eutectic growth regime.
\Cref{fig:vjump-ini,fig:vjump-mid,fig:vjump-last} show the results for speeding up a dendritic-eutectic front.
The eutectic slowly grows upwards until it overtakes the dendrite, resulting in a flat eutectic front.
During this process the eutectic becomes finer, as would be expected from theory.
After a flat eutectic front is obtained, the jump is done in the other direction as to test for hysteresis effects on the morphology.
While the eutectic coarsened after the second jump, the eutectic front stayed stable with no dendrites forming.
Thus there is a certain dependence of prior microstructural history on which morphology is observed.
Since the prior simulations always started from a dendrite, the ``easy'' direction of morphological change was available and thus the boundary curves could be confirmed.
However, if the simulations were started from a eutectic front, it is likely that the eutectic range would be extended beyond the theoretical boundary curve.
Usually, primary solidification takes place before the eutectic grows and thus the morphological hysteresis should not play a role for experiments.

The spacing and velocity of the eutectic are analyzed during the whole process and are shown in \cref{fig:vlam-vjump}, with the black vertical line separating the two different imposed velocities.
It is observed that while the velocity begins adjusting almost immediately, the eutectic spacing lags behind.
After the original velocity is reached again, a similar spacing is observed again, confirming that the eutectic spacing is not subject to hysteresis effects\cite{Trivedi1991}.

\begin{figure}
\begin{center}
   \begin{subfigure}[b]{0.3\textwidth}
    \includegraphics[width=\textwidth]{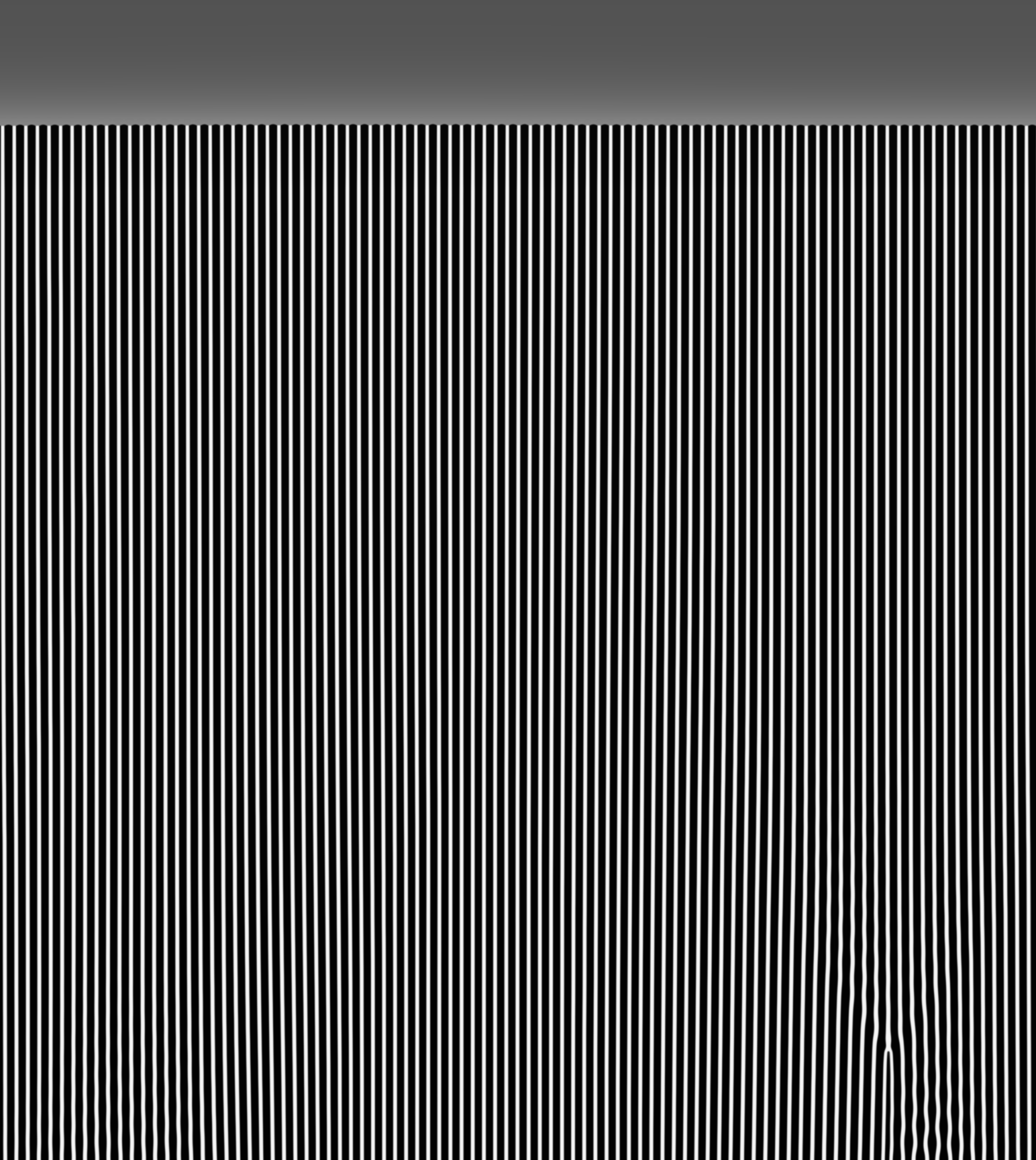}
    \caption{\centering $t = \SI{0}{s}$}
  \end{subfigure}
  \begin{subfigure}[b]{0.3\textwidth}
    \includegraphics[width=\textwidth]{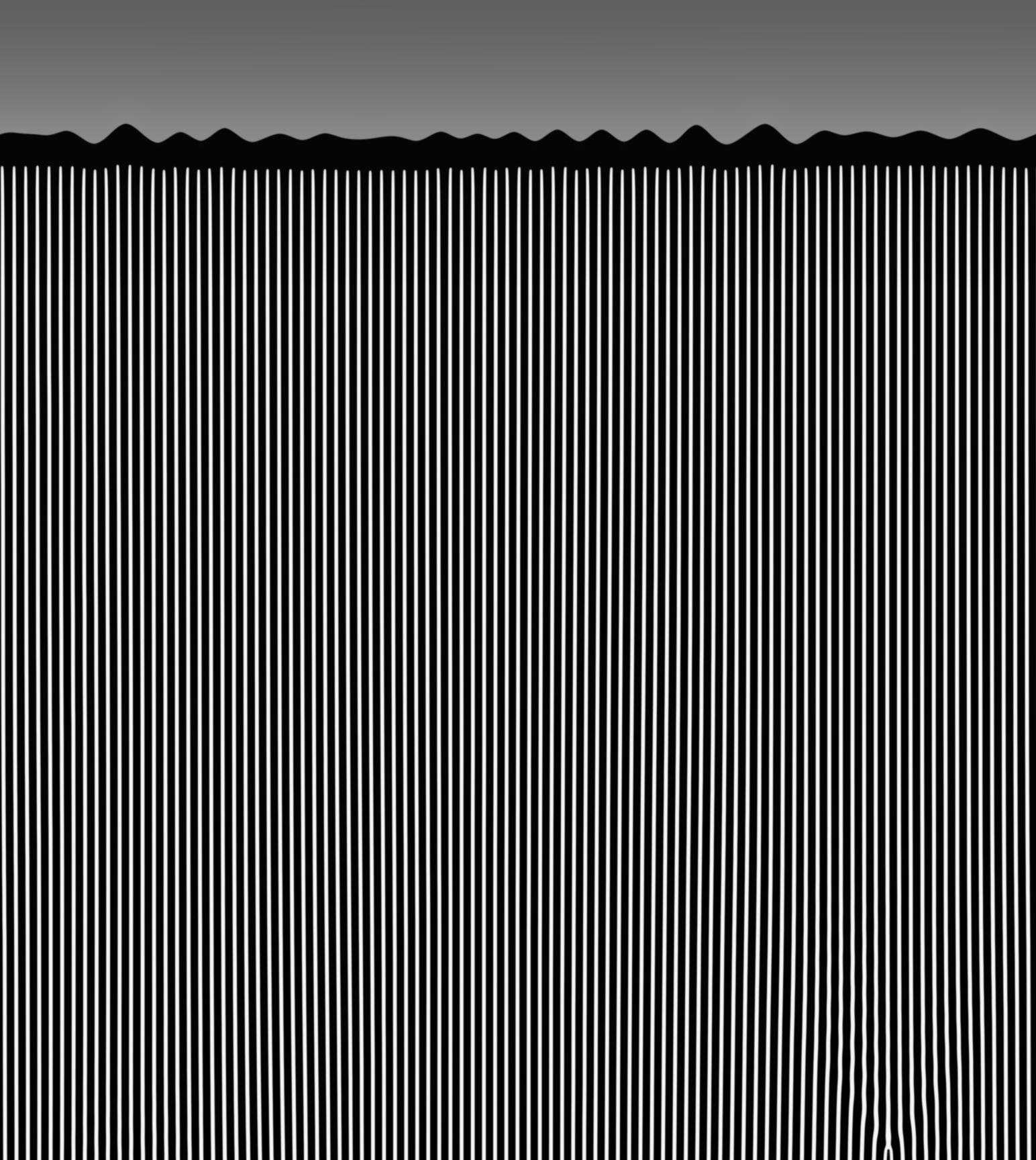}
    \caption{\centering $t = \SI{0.313}{s}$}
  \end{subfigure}
    \begin{subfigure}[b]{0.3\textwidth}
    \includegraphics[width=\textwidth]{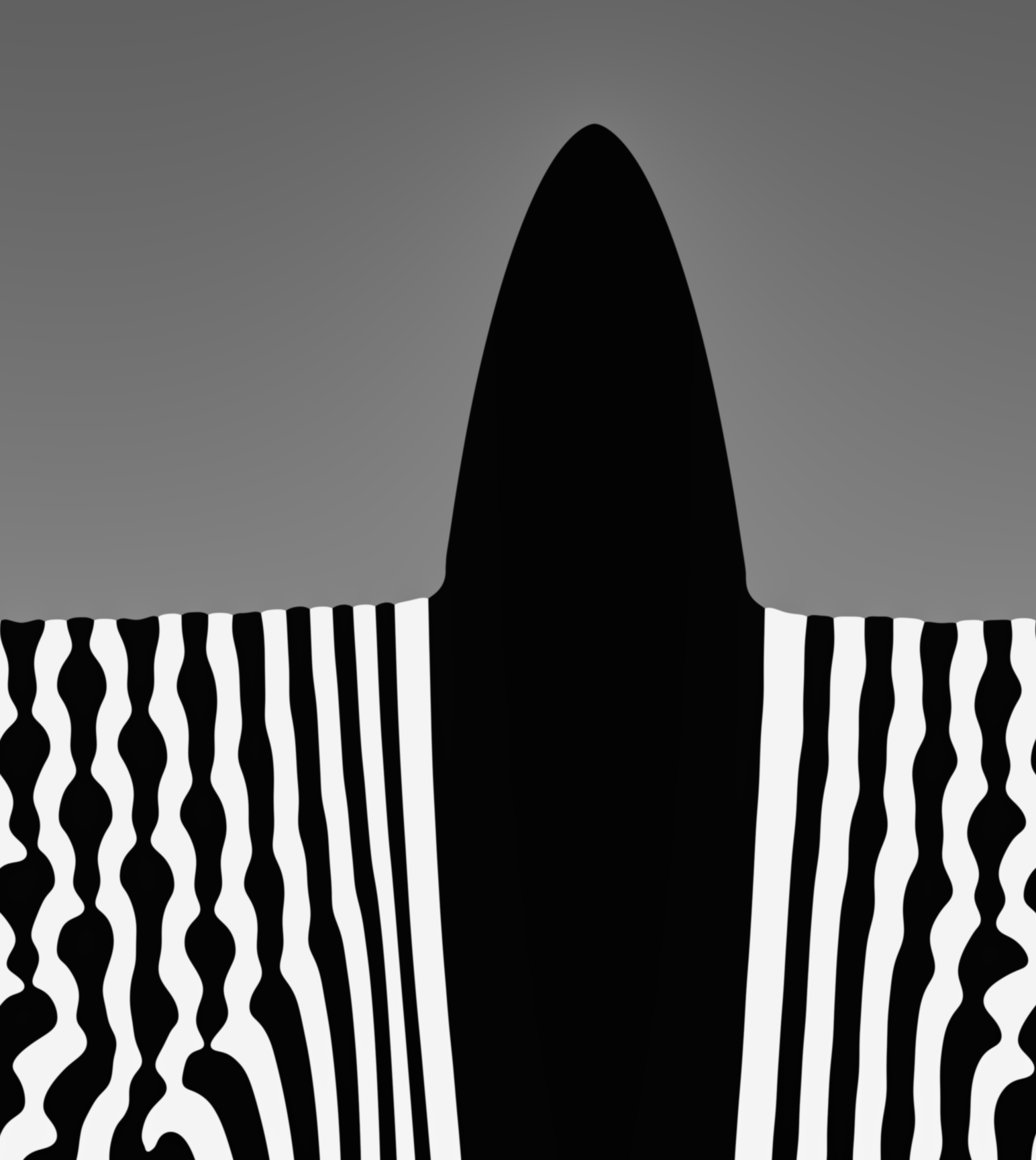}
    \caption{\centering $t = \SI{16.5}{s}$}
    \end{subfigure}
  \caption{
  Intermediate simulation states for a velocity jump from $\SI{320}{\um/s}$ to $\SI{20}{\um/s}$.
  Shortly after the velocity jump a band of $\alpha$ forms above the eutectic front.
  This band undergoes a Mullins-Sekerka instability allowing for a single dendrite to emerge surrounded by coarse eutectic.
  }
  \label{fig:vLower}
  \end{center}
  
\end{figure}

The second transition is for a gradient of $\SI{99}{K/mm}$ and the same melt concentration of $0.12$, with the velocity jump being from $\SI{320}{\um/s}$ to $\SI{20}{\um/s}$, moving a eutectic into the dendritic-eutectic regime.
Due to the priorly observed hysteresis, a much larger velocity jump is employed in this case.
Sufficient space between the solidification front and the boundary is kept by extending the domain height to \SI{1000}{\um}, yielding about 7.5 diffusion lengths.
\Cref{fig:vLower} shows the results for the second case of slowing down a eutectic front.
After a short initial period, $\alpha$ overgrows the eutectic front and forms a band.
This band then undergoes a Mullins-Sekerka type of instability, with $\theta$ nucleating in concave regions.
The convex regions can grow into dendrites.
In the present case only a single dendrite grows, with a coarse eutectic growing around it.
The simulation is not run to convergence as the small velocity would necessitate excessively long simulations.
For this reason and because the eutectic nucleates anew above the destabilized band, the eutectic spacing is not analyzed in this case.

\paragraph{Complete directional solidification}
Three simulations approximating complete directional solidification, from below the liquidus down into the eutectic region, are performed.
The previous simulations start out with the front temperature below the eutectic temperature, in which case there should already have been a dendritic structure for the eutectic to grow into.
For these simulations the moving window technique is deactivated and the domain height is extended to \SI{1500}{\um} and the width to \SI{500}{\um}.
The first two simulations should contain mostly one morphology, with the parameters $v = \SI{320}{\um/s}$, $G=\SI{24.7}{K/mm}$ being employed for both simulations, but two different melt compositions $c_0 \in \{0.08, 0.12\}$ being used.
The former should yield a primarily dendritic structure, with the latter exhibiting a primarily eutectic structure based on the calculated boundary curve.
As an example of a primarily dendritic-eutectic structure, a third simulation with $v = \SI{160}{\um/s}$, $G=\SI{24.7}{K/mm}$ and $c_0 = 0.12$ is conducted.
The starting temperature $T_0 = \SI{836}{K}$ for these simulations is chosen well below the respective liquidus temperatures but above the eutectic temperature.
On one hand this allows a substantial amount of primary solidification while on the other hand cooling below the eutectic temperature is achievable with a reasonable amount of computational effort.

\begin{figure}[p]
\begin{center}
   \begin{subfigure}[b]{0.3\textwidth}
    \includegraphics[width=\textwidth]{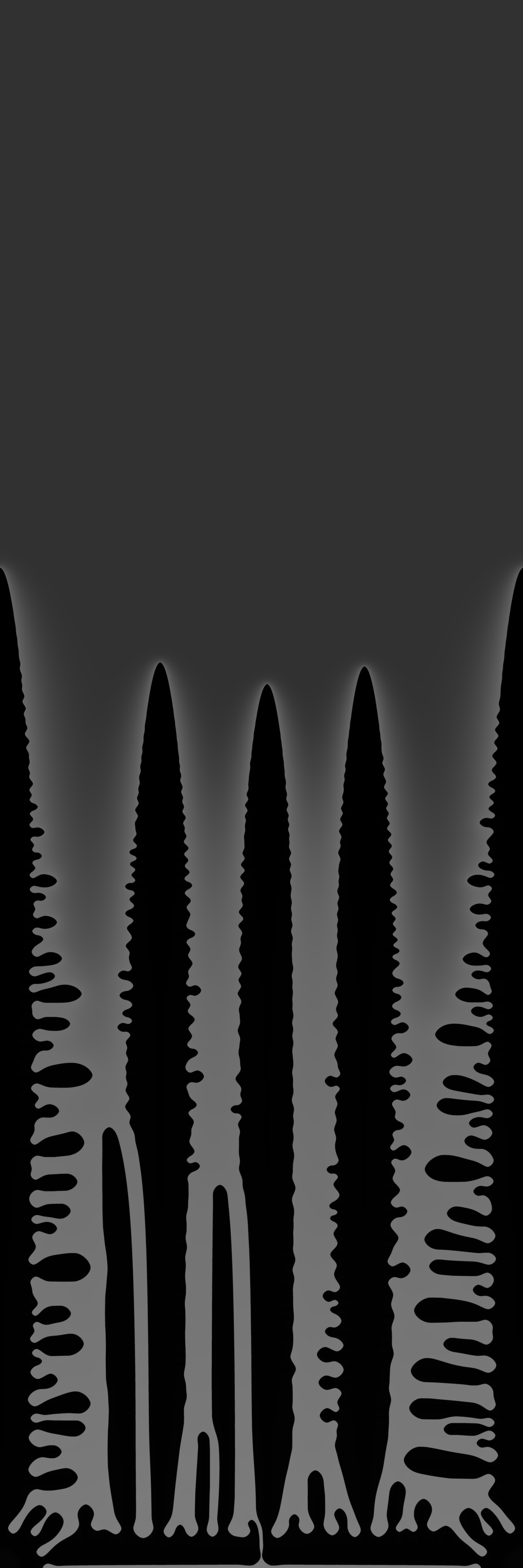}
    \caption{\centering $ t = \SI{1.81}{s}$, $T_b=\SI{821}{K}$}
  \end{subfigure}
  \begin{subfigure}[b]{0.3\textwidth}
    \includegraphics[width=\textwidth]{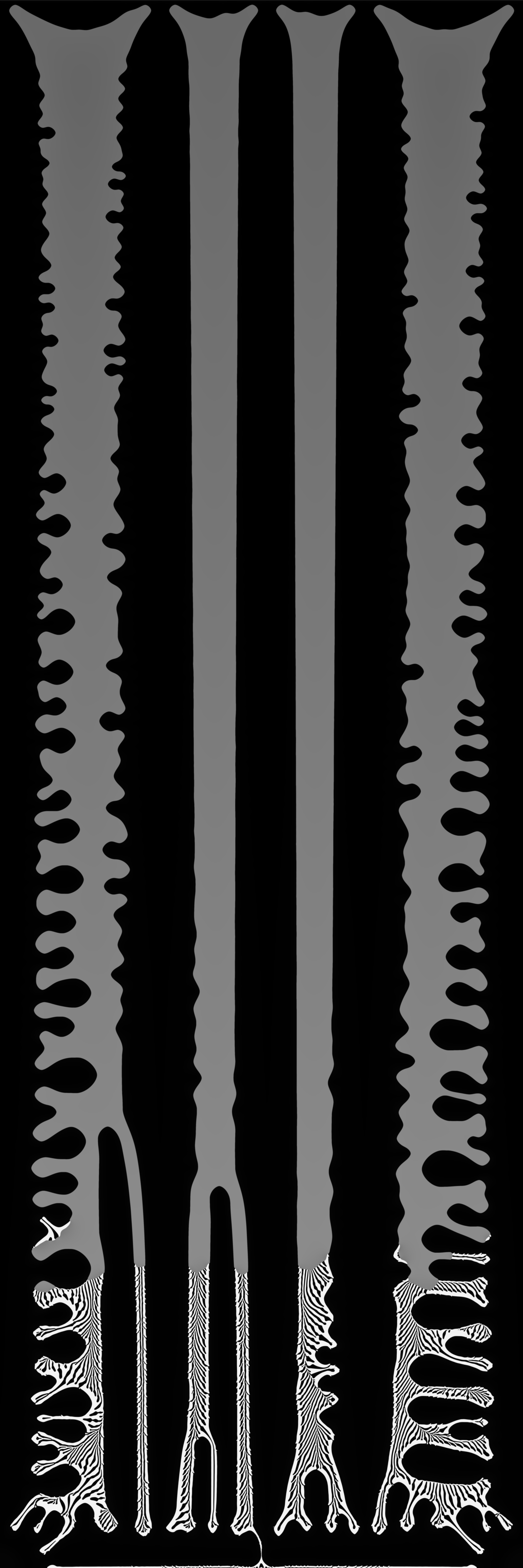}
    \caption{\centering $t = \SI{4.31}{s}$, $T_b=\SI{802}{K}$}
    
  \end{subfigure}
    \begin{subfigure}[b]{0.3\textwidth}
    \includegraphics[width=\textwidth]{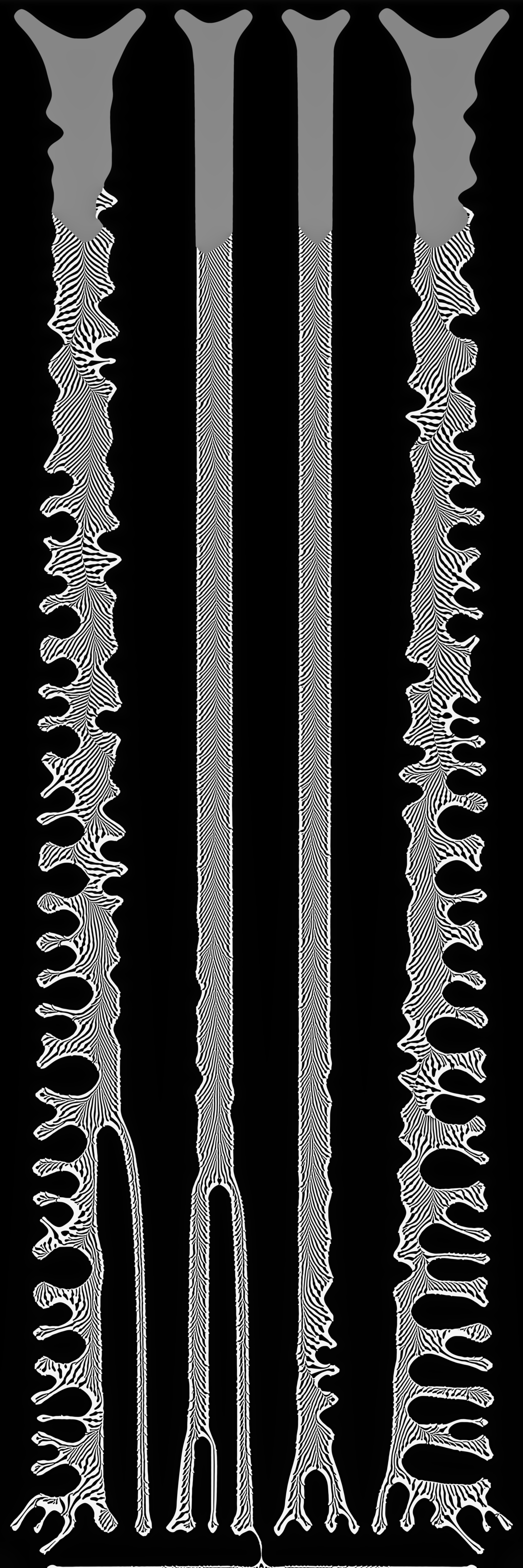}
    \caption{\centering $t = \SI{7.38}{s}$, $T_b=\SI{778}{K}$}
    \end{subfigure}
  \caption{
  Intermediate simulation states for a complete solidification of a Al-8at\%Cu alloy from below the liquidus line across the eutectic line with $v = \SI{320}{\um/s}$.
  First, primary dendrites grow in the direction of the temperature gradient until the top of the domain is reached.
  Afterwards, the dendritic branch structure coarsens and at about \SI{4}{K} below the eutectic temperature the eutectic nucleates near the bottom of the domain.
  This eutectic grows upwards, but new eutectic tends to nucleate faster in the side branch structure than the front can grow.
  Hence different orientations of somewhat lamellar structures are observed.
  }
  \label{fig:complete-sol-d}
    \end{center}
\end{figure}

In \cref{fig:complete-sol-d} the time-resolved microstructure for $c_0 = 0.08$ is shown.
It can easily be observed in (a) that primary solidification occurs via dendrites which grow until they reach the top of the domain (b, c).
Secondary arms are clearly visible (a), but as solidification progresses a significant number of secondary arms retracts towards the primary dendrites (b).
The eutectic starts off nucleating near the bottom of the domain and then grows upwards in the side channels of the dendrites, but this is not the only mode of growth (b,c).
Rather, the eutectic front tends to be nucleated anew in the Cu-rich pockets formed by dendritic sidearms and then grows towards the main channel, partially closing it off to the eutectic growing up from the bottom of the channel.
Thus if an alloy crosses both the primary crystallization regime and the eutectic line during solidification, then eutectics of different dominant orientation should be found around dendritic structures.
One should be mostly aligned with the dendritic growth direction, whereas the other with the growth direction of the side arms.

\begin{figure}[hp]
\begin{center}
   \begin{subfigure}[b]{0.3\textwidth}
    \includegraphics[width=\textwidth]{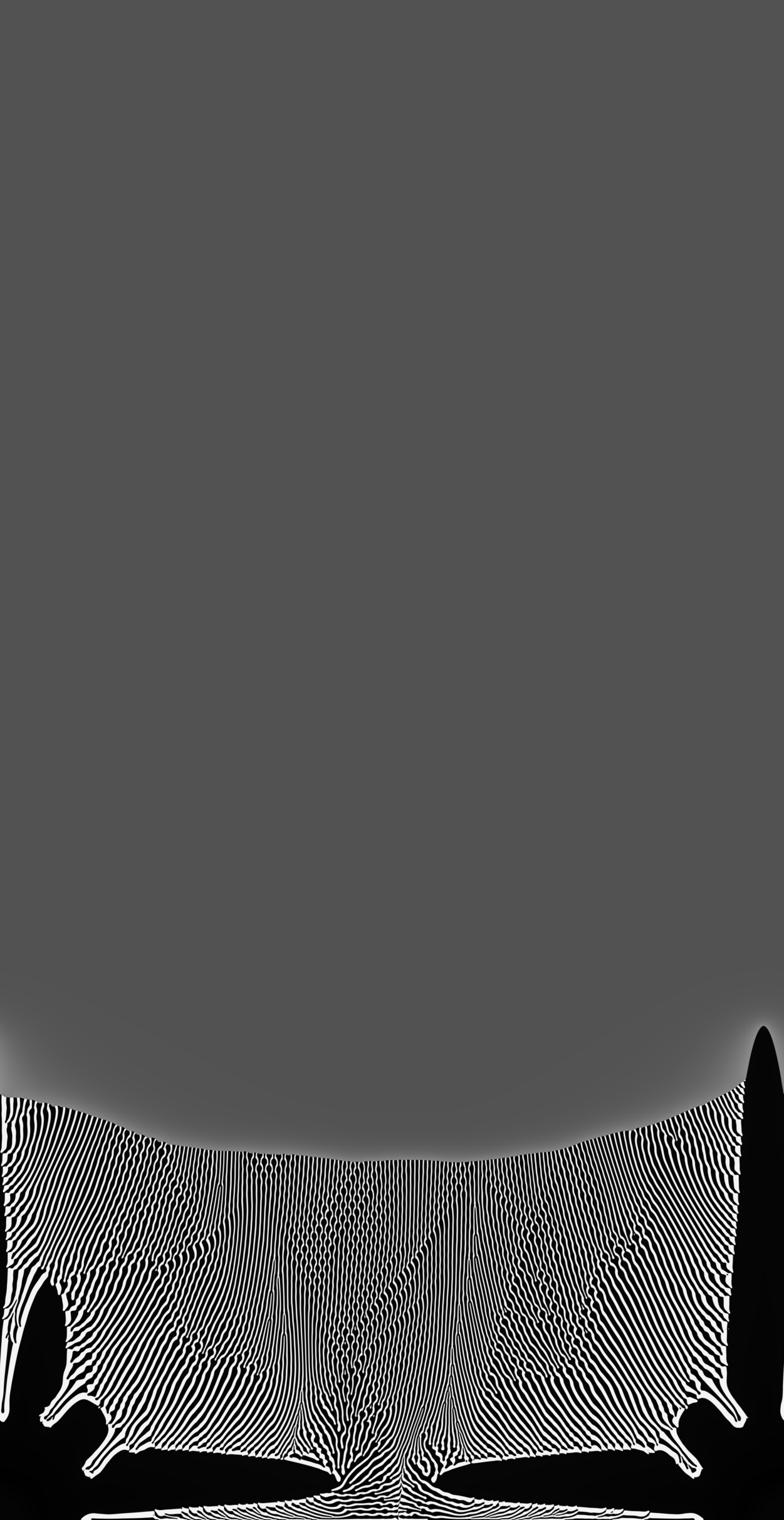}
    \caption{\centering $ t = \SI{4.31}{s}$, $T_b=\SI{802}{K}$}
  \end{subfigure}
  \begin{subfigure}[b]{0.3\textwidth}
    \includegraphics[width=\textwidth]{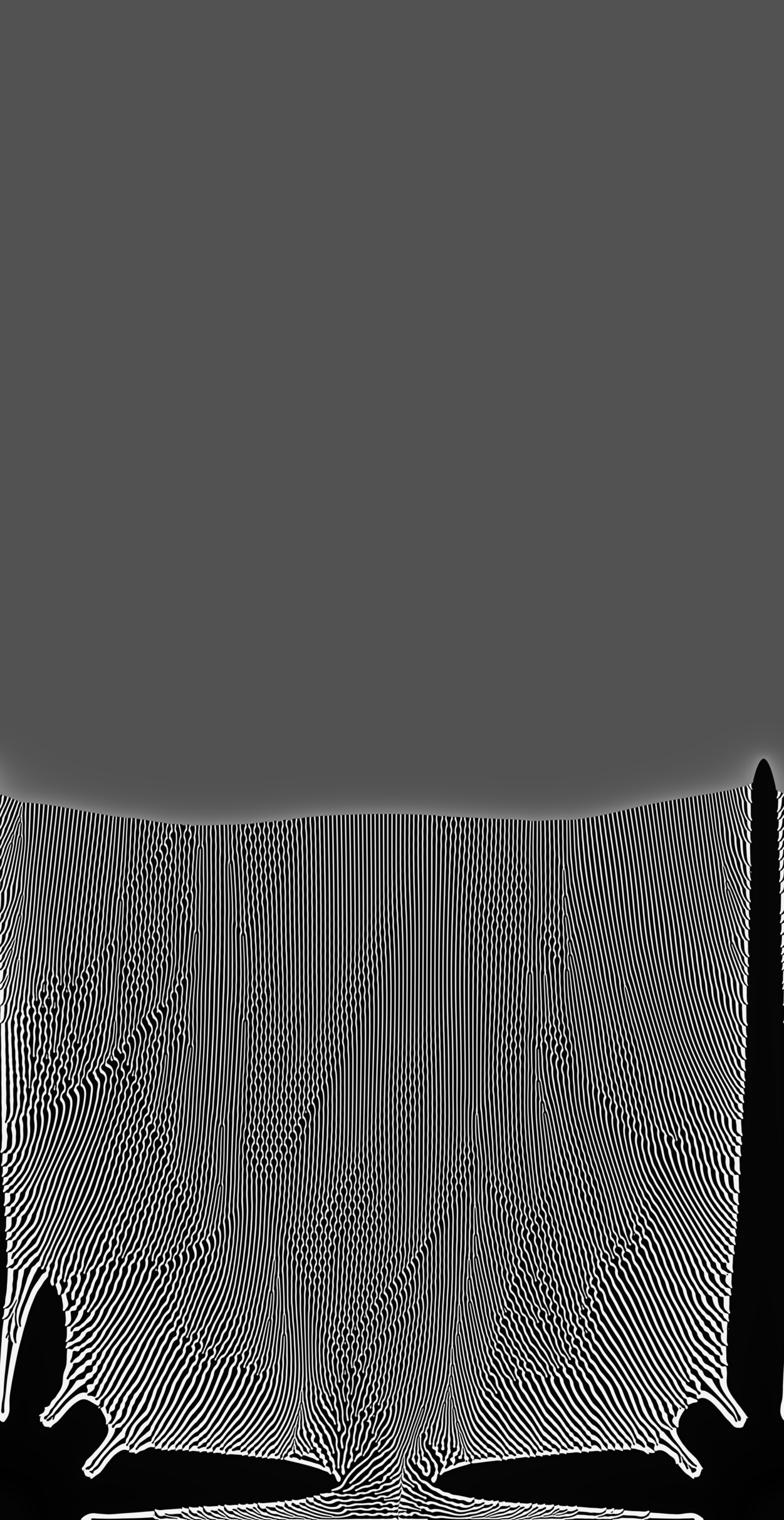}
    \caption{\centering $ t = \SI{4.94}{s}$, $T_b=\SI{797}{K}$}
  \end{subfigure}
    \begin{subfigure}[b]{0.3\textwidth}
    \includegraphics[width=\textwidth]{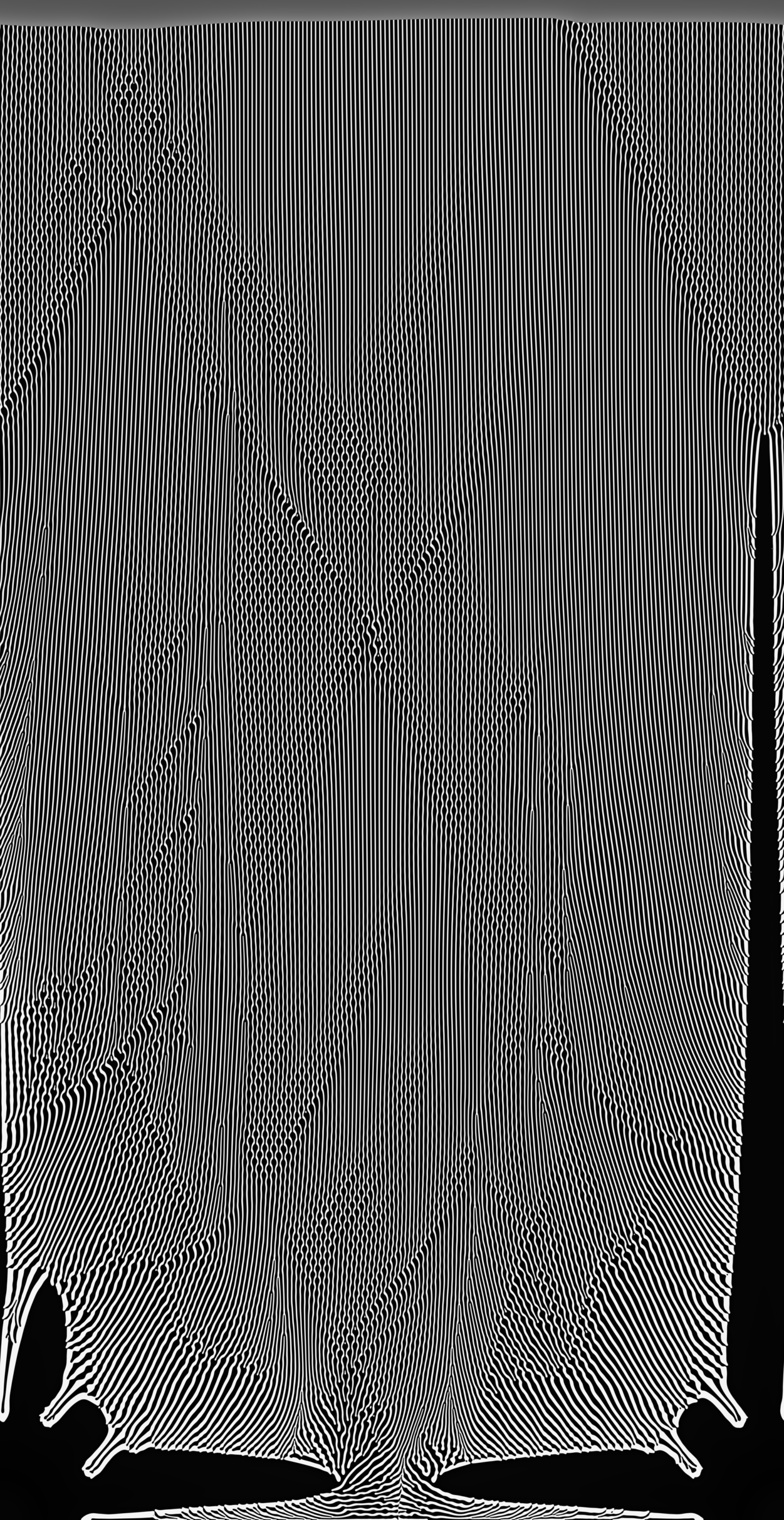}
    \caption{\centering $ t = \SI{6.5}{s}$, $T_b=\SI{785}{K}$}
    \end{subfigure}
  \begin{subfigure}[b]{1.0\textwidth}
   \includegraphics[width=\textwidth]{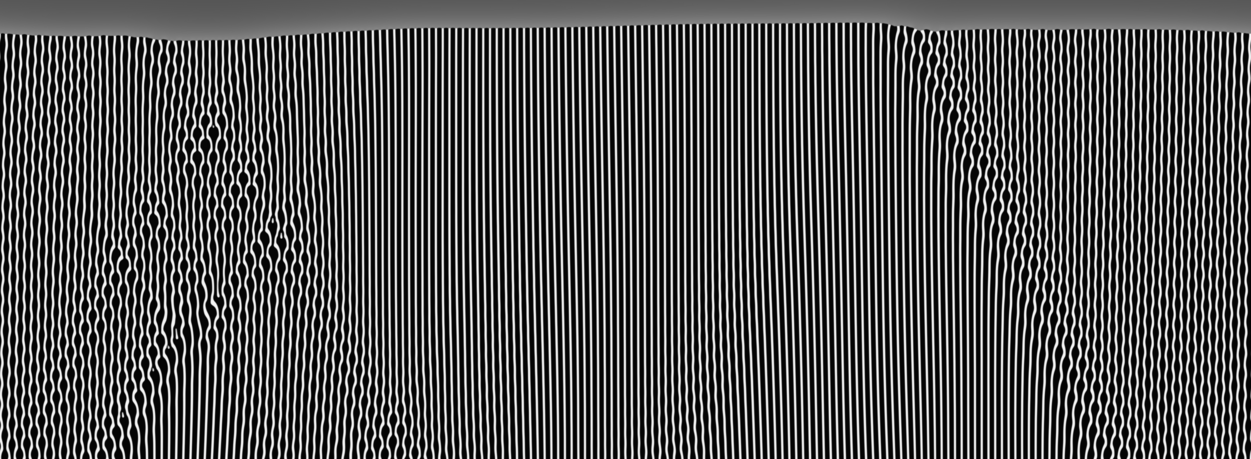}
    \caption{\centering $ t = \SI{6.5}{s}$, $T_b=\SI{785}{K}$, closeup of the eutectic front}
  \end{subfigure}

  \caption{
  Intermediate simulation states for a complete solidification of a Al-12at\%Cu alloy from below the liquidus line across the eutectic line with $v = \SI{320}{\um/s}$.
  The images are cropped to slightly above the final position of the eutectic front, with the remaining size being $\SI{970}{\um} \times \SI{500}{\um}$.
  First, a primary dendrite grows slowly until eutectic starts forming.
  The eutectic creeps up the dendrite, forcing the dendrite to taper off until overgrown.
  Oscillations which travel across the eutectic structure are clearly visible in the closeup.
  Even after the dendrite is eliminated the eutectic front is still observed to be slightly curved.
  }
  \label{fig:complete-sol-e}
  \end{center}
\end{figure}

In \cref{fig:complete-sol-e} the completely eutectic structure is shown.
While a dendrite does grow initially, major parts of it are soon overtaken by the eutectic (a).
The dendrite itself gets progressively thinner as the eutectic grows upwards until it is engulfed by the eutectic.
The eutectic front is observed to be strongly curved during this overgrowth process (a,b), with some curvature still remaining after the overgrowth process (c).
Beyond the initial primary arms, no secondary arms can be observed.
The eutectic structure itself tends to contain oscillating waves (d) which travel across the structure at a roughly \SI{30}{\degree} offset from the growth direction.
This kind of travelling oscillatory wave was also found experimentally in \cite{Zimmermann1990} with a \SI{35}{\degree} offset from the growth direction.
These are also sometimes observed in the simulations with the moving window technique.
It should be noted that regions with oscillating lamellas tend to grow at a slightly lower temperature compared to those with straight lamellas.
Hence there is likely a correlation between the front curvature and the oscillating lamellas, though the determination of cause and effect of this correlation will be the topic of further research.

\begin{figure}[hp]
\begin{center}
   \begin{subfigure}[b]{0.3\textwidth}
    \includegraphics[width=\textwidth]{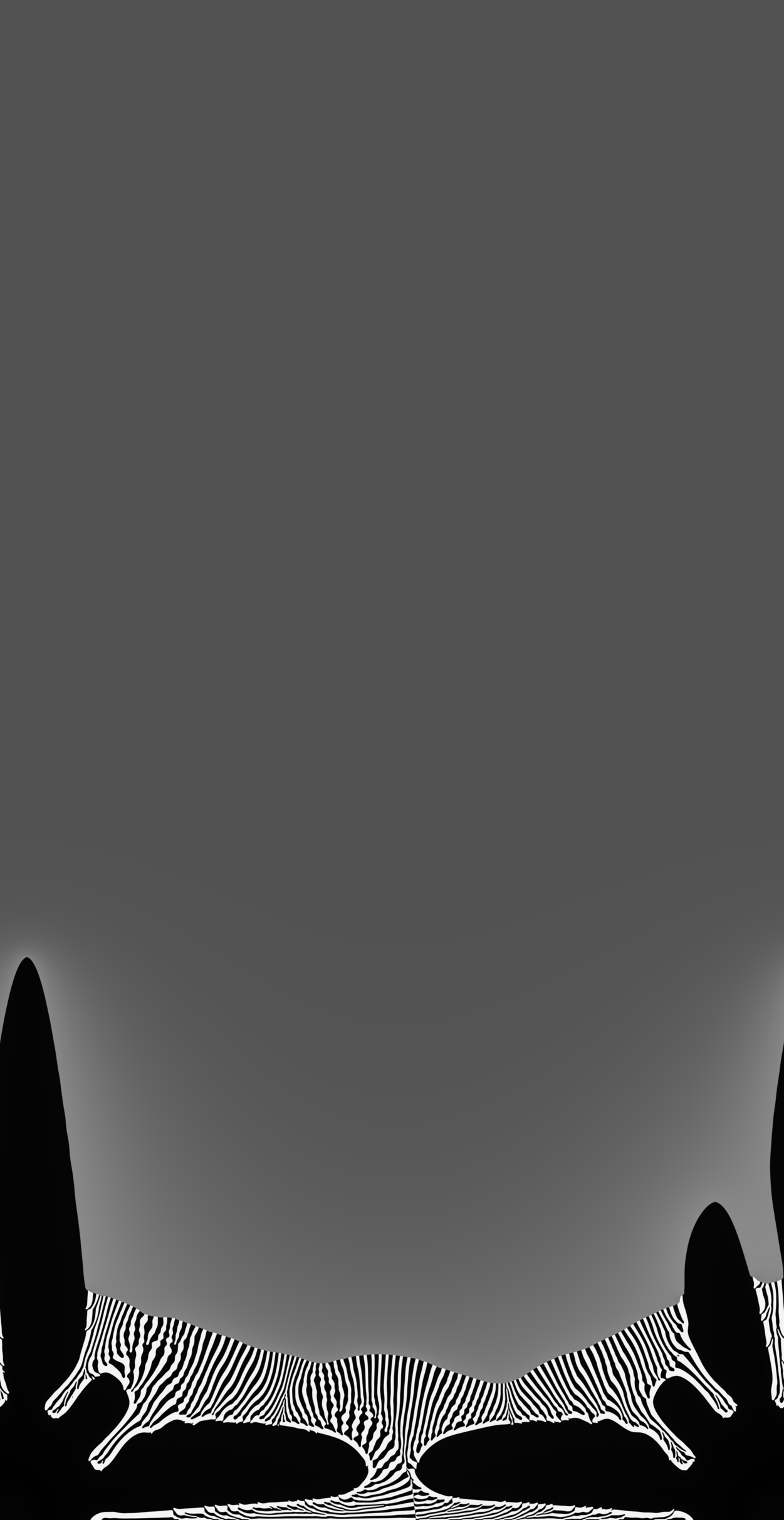}
    \caption{\centering $ t = \SI{7.44}{s}$, $T_b=\SI{806}{K}$}
  \end{subfigure}
  \begin{subfigure}[b]{0.3\textwidth}
    \includegraphics[width=\textwidth]{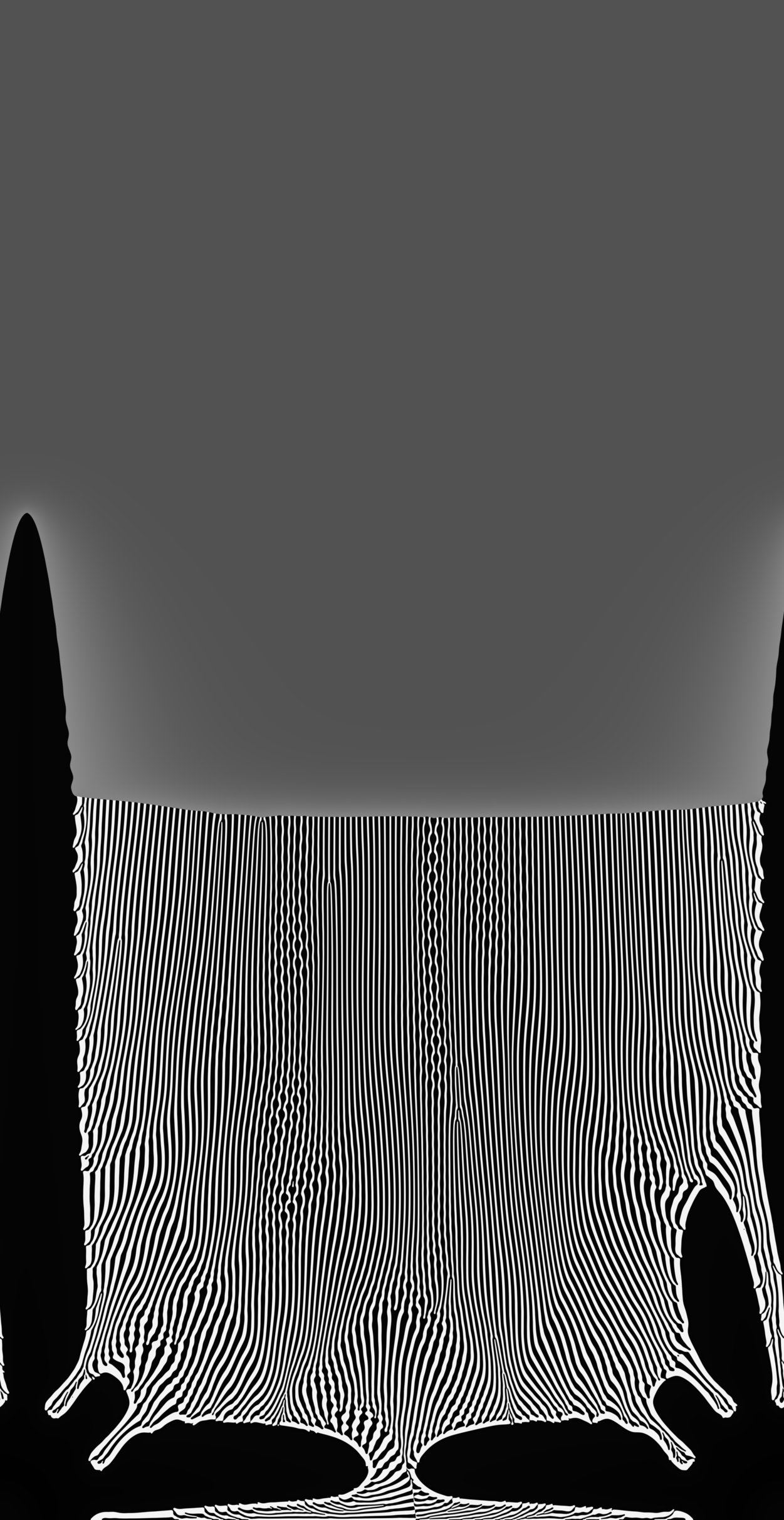}
    \caption{\centering $ t = \SI{9.31}{s}$, $T_b=\SI{799}{K}$}
  \end{subfigure}
    \begin{subfigure}[b]{0.3\textwidth}
    \includegraphics[width=\textwidth]{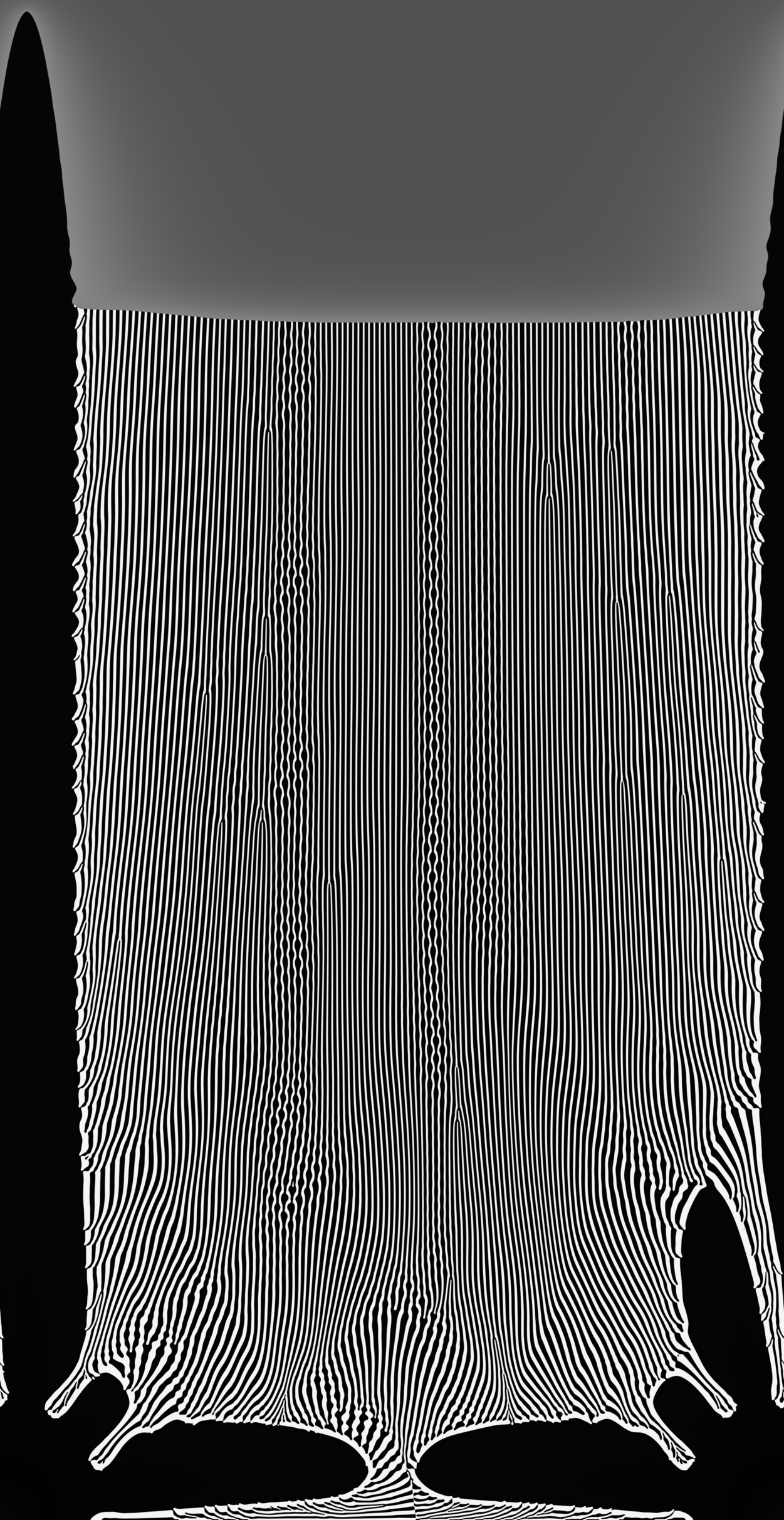}
    \caption{\centering $ t = \SI{11.25}{s}$, $T_b=\SI{792}{K}$}
    \end{subfigure}
    \begin{subfigure}[b]{1.0\textwidth}
    \includegraphics[width=\textwidth]{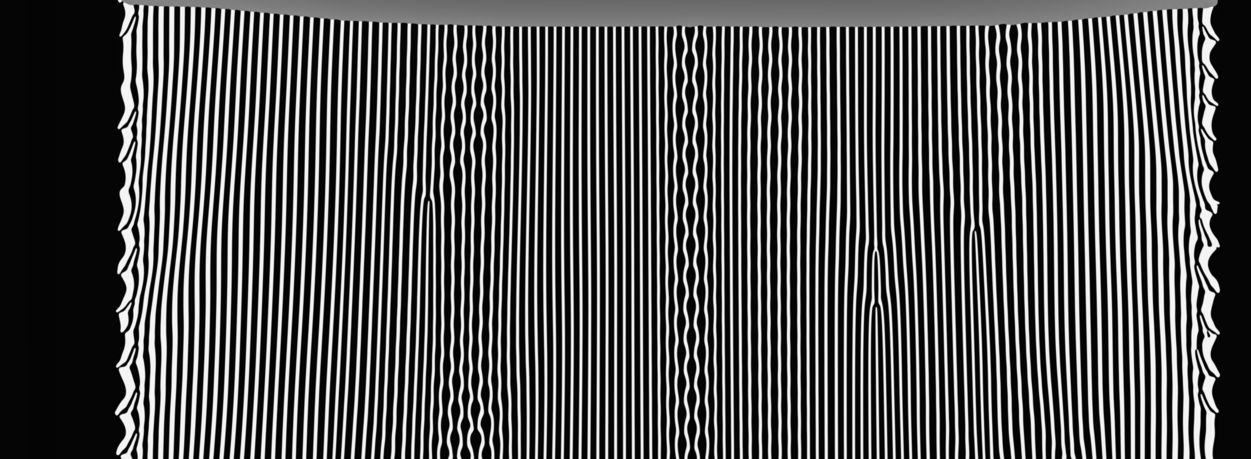}
    \caption{\centering $ t = \SI{11.25}{s}$, $T_b=\SI{792}{K}$, closeup of the eutectic front}
    \end{subfigure}
  \caption{
  Intermediate simulation states for a complete solidification of a Al-12at\%Cu alloy from below the liquidus line across the eutectic line $v = \SI{160}{\um/s}$.
  The images are cropped to slightly above the final position of the dendrite, with the remaining size being $\SI{970}{\um} \times \SI{500}{\um}$.
  First, a primary dendrite grows slowly until eutectic starts forming.
  The eutectic creeps up the dendrite, overgrowing secondary arms but is unable to reach the dendrite tip.
  A constant distance between the eutectic front and the dendrite tip is observed in the later stages.
  The eutectic front is observed to be curved when close to the dendrite.
  }
  \label{fig:complete-sol-de}
  \end{center}
\end{figure}

The last complete directional solidification simulation is shown in  \cref{fig:complete-sol-de}.
Similarly to the dominantly eutectic one, the dendrite grows first followed by eutectic.
However, a constant distance between the dendrite tip and the eutectic front is established and the two morphologies continue to grow in parallel.
The primary dendrite does not develop significant side arms, with the bumps quickly being covered by the eutectic.
While there are again oscillations in the eutectic structure, these do not travel across the structure and are rather localized.
In the closeup (d), the eutectic front can now also be observed to be curved close to the dendrite.
In the previous simulations with the moving window technique, only the lamellas directly adjacent to the dendrites were observed to grow at a different temperature.
This was assumed to have negligible effect on the structure as a whole but might be part of the structural influence leading to the observed refinement between eutectic and dendritic-eutectic structures.

%

\begin{figure}
\begin{subfigure}[b]{0.45\textwidth}
     \includegraphics[width=\textwidth]{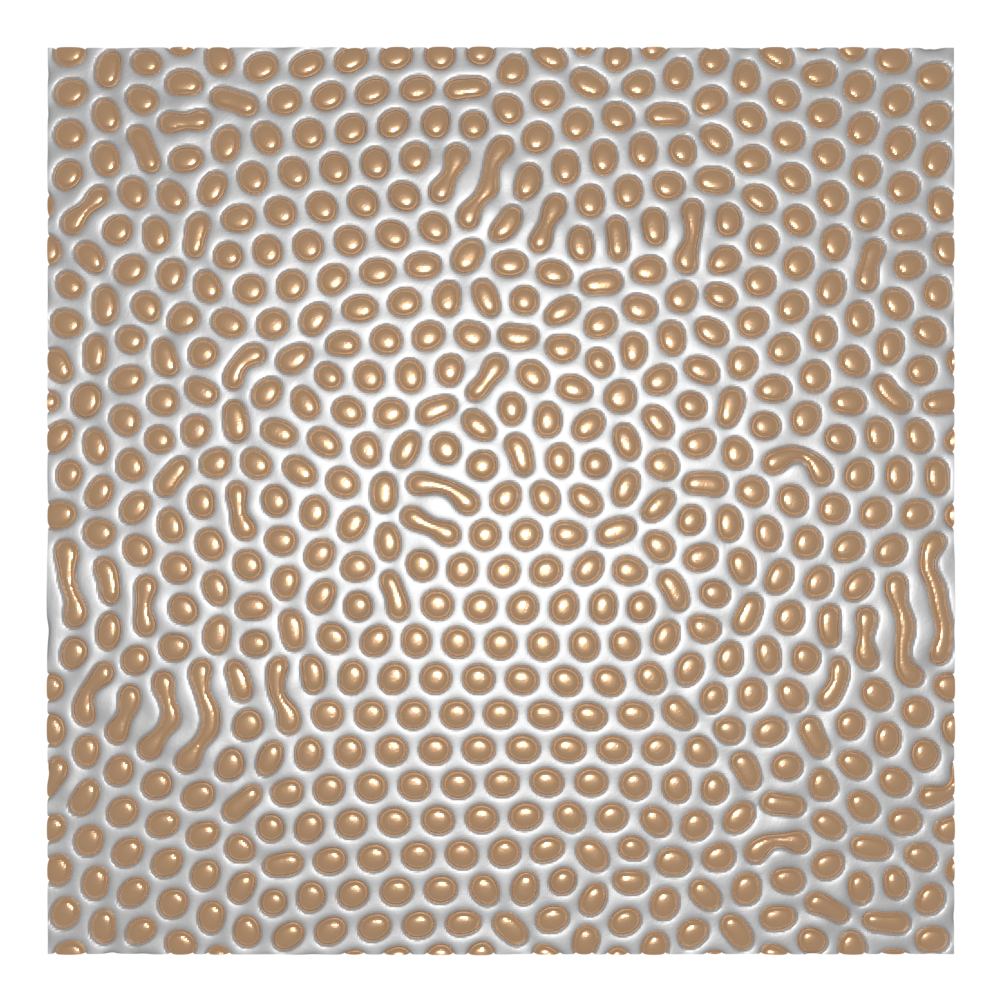}
    \caption{Starting from an eutectic seed with isotropic $\alpha$-Al and isotropic $\theta$ results in a matrix-fiber structure.}
    \label{fig:3dsim-iso}
  \end{subfigure}
  \begin{subfigure}[b]{0.45\textwidth}
    \includegraphics[width=\textwidth]{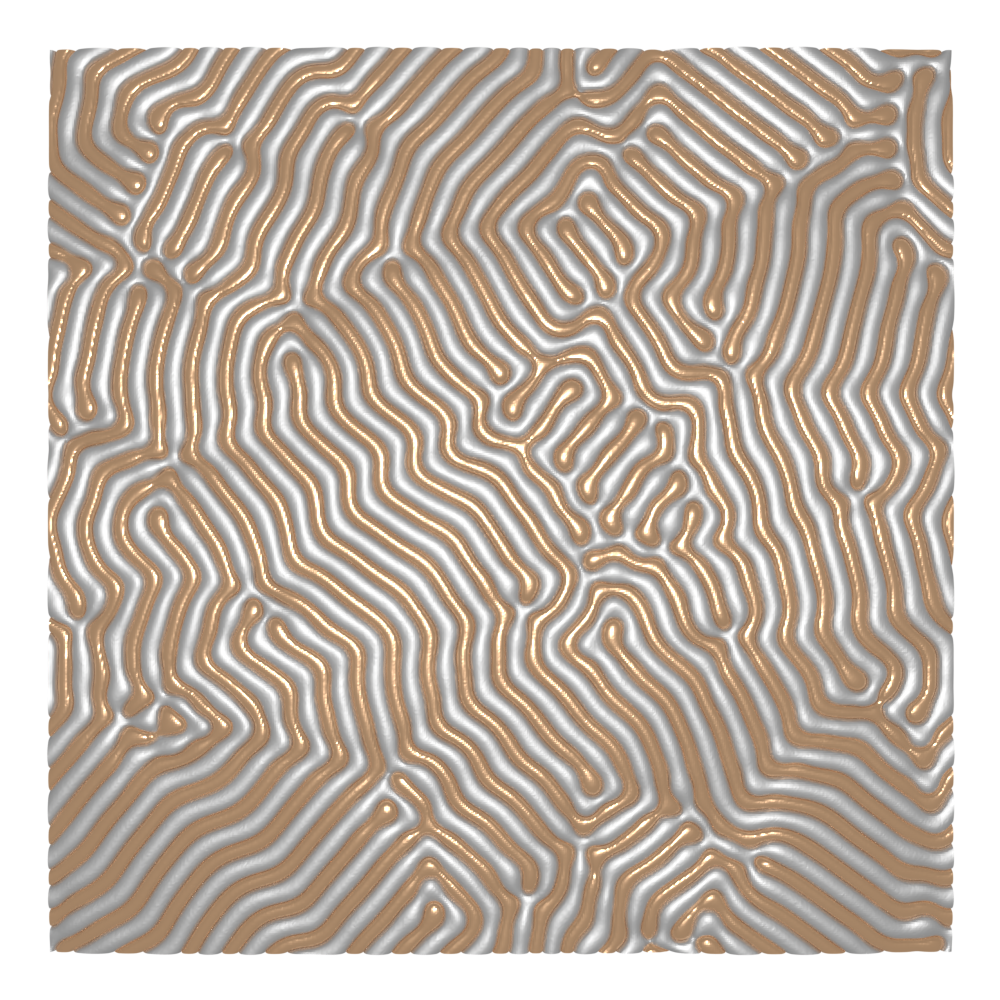}
    \caption{Starting from an eutectic seed with anisotropic $\alpha$-Al and isotropic $\theta$ results in a lamellar structure.}
    \label{fig:3dsim-aniso}
  \end{subfigure}
    \begin{subfigure}[b]{0.9\textwidth}
    \includegraphics[width=\textwidth]{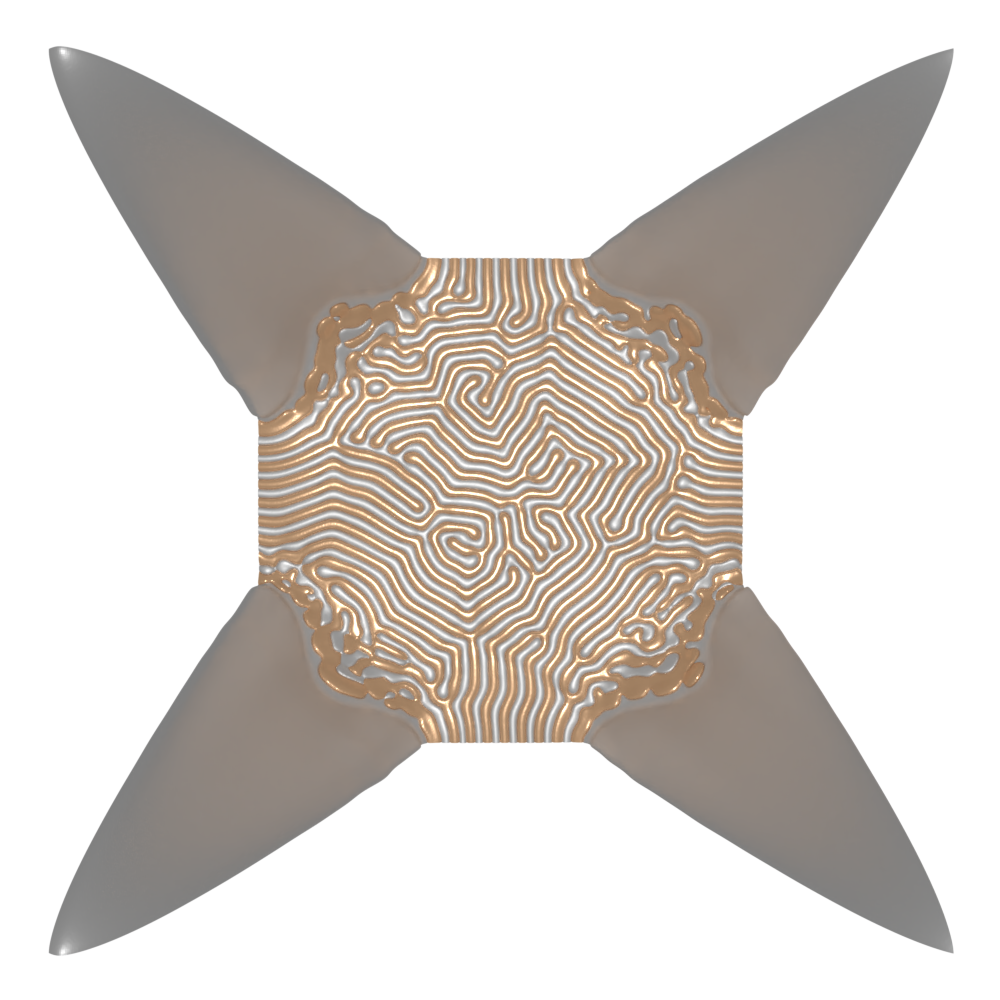}
    \caption{Starting from a dendritic seed and an eutectic seed with isotropic $\alpha$-Al and isotropic $\theta$ results in a lamellar eutectic being observed between the dendrite.}
    \label{fig:3dsim-dendr}
  \end{subfigure}
  \caption{Final states of 3D simulations, showing the distribution of solid phases in the entire domain.}
  \label{fig:3dsims}
\end{figure}

\paragraph{Eutectic morphology in 3D dendritic-eutectic growth}
Finally, the influence of the dendritic-eutectic growth on the eutectic morphology is investigated.
Since the two-dimensional simulations can only show lamellar eutectics, a set of three qualitative three-dimensional simulations is conducted.
The three simulations differ only in their initial conditions:
One starts with a Voronoi tesselation of the isotropic $\alpha$-Al and $\theta$ phases, the second with a Voronoi tesselation of the anisotropic $\alpha$-Al and the isotropic $\theta$ phases.
The last one starts with a periodic anisotropic $\alpha$-Al sphere as a dendrite seed together with a Voronoi tesselation of the isotropic $\alpha$-Al and $\theta$ phases as a eutectic seed.
With this, the effect of the anisotropy on the eutectic can be separated from that of the dendrite, as the morphological hysteresis will force the simulations without an initial dendrite seed into a purely eutectic structure.
The previous two-dimensional simulations were ran at a grid spacing $\Delta x$ of 1, which would lead to excessive computational effort in three dimensions.
Thus a grid spacing of $2$ is employed and the interfacial width is increased to $6$ to keep a diffuse profile.
These steps are taken to reduce the computational effort which will lead to mainly qualitative simulations.
The simulation box size is $700\times500\times500$ cells, corresponding to real dimensions of $\SI{140}{\um}\times\SI{100}{\um}\times\SI{100}{\um}$, with periodic boundary conditions on the basal plane, a no-flux condition on the bottom and a Dirichlet condition at the top.
The processing parameters are $v = \SI{160}{\um/s}$, $G = \SI{99}{K/mm}$ and $c_0 = 0.14$.
The composition is taken to be higher than would be expected to form a dendritic-eutectic structure, as three-dimensional dendrites grow more quickly at the same undercooling compared to their 2D counterparts, whereas a dimensional change has little effect on the eutectic.
%
The mass fractions at the composition $c_0$ are $\SI{60.9}{\percent}$ $\alpha$ and $\SI{39.1}{\percent}$ $\theta$, which suggests that both lamellar and $\alpha$-matrix-$\theta$-fiber structures should be found\cite{Parisi2010}.
The results for the two eutectic simulations are shown in \cref{fig:3dsim-iso,fig:3dsim-aniso}.
The $\alpha$ phase is represented as metallic silver, with the $\theta$ phase as metallic orange.
The isotropic eutectic shows a mostly matrix-fiber structure with a few small lamellas remaining.
However, the anisotropic variant shows only lamellas, as also observed by \cite{Ghosh2017}, although in the present case only one of the solid-liquid phases is anisotropic.
The mass fraction of $\alpha$-Al in the isotropic variant is $\SI{59.2}{\percent}$ and $\SI{60.0}{\percent}$ for the anisotropic variant.
While close to the  lever rule, the remaining difference is likely due to capillary and far-field effects as there is a significant composition gradient left in the system.
In \cref{fig:3dsim-dendr} the final state of the 3D simulation starting with a dendritic seed is shown.
During growth, $\theta$ is primarily nucleated in the concave parts of the dendrite.
As growth proceeds, these $\theta$ patches meet the main eutectic, forming new pairs of anisotropic $\alpha$-Al and isotropic $\theta$ lamellas.
These eventually overtake the isotropic eutectic seed, resulting in the observed lamellar structure.
The $\alpha$-Al mass fraction within the eutectic only is $\SI{53.0}{\percent}$ and thus significantly lower than for the eutectic morphologies.
It is likely that if an isotropic or a much more weakly anisotropic interface were present, this would cause a shift to a more lamellar morphology, instead of it being due to the anisotropic interface.
Furthermore, the mass fraction of $\theta$ is also enriched around the dendrite compared to the middle of the domain.
The average lamellar spacing can be roughly estimated by dividing the volume of the region of interest by the surface area of the lamellas.
The former is directly obtained by geometry, with the latter being related to the integral of the solid interphase boundary $\int_V \phi_\alpha \phi_\theta dV$.
This yields a spacing of \SI{3.49}{\um} for the dendritic-eutectic structure and a spacing of \SI{3.56}{\um} for the purely eutectic structure, which compares well with the two-dimensional eutectic spacing results at the same velocity \cor{($\lambda_{2D} \sim \SI{3.5}{\um}$)}.
The difference is even smaller than for the two-dimensional simulations and thus deemed to be insignificant.


\section{Conclusion}
In this work dendritic, eutectic as well as dendritic-eutectic growth are simulated.
This is achieved by combining a grand potential type of phase-field model with an empirical nucleation mechanism based on the local grand potential difference.
It is validated by showing that a eutectic system with nucleation yields a Jackson-Hunt curve close to that of a system without nucleation.
The dendritic growth is shown to qualitatively match an approximate undercooling model.
Based on both of these validations, an approximate boundary curve between dendritic-eutectic growth and eutectic growth is determined.
This curve is used to determine the processing conditions for simulations to show either dendritic-eutectic growth or pure eutectic growth.
In each case, the observed simulated microstructure is found to agree with the prediction of the boundary curve, with the undercooling-velocity relationship not being appreciably changed by dendritic-eutectic growth.
By analyzing the spacing of the eutectic in the dendritic-eutectic simulations, a slightly refined spacing relative to pure eutectic structure at the same speed is found.
Close to the $\alpha$ dendrite, the $\theta$ eutectic lamellas are found to be significantly thicker.
Going further, the stability of the dendritic-eutectic regime is investigated by employing velocity jumps.
When increasing the solidification speed of a dendritic-eutectic simulation, the eutectic regime is easily entered.
During this increase, the eutectic continuously refines its spacing.
Decreasing the speed back to the original value however does not yield a dendritic-eutectic structure, but rather only a coarsened eutectic with a spacing similar to that of the original dendritic-eutectic simulation.
Thus the spacing is not significantly affected by processing history, but the morphology is observed to depend on the prior processing history.
The effect of significant primary crystallization is investigated in another set of simulations.
Depending on the processing conditions, different dominant microstructures are observed:
In the case of a primarily eutectic structure, an initial primary dendrite is observed but eventually overgrown by the eutectic.
Within the eutectic structure travelling oscillations are observed which also can be found in experimental micrographs, with the entire front being slightly curved.
Reducing the velocity allows the dendrite to grow at a constant distance from the eutectic, resulting in coupled growth of both microstructures.
Oscillations within the eutectic still occur, but do not travel across the microstructure, and the curvature of the front is concentrated to regions close to the dendrite.
At the same time, side branching of the primary dendrite is suppressed.
In contrast to this, keeping the same velocity and decreasing the concentration of copper increases the distance between the primary dendrite tip and the eutectic significantly.
This leads to significant side branching and the formation of melt channels between the dendrites into which the eutectic grows afterwards.
It is observed that the eutectic does not grow as a uniform front but rather tends to nucleate anew in solute rich regions between side branches.
This implies that eutectics of different dominant orientation should be observed around dendrites, which would serve as an experimental test of the present nucleation mechanism.
Finally, qualitative 3D simulations showed that the eutectic morphology is strongly influenced by the presence of interfacial anisotropy.
For the same solidification conditions, isotropic interfaces yielded a fiber-matrix morphology, whereas if even one phase has a four-fold interfacial anisotropy, a lamellar structure is observed.
This extends to the dendritic-eutectic case, in which a lamellar structure between primary dendrites is observed.
While the lamellar spacing did not differ significantly between a 3D lamellar eutectic and the 3D dendritic-eutectic, the mass fractions of $\alpha$-Al and $\theta$ within the eutectic are observed to differ significantly.
Furthermore, the presence of the dendrite changes the spatial distribution of phase widths, with these differing significantly close to the dendrite compared to the bulk of the eutectic, suggesting significant spatial heterogeneity of properties if coupled dendritic-eutectic growth occurs.
In total this paper lays the groundwork for further investigations into solidification microstructures containing different kinds of morphologies evolving at different length scales.


\section*{Data availability \& supplementary material}
Video files of several simulations are available at \url{https://zenodo.org/record/7516370}.
The raw data required to reproduce these findings cannot be shared at this time as the data also forms part of an ongoing study.

\section*{Declaration of Competing Interest}
The authors declare that they have no known competing financial interests or personal relationships that could have appeared to influence the work reported in this paper.

\section*{Acknowledgements}
This work was partially performed on the national supercomputer Hawk at the High Performance Computing Center Stuttgart (HLRS) under the grant number pace3d.
The authors gratefully acknowledge financial support by the DFG under the grant number NE 822/31-1 (Gottfried-Wilhelm Leibniz prize), the Science Data Center ``MoMaF'', funded by the Ministry of Baden-Württemberg and the ``Future Field'' project ``ACDC'' of the strategy of excellence of the Karlsruhe Institute of Technology (KIT).
Special thanks goes to Johannes Hötzer for the helpful discussions and his support.

\section*{References}
\bibliography{literatur}

 



\end{document}